\documentclass[aps,prb,twocolumn,superscriptaddress,showpacs,floatfix,altaffilletter,longbibliography]{revtex4-2}
\usepackage{graphicx}
\usepackage{grffile}
\usepackage{amsmath}
\usepackage{amssymb}
\usepackage{bm}
\usepackage{color}
\usepackage[dvipsnames]{xcolor}
\usepackage{amsmath,amssymb,amsfonts}
\usepackage{epsfig}
\usepackage{times}
\usepackage[colorlinks,bookmarks=false,citecolor=blue,linkcolor=blue,urlcolor=blue]{hyperref}
\usepackage{dsfont}
\usepackage{multirow}
\usepackage{mathrsfs}

\usepackage[customcolors,shade]{hf-tikz}

\usepackage{colortbl}

\newcommand{\ket}[1]{\ensuremath{| #1 \rangle}}
\newcommand{\fref}[1]{Fig.~\ref{#1}}
\newcommand{\eref}[1]{Eq.~(\ref{#1})}
\newcommand{\sref}[1]{Sec.~\ref{#1}}
\newcommand{\tref}[1]{Table~\ref{#1}}

\usepackage{fixfoot}

\usepackage{mathtools}
\usepackage{pifont}

\begin{document}

\title{
    Superconductivity in Correlated Multi-Orbital Systems with Spin-Orbit Coupling: \\ Coexistence of Even- and Odd-Frequency Pairing and the Case of Strontium Ruthenate
}

\author{O.~Gingras}
\email{ogingras@flatironinstitute.org}
\affiliation{Center for Computational Quantum Physics, Flatiron Institute, 162 Fifth Avenue, New York, New York 10010, USA}
\affiliation{D\'epartement de Physique, Institut quantique, Universit\'e de Sherbrooke, Sherbrooke, Qu\'ebec J1K 2R1, Canada}
\affiliation{D\'epartement de Physique, Universit\'e de Montr\'eal, C. P. 6128, Succursale Centre-Ville, Montr\'eal, Qu\'ebec H3C 3J7, Canada}

\author{N.~Allaglo}
\affiliation{D\'epartement de Physique, Institut quantique, Universit\'e de Sherbrooke, Sherbrooke, Qu\'ebec J1K 2R1, Canada}

\author{R.~Nourafkan}
\affiliation{D\'epartement de Physique, Institut quantique, Universit\'e de Sherbrooke, Sherbrooke, Qu\'ebec J1K 2R1, Canada}

\author{M.~C\^ot\'e}
\affiliation{D\'epartement de Physique, Universit\'e de Montr\'eal, C. P. 6128, Succursale Centre-Ville, Montr\'eal, Qu\'ebec H3C 3J7, Canada}

\author{A.-M.~S.~Tremblay}
\affiliation{D\'epartement de Physique, Institut quantique, Universit\'e de Sherbrooke, Sherbrooke, Qu\'ebec J1K 2R1, Canada}

\date{\today}% It is always \today, today,

\begin{abstract}
The superconducting order parameter of strontium ruthenate is the center of a lasting puzzle calling for theoretical studies that include the seldom-considered effects of spin-orbit coupling and the frequency-dependence of the order parameters.
Here we generalize the frequency-dependent theory of superconductivity mediated by spin and charge fluctuations to include spin-orbit coupling in multi-orbital systems and we characterize the superconducting states using the spin-parity-orbital-time $SPOT$ quantum numbers, group theory, and phase distributions in the complex plane.
We derive a pseudospin formulation that maps the inter-pseudospin sector of the normal state Eliashberg equation to a pseudospin-diagonal one. 
Possible superconducting order parameters for strontium ruthenate are obtained starting from a realistic density-functional-theory normal state.
We find that spin-orbit coupling leads to ubiquitous entanglement of spin and orbital quantum numbers, along with notable mixing between even- and odd-frequency correlations.
We propose a phase diagram obtained from the temperature dependence of the leading and subleading symmetries in the pseudospin-orbital basis.
An accidental degeneracy between leading inter-pseudospin symmetries in strontium ruthenate, B$_{1g}^+$ and A$_{2g}^-$, could resolve apparent experimental contradictions.

% The symmetry of the superconducting order parameters in multi-orbital systems involves many quantum numbers.
% Pairing mediated by electronic correlations being retarded, the frequency structure of such an order parameter bears important information.
% Here we generalize the frequency-dependent theory of superconductivity mediated by spin and charge fluctuations to systems with spin-orbit coupling.
% Taking advantage of pseudospin and inversion symmetries, the inter-pseudospin sector of the normal state Eliashberg equation is mapped to a pseudospin-diagonal one. 
% This formulation is applied to strontium ruthenate, with a normal state obtained using density functional theory.
% The superconducting order parameter of this system is the center of a lasting puzzle which calls for theoretical studies that include these seldom-considered effects.
% We find that spin-orbit coupling leads to ubiquitous entanglement of spin and orbital quantum numbers, along with notable mixing between even- and odd-frequency correlations.
% We present the phase diagrams for leading and subleading symmetries in the pseudospin-orbital basis.
% The states are characterized using $SPOT$ contributions, group theory, phase distributions in the complex plane and temperature dependence.
% An accidental degeneracy between leading inter-pseudospin symmetries in strontium ruthenate, B$_{1g}^+$ and A$_{2g}^-$, could resolve apparent experimental contradictions.
\end{abstract}

\maketitle

\section{Introduction}
For several decades, the paradigm of $s$-wave superconductivity has been found inadequate to describe superconductivity in correlated systems.
Instead, these materials host a rich variety of superconducting order parameters (SCOP).
In these so-called unconventional superconductors, many SCOPs have already been identified and understood as highly dependent on the underlying electronic structure and on electronic interactions.
For example, the SCOP in high-temperature cuprate superconductors has been shown to have $d$-wave, or more accurately, B$_{1g}$ symmetry, emerging from the interplay between the electronic kinetic energy and single orbital Coulomb repulsion~\cite{Van_Harlingen_d_wave,scalapino_case_1995,scalapino_superconductivity_1999, Maier_Jarrell_2005, kyung_pairing_2009, armitage_progress_2010,Millis_Parcollet_2013, Fratino_Semon_Sordi_Tremblay_2016,Kowalski_Dash_Semon_Senechal_Tremblay_2021}.
In Hund's metals with $3d$ electrons such as the iron-based superconductors, the local Coulomb repulsion at the Fermi level acts on multiple orbitals~\cite{stewart_superconductivity_2011}.
Pnictides can exhibit $s^\pm$-wave, an A$_{1g}$ symmetry with a sign change between different Fermi surfaces~\cite{wu_boundary-obstructed_2020}.

Usually, one would think that the symmetry of a system's SCOP should be identified rather quickly.
It is thus surprising that, after several decades of work leading to remarkable progress of experimental probes and numerical methods, the symmetry of the SCOP of strontium ruthenate (Sr$_2$RuO$_4$, SRO) has not yet been unambiguously established~\cite{mackenzie_even_2017, huang_review_2021}.
The reason is that certain measurements appear contradictory.
This is not only an experimental challenge but also one for theories of strong electronic correlations in $4d$ / $5d$ multi-orbital system with important spin-orbit coupling (SOC).
A large variety of symmetries are possible in this materials, thus making the prediction of SCOPs more challenging~\cite{kaba_group-theoretical_2019, PhysRevB.100.134506}.

In SRO, initially reckoned a spin-triplet state due to its constant Knight-shift~\cite{ishida_spin-triplet_1998, mackenzie_superconductivity_2003}, independent verifications have highlighted a heating effect so that a dominantly spin-singlet state appears more credible~\cite{pustogow_constraints_2019, ishida_reduction_2020, chronister_evidence_2021}.
Another experiment probing spins using polarized neutrons met a similar fate~\cite{duffy_polarized-neutron_2000, petsch_reduction_2020}.
Previously in contradiction with evidences for Pauli limiting~\cite{mackenzie_even_2017, steppke_strong_2017}, these experiments now agree. 

Another critical characteristic of SRO is its two-component nature inferred by evidences of time-reversal (TR) symmetry breaking~\cite{luke_time-reversal_1998, xia_high_2006}.
Ultrasound experiments also support a two-component SCOP that couples to the B$_{2g}$ shear mode~\cite{benhabib_ultrasound_2021, ghosh_thermodynamic_2021}.
Additionally, the enhancement of the critical temperature under uniaxial pressure~\cite{steppke_strong_2017} not only provides strong evidence for an even-parity (\textit{e-p}) SCOP~\cite{sunko_direct_2019}, it is also a useful knob to study this two-component property. 
Indeed, muon spin relaxation ($\mu$SR) measurements observed two transition temperatures under pressure, indicative of a lifted degeneracy between the two components~\cite{grinenko_split_2021}.
Surprisingly however, specific heat measurements, known to be extremely sensitive to superconducting transitions, have detected only a single transition temperature~\cite{li_high-sensitivity_2021}.

Consequently, various theoretical proposals have been formulated in replacement to the initial chiral $p$-wave state~\cite{Kallin_2009}, including domain-wall physics and inhomogeneities~\cite{ghosh2021strong, Schmalian_Inhomogeneous_2021}.
A two-component character can be realized in two ways.
First, the components can be degenerate by symmetry if the SCOP transforms like a two-dimensional (2D) irreducible representation (irrep) of the $D_{4h}$ point group.
The only \textit{e-p} such possibility is the E$_g$ irrep.
One such proposed state is the $d_{xz}+id_{yz}$ which could originate from momentum-dependent $\textbf{k}$-SOC~\cite{suh_stabilizing_2020, clepkens_shadowed_2021}.
However, density-functional theory (DFT) expects this coupling to be negligibly small in SRO, known to have a quasi-2D character~\cite{hussey_normal-state_1998, PhysRevLett.101.026406}.
Moreover, these E$_g$ states under uniaxial stress should generate two transitions in specific heat.

Another possibility for two components is that they are degenerate by accident and transform like different irreps.
The most natural of the two components is a $d_{x^2-y^2}$ B$_{1g}$ state, since thermal conductivity and scanning tunneling microscopy point in this direction~\cite{hassinger_vertical_2017, sharma_momentum-resolved_2020} and it should originate from antiferromagnetic fluctuations predicted by DFT~\cite{gingras_superconducting_2019} and from the strong spin fluctuations caused by the nesting of the quasi-one-dimensional bands~\cite{sidis_evidence_1999}.
Such a symmetry was well studied in the context of the cuprates~\cite{scalapino_case_1995,Van_Harlingen_d_wave}.
For the second component, some works have proposed an extended $s$-wave~\cite{raghu_theory_2013, romer_knight_2019, romer_fluctuation-driven_2020} or odd-parity (\textit{o-p}) states~\cite{PhysRevB.89.220510, PhysRevLett.121.157002, PhysRevResearch.1.033108,  scaffidi_degeneracy_2020, PhysRevResearch.3.L042002} also originating from this strong nesting vector.
Unfortunately, these combination would not couple to the B$_{2g}$ shear mode.
Other works proposed $g_{xy(x^2-y^2)}$ A$_{2g}$, a higher angular momentum version of $d_{x^2-y^2}$~\cite{kivelson_proposal_2020, yuan_strain-induced_2021, clepkens_higher_2021, Wagner_GL_2021}.
The similar nodal structures of $d_{x^2-y^2}$ and $g_{xy(x^2-y^2)}$ could reduce the signature on specific heat, but not remove it entirely.
It has been proposed theoretically that this accidental degeneracy is more consistent with ultrasound experiments than the other symmetry-protected $d_{xz}+id_{yz}$ proposal~\cite{sheng2021superconducting}. 
Moving away from the $d_{x^2-y^2}$ state, a $d_{xy}\pm i s^*$ state was proposed also proposed~\cite{romer_superconducting_2021,bhattacharyya2021superconducting}.
However, accidental degeneracies should be lifted by small perturbations.
Although not definitive, $\mu$SR measurements under isotropic conditions did not observe a split in the critical temperatures~\cite{grinenko_unsplit_2021}.
Additionally, disorder by non-magnetic impurities could help split the transition temperatures~\cite{zinkl_impurity-induced_2021}.

These seemingly inconsistent results suggest that the set of SCOP considered up to now might be incomplete. 
To guide potential SCOP candidates, we need to first construct an accurate normal state electronic structure on which to build many-body correlations that should generate a broad richness of possible superconducting states in SRO.
The dominant mechanism explaining its superconductivity is pairing mediated by the exchange of spin and charge fluctuations~\cite{Bourbonnais_1986,Scalapino_Hirsch_1986,scalapino_case_1995, esirgen_fluctuation_1998}. The Bardeen-Cooper-Schrieffer~\cite{bardeen_theory_1957} and Ginzburg-Landau theories~\cite{ginzburg_phenomenological_1950} neglect the frequency dependence of this interaction, or replace it by a single even in frequency sharp cutoff at the Debye frequency. 
Doing so constrains the Gorkov function $\left< T_\tau \psi (\tau) \psi \right>$ to be purely even in imaginary time and correspondingly in frequency (even-frequency). 
These approaches hide not only the frequency structure of possible SCOPs but also neglect a whole set of superconducting states that are purely odd in fermionic frequency (odd-frequency)~\cite{linder_odd-frequency_2019}. This type of superconductivity was first proposed by Berezinskii in the context of $^3$He~\cite{berezinskii_new_1974} and then in the strong coupling limit of the electron-phonon interaction~\cite{balatsky_new_1992, abrahams_interactions_1993, abrahams_properties_1995}.
It is now considered ubiquitous at interfaces~\cite{PhysRevB.90.224519, di_bernardo_signature_2015,krieger_proximity-induced_2020}, impurity sites~\cite{perrin_unveiling_2020} and in some multi-orbital systems~\cite{taylor_intrinsic_2012,black-schaffer_odd-frequency_2013,triola_role_2019}.
The finite Kerr effect observed in SRO was linked with the presence of odd-frequency correlations~\cite{komendova_odd-frequency_2017}.
In previous works~\cite{gingras_superconducting_2019, kaser_inter-orbital_2021}, pure odd-frequency solutions were found when examining leading superconducting instabilities of SRO without SOC.
These seldom-considered odd-frequency pairing correlations could help understand the ambiguous behavior of SRO and we are interested to study how they arise in the presence of SOC.

In this paper, we generalize the frequency-dependent formulation of superconductivity mediated by spin and charge fluctuations in multi-orbital systems~\cite{esirgen_fluctuation_1998, bickers_self-consistent_2004, nourafkan_nodal_2016, gingras_superconducting_2019} to include SOC. Although the approach developed is general, in the present study we apply it solely to the SRO case. 

We present in \sref{sec:normal-state} results from density functional theory (DFT) after a qualitative discussion of the electronic structure of SRO. We explain the effects of SOC and how a pseudospin symmetry is respected for the $t_{2g}$ orbitals in the quasi-2D approximation. 
In \sref{sec:spin-fluctu}, frequency-dependent two-body vertex corrections are generalized to incorporate SOC. In the particle-hole (\textit{p-h}) channel, vertex corrections are quantified at the level of the random phase approximation (RPA) using Stoner factors. 
We study the effect of these \textit{p-h} fluctuations on the normal state \textit{p-p} scattering through what we refer to as the normal state linearized Eliashberg equations. 
We explain how pseudospin and inversion symmetries simplify numerical calculations to the same level of numerical complexity as in the absence of SOC.

In \sref{sec:supercond}, we present tools to characterize multi-spin-orbital frequency-dependent Gorkov functions.
First, $SPOT$ decomposition~\cite{linder_odd-frequency_2019} allows to quantify the mixing (entanglement) of spin, momentum, orbital and imaginary-time quantum numbers. Second, with group theory we classify SCOPs by irreps of the normal state's space group along with time-reversal (TR) symmetry. 
In particular, we tabulate the irreps describing how the spin-orbital components of the Cooper pairs transform in $t_{2g}$ systems.

In \sref{sec:supercond_lead}, we present the resulting phase diagram of the leading and subleading gap function symmetries that can emerge from the normal state, depending on the interactions.
We find that SOC couples multiple spin-orbital sectors within a specific irrep and naturally intricates $SPOT$ combinations. 
This mixing induces phases differences between components of the gap function but preserves TR symmetry.
As leading symmetries, we find two states with co-existing $SPOT$ contributions in all spin-orbital sectors. One is a B$_{1g}^+$ state whose intra-orbital components are purely even-frequency, compatible with many proposed solutions for SRO. The other is an A$_{2g}^-$ state whose intra-orbital components are purely odd-frequency.
This state is a SOC-generalization of the odd-frequency state found in Ref.~\citenum{gingras_superconducting_2019} when neglecting SOC.
The weighted distributions of the phases of these leading candidates are studied and suggest a complex interference between the different $SPOT$ contributions.
We also study the temperature dependence of a few leading eigenvalues at two points of the phase diagram. Some symmetries exhibit non-monotonous behaviors.
As subleading states, we find other states that have odd-frequency intra-orbital character: an odd-parity E$_u^-$ $p$-wave-like, another A$_{2g}^-$ and an A$_{1g}^-$ state.

\section{\label{sec:normal-state}Normal state electronic structure}
The normal state's electronic structure is essential to understand the emergence of an unconventional superconducting state. Similarly to the cuprates, the dominant interactions at the Fermi level of body-centered tetragonal SRO involve the quasi-2D ruthenium-oxide planes~\cite{hussey_normal-state_1998, PhysRevLett.101.026406}.
In \sref{sec:normal-state_atom}, we discuss the ruthenium (Ru) atomic problem, which has partially filled $4d$ electrons.
Their localized nature leads to sizeable multi-orbital Coulomb repulsion that can generate the potential mediators of superconductivity. 
%We use the rotationally invariant Kanamori-Slater model characterized by the Hubbard $U$ parameter and the Hund's coupling $J$. 
Furthermore, the large atomic number of these atoms implies strong local SOC.
Spin is no longer a conserved quantity but pseudospin is.
In \sref{sec:normal-state_pseudospin}, we introduce the lattice.
The Ru atom is in the center of a tetragonally elongated octahedron of oxygen (O) atoms.
The resulting crystal field $\Delta_{\text{CF}}$ splits the Ru-$4d$ electrons into the $t_{2g}$ and the $e_g$ subsets.
Omitting the unoccupied $e_g$ states, we model local interactions on a DFT Hamiltonian downfolded to the $t_{2g}$ orbitals using projectors.
In this Hamiltonian, the effect of SOC on the O atoms effectively act as $\textbf{k}$-dependent SOC.
In the quasi-2D approximation, the pseudospin basis stays diagonal.
Details and orbital characters of the resulting Fermi surfaces are presented in \sref{sec:normal-state_realistic}.

\subsection{\label{sec:normal-state_atom}Atomic problem.}
In strongly correlated unconventional superconductors, Cooper pairs emerge from strong electron-electron interactions.
The simplest interacting model showcasing this property is the one-band Hubbard Hamiltonian~\cite{hubbard1963electron, hubbard1964electron2, hubbard1964electron3,Kanamori:1963,Gutzwiller:1963}. 
Its multi-orbital generalization is the Kanamori-Slater Hamiltonian (KSM)~\cite{georges_strong_2013}.
Its rotationally invariant formulation is detailed in \sref{sec:kanamori} and its expression given in \eref{eq:kanamori-ham}.
It depends on two parameters: on-site Coulomb repulsion $U$ and Hund's coupling $J$. 
In SRO, works based on realistic electronic structures have shown that strong electronic correlations improve considerably the quantitative description of the one- and two-body propagators characterizing its normal state~\cite{mravlje_coherence-incoherence_2011,zhang_fermi_2016,kim_spin-orbit_2018, tamai_high-resolution_2019, strand_magnetic_2019, kugler_strongly_2020}.
In this work, we include strong electronic correlations only at the two-body level through spin and charge fluctuation theory described in \sref{sec:spin-fluctu} and detailed in Appendices~\ref{sec:free-energy} and \ref{sec:spin-charge_vertices}.

\begin{figure}
    \centering
    \includegraphics[width=\linewidth]{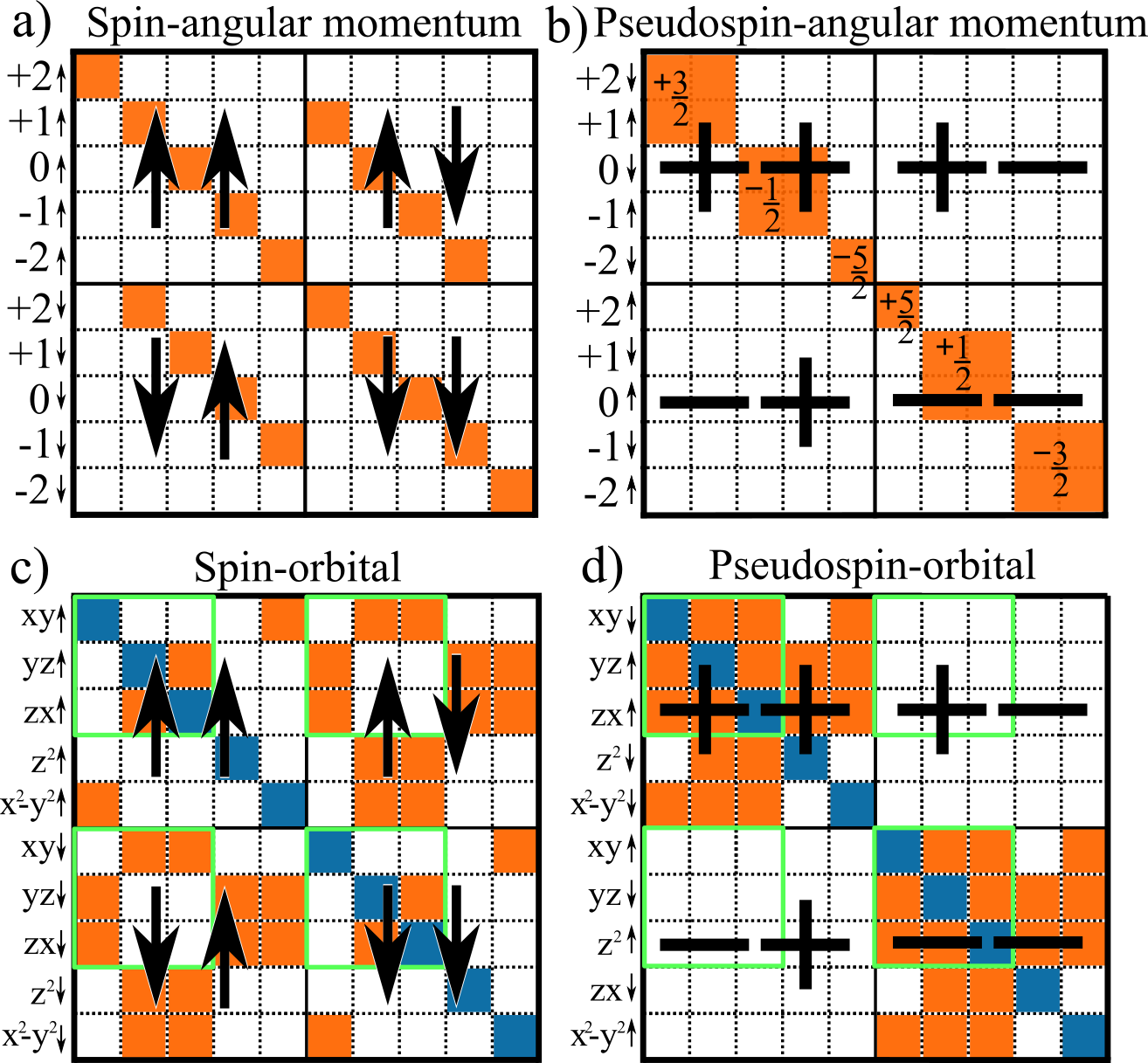}
    \caption{(Color online) Non-zero components of the Hamiltonian projected on the Ru-$4d$ shell in different basis sets, that is the a) spin-angular momentum, b) pseudospin-angular momentum with the $J_z$ labels, c) spin-orbital and d) pseudospin-orbital basis. All basis include atomic SOC components in orange (grey) and the effects of the crystal field  are highlighted in blue (dark grey) for the orbital basis. The lime (light grey) squares highlight the $t_{2g}$ subset in the orbital basis.}
    \label{fig:ham_soc_full}
\end{figure}

Another key physical mechanism affecting the Ru-$4d$ electrons is SOC~\cite{PhysRevLett.112.127002}.
This relativistic effect induced by the important electrostatic potential generated by Ru's large nucleus couples the electronic spin $\bm S$ and angular momentum $\bm L$ together.
When acting on Ru-$4d$ states, these operators respectively have norms $l=2$ and $s=\frac{1}{2}$.
We use a basis that is diagonal in the projections $L_z$ and $S_z$ along the $\hat{z}$-axis. 
The atomic SOC is purely local and thus transforms like the A$_{1g}$ irrep of the $D_{4h}$ space group.
It has the form $H_{\text{SOC}}^{\text{A}_{1g}} = \lambda^{\text{A}_{1g}}_{\text{SOC}} \bm L \cdot \bm S$ which can be recast into ladder operators that couple states with quantum numbers $L_z = l_z$ and $S_z = \frac{1}{2}$ with those having $L_z = l_z+1$ and $S_z = -\frac{1}{2}$.
This coupling is thus block tridiagonal in spin-angular momentum space as represented in \fref{fig:ham_soc_full}~a).
Consequently, the electrons are better described by the total angular momentum quantum operator $\bm J = \bm L + \bm S$ with the eigenstates's quantum numbers $\ket{j, J_z}$ where $j$ is a half-integer between $\frac{1}{2}$ and $2+\frac{1}{2}$ and $J_z$ can go from $-j$ to $j$.
As a result, the ten-fold degeneracy of $4d$ electrons with $l, s = 2, \frac{1}{2}$ is broken into a four- and a six-fold subsets respectively characterized by the total angular momenta $j=\frac{3}{2}$ and $\frac{5}{2}$. \fref{fig:ham_soc_full}~b) shows which spin-angular momentum states are necessary to combine to construct a state with a given $J_z$.

\begin{figure*}[btp]
    \centering
    \includegraphics[width=\linewidth]{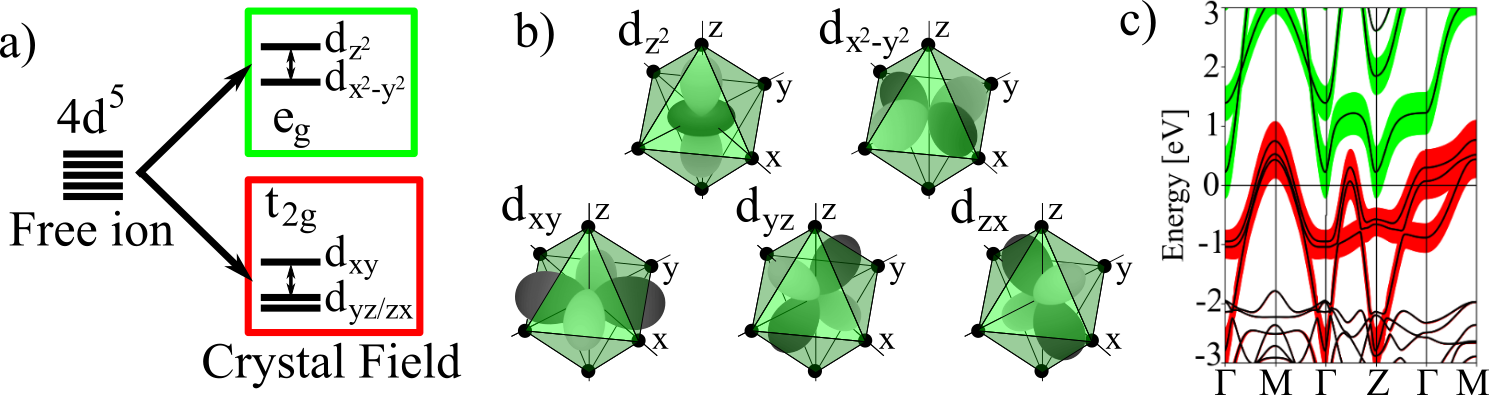}
    \caption{(Color online) Effect of the crystal field generated by the octahedron of oxygen atoms around the Ru-$4d$ electrons. a) Cartoon of the splittings. In a perfect octrahedral environment, the t$_{2g}$ and e$_g$ electrons are split. Including the elongation of the octrahedron generates further splittings within these subsets. b) Real spherical harmonics associated with the eigenstates of the crystal field potential. c) Projection of a realistic electronic band structure on the t$_{2g}$ states in red (dark grey) and the e$_g$ states in green (light grey). There are no e$_g$ states at the Fermi level, which is shifted to 0 in this figure. The labels of the high-symmetry points are shown in \fref{fig:DFT_FS}.}
    \label{fig:crystal_field}
\end{figure*}

\subsection{Crystal field and pseudospin basis.}\label{sec:normal-state_pseudospin}
The crystal structure in which the atoms are embedded breaks the spherical symmetry of the atomic limit.
In many correlated superconductors, the solutions to the local crystal field are real spherical harmonics also called orbitals.
In SRO, the space group is $D_{4h}$ and the Ru atom sits in the center of a tetragonal octahedron of O atoms.
The hybridization between Ru-$4d$ and O-$p$ electrons separates the $4d$-shell into the t$_{2g}$ and e$_g$ subsets, depicted in \fref{fig:crystal_field}~a).
The elongated nature of the octahedron introduces further splittings within these subsets.

The t$_{2g}$ subset includes a quasi-2D $d_{xy}$ orbital along with two symmetry related quasi-one-dimensional $d_{yz}$ and $d_{zx}$ orbitals while $d_{z^2}$ and $d_{x^2-y^2}$ form the e$_g$ subset.
These orbitals are shown in \fref{fig:crystal_field}~b) and the transformation from the angular momenta to the orbitals is given by
\begin{align}
    \nonumber 
    \ket{z^2} = \ket{l_z = 0}
    , \quad
    \begin{array}{c}
        \ket{yz} \\ \ket{zx}
    \end{array} = 
    \left[ \begin{array}{cc}
        \frac{i}{\sqrt{2}} & \frac{i}{\sqrt{2}} \\ \frac{i}{\sqrt{2}} & \frac{-i}{\sqrt{2}}
    \end{array} \right]
    \begin{array}{c}
        \ket{l_z = -1} \\ \ket{l_z = +1}
    \end{array} \\
    \begin{array}{c}
        \ket{xy} \\ \ket{x^2-y^2}
    \end{array} = 
    \left[ \begin{array}{cc}
        \frac{1}{\sqrt{2}} & \frac{-1}{\sqrt{2}} \\ \frac{1}{\sqrt{2}} & \frac{1}{\sqrt{2}}
    \end{array} \right]
    \begin{array}{c}
        \ket{l_z = -2} \\ \ket{l_z = +2}
    \end{array}. \quad \quad \label{eq:crystal_field}
\end{align}
The non-vanishing crystal field and SOC components of the Hamiltonian in the spin-orbital basis are depicted in \fref{fig:ham_soc_full}~c).

The orbitals diagonalize the CF Hamiltonian but not $H_{\text{SOC}}^{\text{A}_{1g}}$.
Using the total angular momentum instead, it can be diagonalized into two blocks, as shown in \fref{fig:ham_soc_full}~d).
As one can see, the pseudospin is an orbital dependent spin defined as
% \begin{equation}
%     \rho = \pm \quad \text{for} \ J_z \in \bigcup_{i=0}^l \left(-l\pm \frac{1}{2} + 2i\right).
% \end{equation}
\begin{align}
    \rho = \pm \quad \text{for} \ J_z \in \left\{ \pm \frac{1}{2}, \mp \frac{3}{2}, \pm \frac{5}{2} \right\}.
\end{align}

Although predominantly local, SOC can have additional contributions to the atomic part.
For example, an electron from the Ru-$4d$ shell could hop to a neighbouring O, flip its spin because of SOC and hop back with a different spin.
Thus, when downfolding to the Ru-$t_{2g}$ orbitals, this SOC process appears as being momentum dependent and is known as $\textbf{k}$-SOC.
In Ref.~\citenum{clepkens_shadowed_2021}, they find that different hoping sequences lead to three such effective couplings: the intra-layer $H_{\text{SOC}}^{\text{B}_{1g}}$ and $H_{\text{SOC}}^{\text{B}_{2g}}$, along with the inter-layer $H_{\text{SOC}}^{\text{E}_g}$ terms.
In the $t_{2g}$ subset, $H_{\text{SOC}}^{\text{B}_{1g}}$ and $H_{\text{SOC}}^{\text{B}_{2g}}$ connect electrons on the $xy$ orbital having a spin $\sigma$ with electrons on the $yz$ and $zx$ orbitals having a spin $-\sigma$.
These contributions are intra-pseudospin.
$H_{\text{SOC}}^{\text{E}_g}$, on the other hand, is inter-pseudospin since it connects an electron in the $xy$ orbital with one in the $yz$ and $zy$ orbitals without affecting its spin.
Fortunately, SOC associated to O-$p$ electrons is small, as is also inter-layer hoping in SRO.
Consequently, DFT predicts this inter-pseudospin potential to be negligibly small~\cite{clepkens_shadowed_2021}, as we find.
In the present paper, we use the quasi-2D character of SRO to simulate the k$_z=0$ and k$_z=2\pi/c$ planes for which the $H_{\text{SOC}}^{\text{E}_g}$ coupling completely vanishes.
Thus the non-interacting Hamiltonian is block diagonal, a useful property to simplify the problem.

\begin{figure*}
    \includegraphics[width=\linewidth]{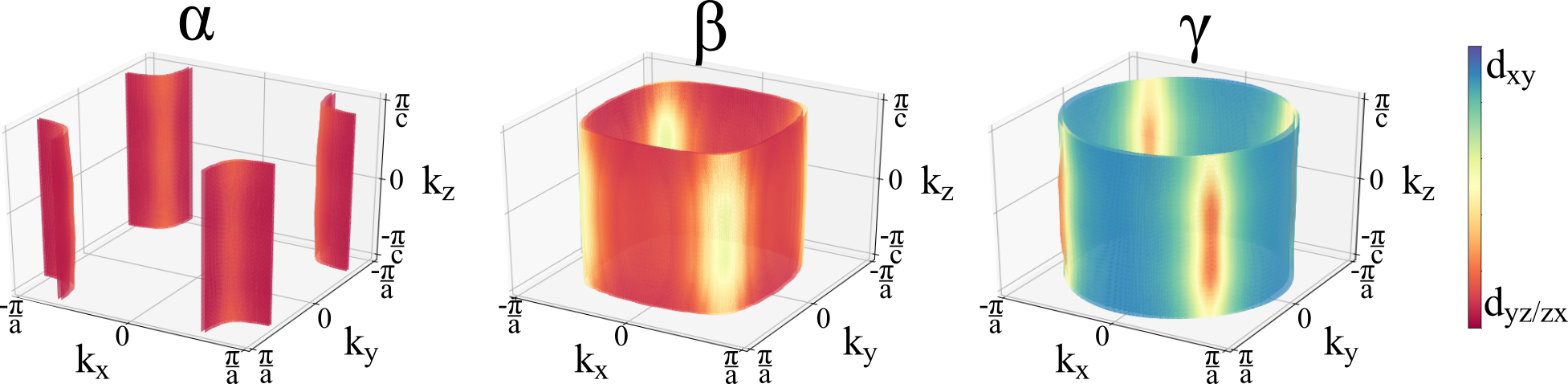}
    \caption{\label{fig:orb_char}
    Orbital character projected on the $\alpha$, $\beta$ and $\gamma$ Fermi sheets as also calculated in Ref.~\citenum{PhysRevLett.112.127002}. 
    The blue color corresponds to $xy$ orbital character while the red color corresponds to $yz$ and $xz$ orbital characters without distinction between these two. }
\end{figure*}

\subsection{\label{sec:normal-state_realistic}Realistic electronic structure.}
In spin and charge fluctuation theory, the superconductivity is mediated through partially filled orbitals of the correlated atoms only.
Devising a tight-binding model of the appropriate set of local orbitals requires to fit many parameters.
Instead, we start from a realistic electronic structure that we downfold onto the appropriate set of local orbitals.
We use DFT from the ABINIT package~\cite{gonze_abinit_2020, romero_abinit_2020, amadon_plane-wave_2008} with the projected augmented wavefunction pseudopotentials~\cite{blochl_projector_1994, torrent_implementation_2008}.

In $4d$ electron systems, the crystal-filed splitting is sufficiently larger than SOC such that the latter can be considered as a perturbation with respect to the former.
In SRO, there are only four electrons that partially fill the $4d$-$t_{2g}$ shell of the Ru atom and the $4d$-$e_g$ are states too far above the Fermi level to mix with the $t_{2g}$ orbitals.
This is confirmed by looking at the orbital character of the band structure obtained using DFT as shown in \fref{fig:crystal_field}~c).
Thus we omit the $e_g$ states and project the Hamiltonian on the $t_{2g}$ subset only.

The large SOC on the Ru atom couples the $xy$ with the $xz$ and $yz$ orbitals with opposite spins.
Without SOC, spin conservation makes the normal-state Hamiltonian diagonal in spins while crystal-field symmetry preserves the block diagonal form of the $xz,yz$ sector.
With SOC, the spin and orbital sectors are coupled~\cite{PhysRevLett.112.127002}.
The intra-pseudospin block of the normal-state Hamiltonian shown in \fref{fig:ham_soc_full}~d) would be diagonal in the band basis, but we work in the orbital basis since it is the natural choice to study spin and charge fluctuations originating from local interactions.
The colors in \fref{fig:orb_char} represent the orbital characters of the resulting Fermi sheets in the first Brillouin zone.
The $\alpha$ and $\beta$ sheets mainly comes from quasi-one-dimensional bands with $xz$ and $yz$ orbital characters, while the $\gamma$ sheet is mostly a quasi-2D band with $xy$ orbital character.
The color code clearly shows that SOC introduces spin-orbital entanglement around the k$_x = \pm \text{k}_y$ diagonals, most obvious there because the bands are degenerate at these points in the absence of SOC.

\begin{figure}[b]
    \centering
    \includegraphics[width=\linewidth]{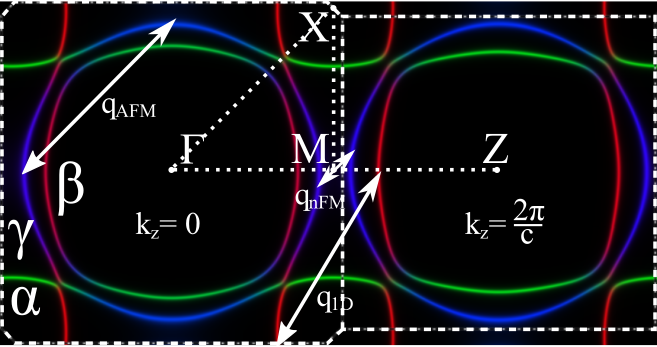}
    \caption{Spectral function in the orbital basis of the non-interacting system obtained by DFT with SOC for in-plane momenta k$_z=0$ and k$_z=\frac{2\pi}{c}$.
    The resulting $\alpha$, $\beta$ and $\gamma$ bands form the Fermi surface.
    The $xy$, $yz$ and $zx$ orbitals of the Ru atom are projected on the Fermi surface with orbital character identified by the blue, green and red colors respectively.
    High-symmetry points are label and connected by dots to show a specific path.
    Dashed lines highlight the Brillouin zone, characterized by its body-centered tetragonal nature.
    The arrows label different dominant nesting vectors of the Fermi surface defined in \fref{fig:barechi_ph_pp}~a).}
    \label{fig:DFT_FS}
\end{figure}

Similarly, \fref{fig:DFT_FS} shows the projection of the $xy, yz, zx$ orbital characters on the $\alpha$, $\beta$ and $\gamma$ bands at the Fermi surface for cuts k$_z=0$ and k$_z=2\pi/c$.
These $k_z$ momenta are connected by in-plane vectors because of the body-centered nature of the $D_{4h}$ space group.
We use the downfolded spin-orbital Hamiltonian to construct $\bm G_{Kl_1l_2}^{\rho}$, the Green functions that characterize the propagation of a normal state's quasi-particle with pseudospin $\rho$ from orbital $l_1$ to $l_2$ with energy-momentum $K \equiv (i\omega_m, \textbf{k})$.
This model Green function is connected to the general Green function in \sref{sec:one-p-G_conn} and its properties are given in \sref{sec:one-p-G_prop}.

Now starting from the normal state, we study instabilities towards ordered states.
In SRO, the Fermi liquid (FL) state preceding the superconducting state appears below $T_{\text{FL}} \sim 25$~K~\cite{hussey_normal-state_1998, bergemann_normal_2001}.
By using DFT, we have a well defined Fermi liquid which allows us to study the role of SOC but neglects strong electronic correlations at the one-body level.
Although various works have shown their importance for SRO~\cite{mravlje_coherence-incoherence_2011, zhang_fermi_2016, kim_spin-orbit_2018, tamai_high-resolution_2019, strand_magnetic_2019, kugler_strongly_2020}, none of them was able to correctly account for electronic correlations, SOC and temperatures below $T_{\text{FL}}$ at the same time.
The exactitude of SRO's normal state in its Fermi liquid regime is an ongoing challenge.
In this context, we decided to focus on generalizing spin and charge fluctuation mediated superconductivity to multi-orbital systems with SOC.
This formalism will remain valid once it will be possible to include all ingredients to obtain the normal state of SRO.

\section{\label{sec:spin-fluctu}Spin and charge fluctuation theory}
Since interactions are diagonal in the basis of isolated atoms, they do not commute with the band Hamiltonian.
In perturbation theory, even the ground state is a linear combination of Slater determinants.
With long-range Coulomb repulsion being screened, the atomic interactions that are left lead to rather large energy denominators.
Instead a band electron creates electron-hole pairs best described as spin and charge-density waves.
These influence other electrons, a process known as electron-electron scattering by exchange of spin- or charge-density fluctuations.
They are the lowest-energy excitations and hence lead to small energy denominators in perturbation theory. 
The resulting phase diagrams exhibit a rich variety of competing ordered states with associated order parameters.
Isoelectronic doping suggests that SRO lies in the vicinity of magnetic orderings~\cite{carlo_new_2012}, consistent with the important spin fluctuations found by neutron scattering~\cite{sidis_evidence_1999,iida_inelastic_2011,steffens_spin_2019, jenni_neutron_2021}. 
In addition, the well established correlated character of Ru $t_{2g}$ electrons~\cite{mravlje_coherence-incoherence_2011, Rozbicki_2011, zhang_fermi_2016,kim_spin-orbit_2018, strand_magnetic_2019, kugler_strongly_2020, tamai_high-resolution_2019} makes SRO the archetypal representative of superconductivity mediated by spin and charge fluctuations. 
In this section we summarize the basic ideas, which are formulated in more details in Appendices~\ref{sec:free-energy} and \ref{sec:spin-charge_vertices}.

In \sref{sec:spin-charge_2p-susc}, two-body susceptibilities $\bm \chi_{\alpha}$ are introduced from a free-energy perspective for $\alpha$ either the \textit{p-h} or the \textit{p-p} channels.
They can signal phase transitions and are expressed in terms of bare susceptibilities $\bm \chi_{\alpha}^0$, complemented with vertex corrections.
We present the dominant components of the bare susceptibilities and discuss the origin of their momentum and orbital structure.
We comment on the interplay between both channels through Parquet equations.

In \sref{sec:spin-charge_ph}, the enhancement of spin (charge) fluctuations due to two-body interactions in the \textit{p-h} channel is quantified using the magnetic (density) Stoner factor $S^m$ ($S^d$).
We model the \textit{p-h} vertex corrections using the Kanamori-Slater Hamiltonian (KSH), incorporated within the random phase approximation (RPA).
We verify a scaling relation observed in the bare vertex and justify our choice of parameter sampling.

Now, these spin and charge fluctuations generated by local interactions can mediate pairing in the \textit{p-p} channel.
Ultimately, we are interested in the structure of resulting SCOPs, solutions to the linearized Eliashberg equation.
Applying the frequency dependent formulation of the pairing interactions to multi-orbital systems with SOC is extremely challenging and requires huge numerical capabilities.
In \sref{sec:spin-charge_pp}, we show how the linearized Eliashberg equation for systems with inversion and pseudospin symmetries can be effectively reduced to a problem that is as computationally expensive as the spin-diagonal case encountered when there is no SOC.

\subsection{\label{sec:spin-charge_2p-susc}Two-particle susceptibilities.}
As derived in Appendix~\ref{sec:free-energy}, instabilities can be found from responses to source fields.
At the two-body level, they are either number-conserving or pairing fields, respectively taking the forms
\begin{equation}
    \label{eq:fields}
    \psi^{\dag} (\bm 2) \bm \phi_{11}(\bm 2; \bm 1) \psi (\bm 1)
    \quad \text{and} \quad 
    \psi(\bm 2) \bm \phi_{21}(\bm 2; \bm 1) \psi(\bm 1)
\end{equation}
with corresponding Hermitian conjugates.
Here we use the superindex $\bm i \equiv (\textbf{k}_i, \mu_i, \tau_i)$ which contains momentum, spin-orbital $\mu_i \equiv (\sigma_i, l_i)$ and imaginary time quantum numbers of an electron created and destroyed, respectively, by $\psi^{\dag}(\bm i)$ and $\psi(\bm i)$.
Taking the first derivative of the free-energy $\mathcal{F}$ with respect to those fields respectively lead to the normal and anomalous Green functions
\begin{equation}
    \bm G(\bm 1; \bm 2) = \beta \frac{\delta \mathcal{F}[\phi]}{\delta \phi_{11}(\bm 2; \bm 1)} \Bigg|_{\phi=0} 
    \hspace{-.3cm} \text{and} \
    \bm F(\bm 1; \bm 2) = \beta \frac{\delta \mathcal{F}[\phi]}{\delta \phi_{21}(\bm 2; \bm 1)} \Bigg|_{\phi=0}.
\end{equation}
In the normal state, $\bm G$ describes the propagation of an electron, 
%as a \textit{p-h} fluctuation, 
with its largest value near the Fermi energy-momentum.
Introducing more electronic correlations at the one-body level makes electrons further away from the Fermi energy contribute more~\cite{gingras_superconducting_2019}.
On the other hand, $\bm F$ does not conserve the number of particles and must vanish in the absence of source-fields in the normal state.
More details and properties of $\bm G$ and $\bm F$ are given in Appendices~\ref{sec:one-p-G_conn} and \ref{sec:one-p-G_prop}.

The second derivatives lead to
two-body susceptibilities $\bm \chi_{\alpha}$.
A transition from the normal to an ordered phase happens when infinitesimally small fields can trigger finite responses, signaled by a diverging $\bm \chi_{\alpha}$ at low temperature.
Neglecting two-body interactions, we find the bare susceptibilities $\bm \chi^0_{\alpha}$ given by
\begin{align}
    \label{eq:bare_ph_susc}
    \left[\bm\chi^0_{ph}(Q)\right]_{KK'}^{\mu_1\mu_2\mu_3\mu_4} & = - \frac{1}{\beta} \bm G_{K+Q}^{\mu_1\mu_3} \bm G_{K}^{\mu_4\mu_2} \delta_{KK'}
    \\
    \label{eq:bare_pp_susc}
    \left[ \bm \chi^0_{pp}(Q) \right]_{KK'}^{\mu_1\mu_2\mu_3\mu_4} & = \frac{1}{2\beta} \bm G_{K+Q}^{\mu_1\mu_3} \bm G_{-K}^{\mu_2\mu_4} \delta_{KK'}
\end{align}
where $K = (\textbf{k}, i\omega_m), K' = (\textbf{k}', i\omega_m')$ are the fermionic four-momentum vectors and $Q = (\textbf{q}, i\nu_n)$ the bosonic one. $\beta$ is the inverse temperature.
In this work, $\bm \chi^0_{\alpha}$ are constructed from a DFT calculation with SOC.
Several relations for $\bm \chi^0_{pp}$ are given in Apprendix~\ref{sec:susc_props}.

\begin{figure}[t!]
    \centering
    \includegraphics[width=\linewidth]{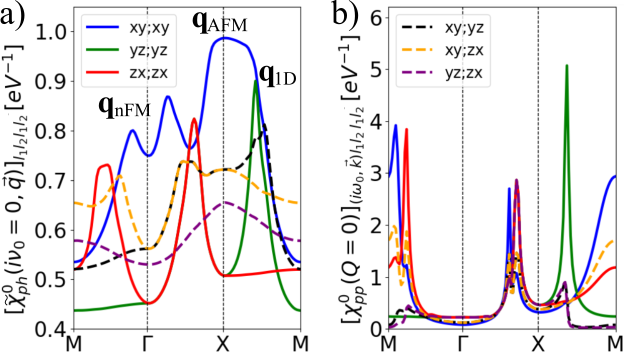}
    \caption{(Color online) Real part of the dominant intra-spin components of the a) \textit{p-h} and b) \textit{p-p} bare susceptibilities in the orbital basis along a high-symmetry path.
    Full lines (dashed lines) are attributed to the intra-orbital (inter-orbital) components.
    In a), we show the RPA susceptibility $\tilde{\bm \chi}^0_{ph}$ defined in \eref{eq:ph_rpa} and important peaks associated to nesting vectors in \fref{fig:DFT_FS} are identified.
    We verified that the inter-spin components are negligible compared to the intra-spin ones.}
    \label{fig:barechi_ph_pp}
\end{figure}

In \fref{fig:barechi_ph_pp}~a), we show the real part of the dominant intra- and inter-orbital components of $\tilde{\bm \chi}^0_{ph} \equiv \frac{1}{N\beta}\sum_K [\bm \chi^0_{ph}]_{KK}$ at $i\nu_0=0$ with $N$ the number of $K$-points.
The summation is perform at the bare level, justified by the subsequent RPA dressing explained in \sref{sec:spin-charge_ph}.
We only show intra-spin components as they dominate.
Because the DFT propagators characterize mostly states at the Fermi level, the $\textbf{q}$-vectors for which the susceptibility has peaks are associated with the overlapping of a part the Fermi sheets and another that is shifted by $\textbf{q}$.
These peaks are called nesting vectors and, for SRO, some are labelled in \fref{fig:barechi_ph_pp}~a) and are visualized in \fref{fig:DFT_FS}.
They were similarly discussed in Ref.~\citenum{gingras_superconducting_2019}.
The $\textbf{q}_{1D}$ vector that folds nicely the $\beta$ band onto the $\alpha$ one is largest in the $yz$-$zx$ orbitals sector.
$\textbf{q}_{1D}$ was experimentally characterized by various neutron scattering experiments~\cite{sidis_evidence_1999, iida_inelastic_2011, iida_spin_2019, jenni_neutron_2021}.
The two other labelled peaks involve the large pockets of states of the $\gamma$ band, with $xy$ character.
These pockets are near a van Hove singularity where the density of states diverges~\cite{kikugawa_rigid-band_2004, burganov_strain_2016, lee_interplay_2020}.
In certain pressure conditions, the $\gamma$ band reaches the van Hove singularity, leading to a Lifshitz transition~\cite{lifshitz1960anomalies, shen_evolution_2007, hicks_strong_2014, steppke_strong_2017, liu_superconductivity_2018, watson_micron-scale_2018, sunko_direct_2019, li_high_2019, karp_mathrmsr_2mathrmmoo_4_2020}.
The nesting of these pockets produces an antiferromagnetic plateau (AFM) along with a nearly ferromagnetic peak (nFM).
Neglecting SOC, these peaks were similar in height~\cite{gingras_superconducting_2019}.
Including SOC, the AFM peak is significantly favored over to the nFM one.

In \fref{fig:barechi_ph_pp}~b), we show the same components but for the real part of $\bm \chi^0_{pp}$.
We look only at the $Q=0$ bosonic four-momentum because we are interested in superconducting Cooper pairs with no center-of-mass energy-momentum, as discussed in \sref{sec:spin-charge_pp}.
In fermionic Matsubara frequencies, we show only $i\omega_0$ for which the susceptibility is largest.
As understood from inversion symmetry in \eref{eq:bare_pp_susc}, the peaks are associated to $\textbf{k}$-points of the Fermi surface.
By inspecting carefully a peak of an intra-orbital component, one can observe that many peaks are split in two because different orbitals contribute to nearly touching Fermi sheets due to SOC.

Now we include two-body interactions captured by the irreducible vertices $\bm \Gamma_{\alpha}$.
Susceptibilities $\bm \chi_{\alpha}$ are expressed in series expansions in the interactions as
\begin{equation}
    \label{eq:bethe-salpeter}\bm \chi_{ph} = \frac{\bm \chi^0_{ph}}{ \textbf{1} - \bm \Gamma_{ph} \bm \chi^0_{ph}}
    \quad \text{and} \quad
    \bm \chi_{pp} = \frac{\bm \chi^0_{pp}}{ \textbf{1} - \bm \Gamma_{pp} \bm \chi^0_{pp}},
\end{equation}
that are commonly called the Bethe-Salpeter equations~\cite{Salpeter_Bethe_1951}.
An instability in the $\alpha$ channel is attained once the largest eigenvalue $\lambda_{\alpha}$ of the operator $\bm V_{\alpha} = \bm \Gamma_{\alpha} \bm \chi^0_{\alpha}$ reaches unity.
Then the associated $\bm \chi_{\alpha}$ diverges and the system reorganizes in a different phase.
These vertices result from electronic interactions and are coupled through the Parquet equations derived in Appendix~\ref{sec:spin-charge_vertices}.
It is this complex interplay that leads to competition between ordered states and a rich variety of emergent phases in correlated systems~\cite{Bourbonnais_1986,Scalapino_Hirsch_1986,scalapino_common_2012}.
The idea of spin and charge fluctuation mediated superconductivity is that although the \textit{p-h} fluctuations can be too weak to induce a transition in the \textit{p-h} channel, they still can mediate interactions in the \textit{p-p} channel that leads to a divergence.
Indeed in this scenario, the SCOPs are very dependent on the details of the \textit{p-h} fluctuations and thus the electronic interactions.
The subtle competition between different nesting vectors in the spin and charge channels has a crucial influence on the type and symmetry of the superconducting states that can arise.

\begin{figure*}
    \centering
    \includegraphics[width=\linewidth]{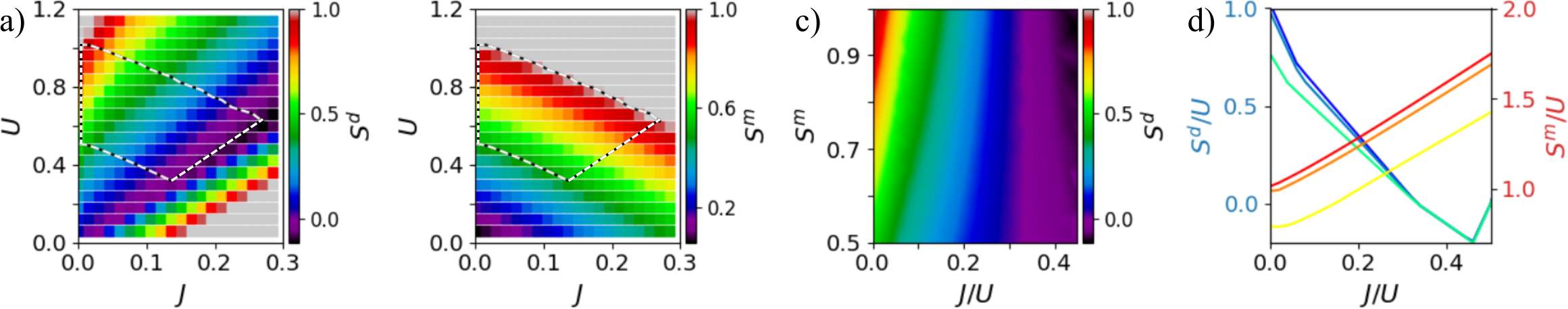}
    \caption[Dependence of the Stoner factors on the parameters of the bare \textit{p-h} vertex.]{
    At $T=250$~K, in the $J$ and $U$ parameter space, a) and b) respectively show the density ($S^d$) and magnetic ($S^m$) Stoner factors defined by \eref{eq:stoners}.
    Instead of $U$ and $J$ as parameters, we prefer to use $S^m$ and $J/U$ as discussed in the main text.
    In c), we show $S^d$ in this new parameter space, constrained by physical intuition.
    This region is highlighted in a) and b) by the doted lines.
    In d), we show the scaling relation \eref{eq:stoner_scaling} for the Stoner factors, with their respective temperature dependence.
    $S^d/U$ is plotted from dark blue to light green and $S^m/U$ from dark red to yellow. The darker color is computed at $T=100$~K, the one in between at $T=250$~K and the lighter one at $T=1000$~K. The kinks are related to changes in dominant nesting vectors.} 
    \label{fig:stoners}
\end{figure*}

\subsection{\label{sec:spin-charge_ph} Stoner factors in the particle-hole channel.}
In the \textit{p-h} channel, the largest eigenvalue of $\bm V_{ph}$ called the Stoner factor never reaches unity.
Doing so would mean we have already fallen into a magnetic or charge order.
We use the Stoner factor as a quantifier of the proximity to a \textit{p-h} ordered state, since it measures the enhancement in \textit{p-h} fluctuations due to two-body interactions.
Without SOC, the system is spin-diagonal and $\bm V_{ph}$ is block diagonal in the so-called density and magnetic channels associated respectively with singlet and triplet particle-hole fluctuations.
In the $S_z=0$ channel, the density/magnetic ($d/m$) block is given as 
\begin{equation}
    \label{eq:vertex_block_diag}
    \bm V_{d/m}^0(Q) \equiv \bm V_{ph}^{\uparrow\uparrow\uparrow\uparrow}(Q) +/- 
    \bm V_{ph}^{\uparrow\uparrow\downarrow\downarrow}(Q)
\end{equation}
where we used $\bm V_{ph}^{\sigma\sigma\sigma'\sigma'}(Q) = \bm V_{ph}^{\sigma'\sigma'\sigma\sigma}(Q)$. The other magnetic channels with $S_z=\pm 1$ are obtained as
\begin{equation}
    \bm V_m^{+1}(Q) \equiv 
    \bm V_{ph}^{\uparrow\downarrow\uparrow\downarrow}
    (Q)
    \quad \text{and} \quad
    \bm V_m^{-1}(Q) \equiv
    \bm V_{ph}^{\downarrow\uparrow\downarrow\uparrow}(Q).
\end{equation}

Including SOC, this block diagonal property of $\bm V_{ph}$ is no longer true and spin and charge fluctuations are coupled together.
In the present case, we observe that these deviations are small enough so that we still refer to charge (spin) fluctuations and we quantify them by looking at the density (magnetic) Stoner factors $S^{d(m)}$ defined as 
\begin{equation}
    \label{eq:stoners}
    S^{d(m)} \equiv \text{max eig}\{ \bm V_{d(m)}^0 (Q) \}.
\end{equation}

Let us now discuss the specific choice of vertex $\bm \Gamma_{ph}$ entering $\bm V_{ph} = \bm \Gamma_{ph} \bm \chi^0_{ph}$.
As discussed in Appendix~\ref{sec:kanamori}, the two-body interactions are modelled using the rotationally invariant formulation of the KSH.
This model is a multi-orbital generalization of the Hubbard model and depends on the on-site Coulomb repulsion $U$, along with the Hund's coupling $J$.
In the \textit{p-h} channel, the spinful \textit{p-h} irreducible vertex function $\bm \Gamma_{ph}$ is taken as the local and static antisymmetrized Coulomb interaction $\bm \Lambda_{ph}$ explicit in \eref{eq:kanamori_vertex}.
Comparing with \eref{eq:ph_vertex}, we miss ladder functions in both in the \textit{p-p} and \textit{p-h} channels.
The \textit{p-p} one can be neglected, since we work in the normal state were \textit{p-p} fluctuations remain small.
The \textit{p-h} one is absorbed in $\bm \Lambda_{ph}$ and renormalizes $U$ and $J$ from the bare values to further screened ones.
This is similar to what is done in the two-particle self-consistent (TPSC) approach~\cite{vilk_non-perturbative_1997}.

Now the local and static properties of $\bm \Lambda_{ph}$ simplifies the dressing of the \textit{p-h} susceptibility in \eref{eq:bethe-salpeter} so that it becomes diagonal in fermionic four-momentum, that is
\begin{equation}
    \label{eq:ph_susc_K-diag}
    \left[\bm \chi_{ph}(Q)\right]_{KK'} = \left[\bm \chi_{ph}(Q)\right]_{KK}\delta_{KK'}.
\end{equation}

Moreover, the \textit{p-h} susceptibility influences pairing through the ladder function \eref{eq:ladder_ph} that characterizes the exchange of a \textit{p-h} fluctuation between to electrons.
Using \eref{eq:ph_susc_K-diag}, it simplifies to
\begin{equation}
    \left[\bm \Phi(Q)\right]^{\mu_1\mu_2\mu_3\mu_4}_{KK'} = \left[\bm \Lambda_{ph} \tilde{\bm \chi}_{ph}(Q) \bm \Lambda_{ph} \right]^{\mu_1\mu_2\mu_3\mu_4} \delta_{KK'}
\end{equation}
where we only need the so-called RPA susceptibility
\begin{equation}
    \label{eq:ph_rpa}
    \left[\tilde{\bm \chi}_{ph}(Q)\right]^{\mu_1\mu_2\mu_3\mu_4} = \frac{1}{N\beta} \sum_K \left[ \bm \chi_{ph}(Q) \right]^{\mu_1\mu_2\mu_3\mu_4}_{KK}.
\end{equation}
This operation can be performed at the bare level, thus we define the RPA bare \textit{p-h} susceptibility as
\begin{equation}
    \label{eq:ph0_rpa}
    \left[\tilde{\bm \chi}^0_{ph}(Q)\right]^{\mu_1\mu_2\mu_3\mu_4} = \frac{1}{N\beta} \sum_K \left[ \bm \chi^0_{ph}(Q) \right]^{\mu_1\mu_2\mu_3\mu_4}_{KK}.
\end{equation}

Several relations for $\tilde{\bm \chi}_{ph}^0$ are given in Appendix~\ref{sec:susc_props}.
This approximation reduces greatly the amount of numerical resources needed for vertex corrections at the cost of lacking a proper description of the energy-momentum dependence of the fluctuations.
In this work, we focus on the effects of SOC.
In another study neglecting SOC for a better treatment of the electronic correlations using dynamical mean-field theory (DMFT), it was shown that the frequency-dependence of the \textit{p-h} vertex leads to a reduction of the AFM peak and an enhancement of the nFM one~\cite{strand_magnetic_2019}.
All of these effects can influence the ordering of the leading superconducting states.

In \fref{fig:stoners}~a) (b), we show the dependence of $S^{m}$ ($S^{d}$) on the parameters $U$ and $J$.
Instead of these parameters, it is useful to employ $J/U$ and $S^m$ because they have more physically relevant interpretations which allows to constrain them.
Moreover, the \textit{p-h} vertex $\bm \Lambda_{ph}$ satisfies a scaling relation \eref{eq:scaling_kanamori}, which directly applies to the Stoner factors as
\begin{equation}
    \label{eq:stoner_scaling}
    S^{d(m)}[U, J] = U \cdot S^{d(m)}[1, J/U].
\end{equation}

To insure repulsive on-site interactions in \eref{eq:kanamori_vertex}, one should satisfy $J/U < 1/3$.
We extend the constraint to $J/U \leq 0.45$ because some work have considered attractive on-site interactions in their study of superconductivity in SRO~\cite{lindquist_distinct_2020, clepkens_higher_2021, lindquist_2021_evolution}.
Moreover, in the spirit of Hund's coupling which favours same spin alignment, the inter-orbital Coulomb repulsion should be stronger in the inter-spin channel $U'= U-2J$ compared to the intra-spin one $U''=U-3J$, imposing $J/U \geq 0$.
As for $S^m$, it quantifies the role of two-body interactions in generating magnetic fluctuations. Isoelectronic doping experiments have shown that there is a lot of magnetic ordering in proximity to SRO~\cite{carlo_new_2012}. We study $S^m \geq 0.5$ to tune the system in the vicinity of a magnetic instability, yet never do we reach $S^m \ (S^d) \geq 1$ since this would imply a magnetic (charge) instability in the \textit{p-h} channel.

As a result, $0 \leq J/U \leq 0.45$ and $0.5 \leq S^m \leq 0.95$ defines the region of parameter space where we study superconductivity in \sref{sec:supercond}.
\fref{fig:stoners}~c) shows $S^d$ in this parameter space, which is also highlighted by the dotted lines in \fref{fig:stoners}~a) and b). 
The charge channel becomes more important at small $J/U$ where Hund's coupling does not force spins to be aligned so that charge can move more freely between local orbitals.

In \fref{fig:stoners}~d), we highlight the scaling relation \eref{eq:stoner_scaling}, which implies the dependence of $S^{d/m}/U$ over $J/U$ and temperature.
The lighter colors corresponds to $T=1000$~K, the middle one to $T=250$~K and the darker to $T=100$~K.
In our study, $U$ and $J$ are kept fixed with temperature and the dependence of the Stoner factors over temperature can be assessed from the height of the bare \textit{p-h} susceptibility's peaks that sharpen with lowering temperatures.
The discontinuities are associated to different leading $\textbf{q}$-vectors depending on the interaction parameters.

\subsection{\label{sec:spin-charge_pp}Particle-particle channel in the pseudospin basis.}
In the \textit{p-p} channel where instabilities lead to superconductivity, we are interested in the eigenvectors $\bm \Delta \equiv \bm\Delta_{pp}(Q=0)$ of the operator $\bm V_{pp} \equiv \bm V_{pp}(Q=0)$ with largest eigenvalue $\lambda \equiv \lambda_{pp}$.
$Q=0$ is chosen because, in the absence of an external magnetic field, the leading instability is usually for Cooper pairs with no net four-momentum since this minimizes the free-energy by having no net superfluid flow.
In the normal state, these eigenvectors characterize instabilities that satisfy the linearized Eliashberg equation
\begin{equation}
    \label{eq:eliashberg}
    \lambda \bm \Delta^{\mu_1\mu_2}_K = - \sum_{K'\mu_3\mu_4} \left[\bm V_{pp}\right]^{\mu_1\mu_2\mu_3\mu_4}_{KK'} \bm \Delta^{\mu_3\mu_4}_{K'}.
\end{equation}

As discussed in Sec.~\ref{sec:normal-state_pseudospin}, the non-interacting DFT Hamiltonian projected on the $4d$-$t_{2g}$ orbitals of the Ru atom is diagonal in spina and orbitals when neglecting SOC.
Including it entangles the spin and orbital quantum numbers of electrons.
Rotating to the pseudospin-orbital basis, the quasi-2D Hamiltonian becomes diagonal in pseudospin, although orbitals and spins stay entangled.
Consequently, the bare susceptibilities Eqs~(\ref{eq:bare_ph_susc}, \ref{eq:bare_pp_susc}) are also diagonal in pseudospin with
\begin{equation}
    \left[\bm \chi^0_{\alpha}\right]^{\rho_1\rho_2\rho_3\rho_4} = \left[\bm \chi^0_{\alpha}\right]^{\rho_1\rho_2\rho_1\rho_2} \delta_{\rho_1\rho_3} \delta_{\rho_2\rho_4}.
\end{equation}

\begin{figure}[b]
    \centering
    \includegraphics[width=.7\linewidth]{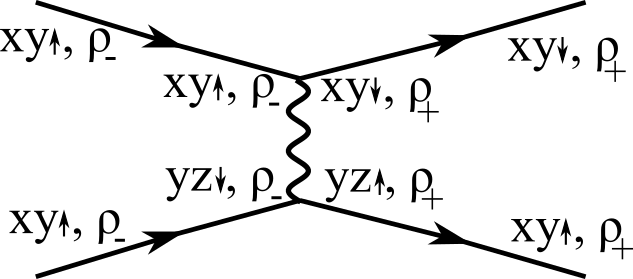}
    \caption{Example of a first order diagram of the dressed \textit{p-h} susceptibility that does not preserve pseudospin.}
    \label{fig:spin_flip}
\end{figure}

As highlighted in \fref{fig:ph_vertex_spin} of Appendix~\ref{sec:kanamori}, our choice of \textit{p-h} vertex function $\bm \Lambda_{ph}$ is spin-diagonal, but not pseudospin-diagonal.
An example of an interaction diagram that does not preserve pseudospin is shown in \fref{fig:spin_flip}, which leads to finite $\left[\bm \Gamma^0_{ph}\right]^{\rho\rho\bar{\rho}\bar{\rho}}$ contributions, with $\bar{\rho} = -\rho$.

However, the dressed susceptibility is still block diagonal, having an intra-pseudospin channel
\begin{equation}
    \bm \chi'_{ph} = \left[
        \begin{array}{cccc}
            \bm \chi^{++++}_{ph} & \bm \chi^{++--}_{ph} \\
            \bm \chi^{--++}_{ph} & \bm \chi^{----}_{ph}
        \end{array}
    \right]
\end{equation}
and an inter-pseudospin channel
\begin{equation}
    \bm \chi''_{ph} = \left[
        \begin{array}{cccc}
            \bm \chi^{+-+-}_{ph} & \bm \chi^{+--+}_{ph} \\
            \bm \chi^{-++-}_{ph} & \bm \chi^{-+-+}_{ph}
        \end{array}
    \right].
\end{equation}
Consequently, the pairing vertex $\bm V_{pp}$ can also be expressed as a block diagonal matrix in the intra- and inter-pseudospin channels.
The solutions of the Eliashberg equation in each of these channels are thus independent.
\eref{eq:eliashberg} restricted to the inter-pseudospin channel leads to the following coupled equations:
\begin{align}
\label{eq:reduced_eliashberg}
    \lambda \left[ \bm \Delta_K \right]^{+-}_{l_1l_2} 
        =
        & - \bm \left[ \bm V_{KK'} \right]^{+-+-}_{l_1l_2l_3l_4}
        \left[\bm \Delta_{K'} \right]^{+-}_{l_3l_4} \nonumber \\
        & - \left[ \bm V_{KK'} \right]^{+--+}_{l_1l_2l_4l_3}
        \left[ \bm \Delta_{K'} \right]^{-+}_{l_4l_3}, \nonumber \\
    \lambda \left[ \bm \Delta_K \right]^{-+}_{l_1l_2} 
        =
        & - \bm \left[ \bm V_{KK'} \right]^{-++-}_{l_1l_2l_3l_4}
        \left[\bm \Delta_{K'} \right]^{+-}_{l_3l_4} \\
        & - \left[ \bm V_{KK'} \right]^{-+-+}_{l_1l_2l_4l_3}
        \left[ \bm \Delta_{K'} \right]^{-+}_{l_4l_3} \nonumber 
\end{align}
where $V_{KK'} \equiv \left[V_{pp}\right]_{KK'}$.
The Pauli principle leads to
\begin{align}
    \left[ \bm \Delta_{K} \right]^{+-}_{l_1l_2} = - \left[ \bm \Delta_{-K} \right]^{-+}_{l_2l_1}
\end{align}
and in systems with inversion symmetry, the solutions are either even- or odd-parity (\textit{e-p} or \textit{o-p}).
They are defined as
\begin{equation}
    \label{eq:gap_function_parity}
    \left[ \bm \Delta_K^{ep/op} \right]_{l_1l_2} = \left[ \bm \Delta_K \right]^{+-}_{l_1l_2} \mp \left[ \bm \Delta_{K^*} \right]^{-+}_{l_2l_1},
\end{equation}
with $K^* \equiv (\textbf{k}, -i\omega_m)$.
In this basis, we found that off-diagonal elements completely vanish when SOC in neglected.
Considering SOC within the pseudospin approximation, we found that they were several orders of magnitude smaller than the diagonal ones so that Eqs~(\ref{eq:reduced_eliashberg}) can be expressed as
\begin{equation}
    \label{eq:effective_eliashberg}
    \lambda \left[
        \begin{array}{c}
            \bm \Delta_K^{ep} \\
            \bm \Delta_K^{op}
        \end{array}
    \right]
    =
    -\frac{1}{2} \left[ \begin{array}{cc}
        \bm V^{ep}_{KK'} & \bm 0 \\
        \bm 0 & \bm V^{op}_{KK'}
    \end{array} \right]
    \left[
        \begin{array}{c}
            \bm \Delta_{K'}^{ep} \\
            \bm \Delta_{K'}^{op}
        \end{array}
    \right]
\end{equation}
where
\begin{align}
    \label{eq:pseudospin_pp_vertices}
    \left[\bm V^{ep/op}_{KK'}\right]_{l_1l_2l_3l_4} 
        & = \left[\bm V_{KK'} \right]^{+-+-}_{l_1l_2l_3l_4} \mp \left[\bm V_{KK'^*} \right]^{+--+}_{l_1l_2l_4l_3} \\
        & \mp \left[\bm V_{K^*K'} \right]^{-++-}_{l_2l_1l_3l_4} + \left[\bm V_{K^*K'^*} \right]^{-+-+}_{l_2l_1l_4l_3}. \nonumber 
\end{align}

Using the relation Eq.~(\ref{eq:chipp0_1221}), the effective pairing vertex can finally be expressed as
\begin{align}
    \bm V^{ep/op}_{KK'} = \left[ \bm \Gamma^{ep/op}_{KK'} \right] \left[ \bm \chi_{pp}^0(K') \right]^{+-+-}
\end{align}
with
\begin{align}
    \left[\bm \Gamma^{ep/op}_{KK'}\right]_{l_1l_2l_3l_4} 
        & = \left[\bm \Gamma_{KK'} \right]^{+-+-}_{l_1l_2l_3l_4} \mp \left[\bm \Gamma_{KK'^*} \right]^{+--+}_{l_1l_2l_4l_3} \\
        & \mp \left[\bm \Gamma_{K^*K'} \right]^{-++-}_{l_2l_1l_3l_4} + \left[\bm \Gamma_{K^*K'^*} \right]^{-+-+}_{l_2l_1l_4l_3}. \nonumber 
\end{align}

For the intra-pseudospin channel, Pauli's exclusion principle is not sufficient to construct an effective pairing vertex and we have to solve the spinful system through \eref{eq:eliashberg} to study these solutions.
They can be understood as having total pseudospin of $1$ and pseudospin projection $\pm 1$.
Comparing effective solutions in the inter-pseudospin channel with those of the spinful system, we not only verify that both give the same answer but also that all pseudo-triplet states are almost degenerate.
This is very fortunate because, while we cannot go to low enough temperatures in the spinful case, the degeneracy between the intra-pseudospin and inter-pseudospin solutions allow us to discuss all possible solutions by only inspecting the inter-pseudospin channel.

\section{\label{sec:supercond}Characterization of gap functions}
Ordered states like superconductivity are characterized by an order parameter that breaks at least one symmetry of the normal state.
As explained in \sref{sec:spin-fluctu}, the gap function $\bm \Delta$ is an eigenvector of the pairing vertex $\bm V_{pp}$.
This is the SCOP that we want to characterize.
In frequency-dependent multi-spin-orbital systems, these SCOP have many degrees of freedom involving various types of classifications.

In \sref{sec:supercond_spot}, we first introduce the $SPOT$ decomposition as a tool to quantify the interplay between these quantum numbers.
In \sref{sec:supercond_group}, we apply group theory to assign a single label, the irrep of the $D_{4h}$ space group, that involves all quantum numbers together.
We decompose spin-orbital basis functions of the $t_{2g}$ subset in terms of irreps and discuss time-reversal symmetry.
Finally in \sref{sec:supercond_lead}, we apply these tools to the inter-pseudospin solutions of the effective Eliashberg equation \eref{eq:effective_eliashberg} for SRO.
As leading eigenvectors, we find B$_{1g}^+$ and A$_{2g}^-$ states.
We discuss how SOC couples different orbital sectors and compare them with experiments, discuss their phase weighted distributions and investigate how their eigenvalues evolve with temperature.
As subleading symmetries, we find another A$_{2g}^-$, a set of degenerate E$_u^-$ and an A$_{1g}^-$.
We discuss why we do not find some of the other proposed solutions for SRO.

\subsection{\label{sec:supercond_spot}SPOT contributions.}
The gap function $\bm \Delta$ transforms like the anomalous Green function or Gorkov function $\bm F$ defined in Appendix~\ref{sec:one-p-G_prop}.
It is given by a time-ordered product of two fermion destruction operators
\begin{equation}
    \bm F^{\mu_1\mu_2}_{K} = \int_0^\beta d\tau \ e^{i\omega_m \tau} \langle T_{\tau} \psi^{\sigma_1}_{\textbf{k} l_1}(\tau) \psi^{\sigma_2}_{-\textbf{k} l_2} \rangle,
\end{equation}
where the two electrons forming a pair have four quantum numbers each: spin-orbitals $\mu_1 = (\sigma_1, l_1)$ and $\mu_2$, along with  energy-momenta in Matsubara frequency $K = (i\omega_m, \textbf{k})$ and $-K$.
$T_{\tau}$ is the imaginary-time-ordering operator.
Using the sign change upon the exchange of two fermions,  this two-fermion object, and thus $\bm \Delta$ as well, satisfies the Pauli principle
\begin{equation}
    \label{eq:pauli}
    \bm \Delta^{\sigma_1\sigma_2}_{\textbf{k}l_1l_2}(i\omega_m) = - \bm \Delta^{\sigma_2\sigma_1}_{-\textbf{k}l_2l_1}(-i\omega_m).
\end{equation}

The exchange of each quantum number is respectively characterized by the $\hat{S}$, $\hat{P}^*$, $\hat{O}$ and $\hat{T}^*$ operators~\cite{linder_odd-frequency_2019}, acting as follows:
\begin{align}
    & \left[ \hat{S} \bm \Delta\right]^{\sigma_1\sigma_2} \equiv \left[\bm \Delta\right]^{\sigma_2\sigma_1}
    & \quad & \left[ \hat{P}^* \bm \Delta \right]_{\textbf{k}} \equiv \bm \Delta_{-\textbf{k}} \\
    & \left[ \hat{O} \bm \Delta \right]_{l_1l_2} \equiv \bm \Delta_{l_2l_1}
    & \quad & \left[\hat{T}^*\bm \Delta\right](i\omega_m) \equiv \bm \Delta (-i\omega_m) \nonumber
\end{align}
where, when omitted, quantum numbers stay untouched.
Note that $\hat{T}^*$ is simply the exchange of relative time, different from time-reversal $\hat{\mathcal{T}}$.
Moreover, there was recently a generalization of these operators to systems with strong SOC involving the exchange of total angular momentum $\hat{J}$ instead of the exchanges of spin $\hat{S}$ and orbitals $\hat{O}$~\cite{PhysRevResearch.3.033255}.
We did not consider it in this work.
In terms of these operators, \eref{eq:pauli} be expressed as $\hat{S}\hat{P}^*\hat{O}\hat{T}^*\bm \Delta = - \bm \Delta$.
Because these operations are idempotent, the eigenvalues of each operators are $+$ and $-1$, associated to even and odd eigenvectors.
The SCOPs aren't necessarily eigenvectors, so we use the $SPOT$ decomposition to characterize gap functions in systems where multiple $SPOT$ eigenvectors are cohabiting.

In centrosymmetric systems like $D_{4h}$, all irreps are eigenvectors of $\hat{P}^*$ and they are labelled by $g$ ($u$) for \textit{e-p} (\textit{o-p}).
They are denoted $^+P$ ($^-P$) and gap functions remain pure in that quantum number.
In spin-diagonal cases such as that obtained when neglecting SOC, gap functions satisfy $\hat{S}\bm \Delta = + (-) \bm \Delta$ for triplet $^+S$ (singlet $^-S$) solutions.
In that case, the singlet channel $\bm \Delta^{\uparrow\downarrow} - \bm \Delta^{\downarrow\uparrow}$ that has total spin $0$ and the three degenerate triplet channels $\bm \Delta^{\uparrow\uparrow}$, $\bm \Delta^{\downarrow\downarrow}$ and $\bm \Delta^{\uparrow\downarrow} + \bm \Delta^{\downarrow\uparrow}$ have total spin $1$ and total spin projections $+1$, $-1$ and $0$, respectively.

Considering SOC in multi-orbital systems introduces spin-flips that entangles those spin channels together and can lead to coexistence of $^+S$ and $^-S$ contributions~\cite{PhysRevLett.112.127002}.
It also introduces inter-orbital interactions that hybridize formerly decoupled orbital sectors.
In other words, it can induce mixing between $^+O$ and $^-O$ contributions.
These hybridizations between non-degenerate orbitals were shown to generate coexistence between $^+T$ and $^-T$ contributions~\cite{black-schaffer_odd-frequency_2013,triola_role_2019}.

In this notation, conventional superconductors are purely $^-S^+P^+O^+T$ with a $s$-wave symmetry in $\textbf{k}$-space.
Similarly, cuprates are known to be spin-singlet one-band $d$-wave superconductors~\cite{scalapino_case_1995}.
On the other hand, the few candidates for spin-triplet superconductivity are usually classified as \textit{o-p} with a pure $^+S^-P^+O^+T$ contribution.
Among potential materials, there are many uranium-based materials like UPt$_3$~\cite{Joynt_Taillefer_2002} and UTe$_2$~\cite{Ran_Eckberg_Ding_Furukawa_Metz_Saha_Liu_Zic_Kim_Paglione_2019}.
Also, the $p_+ \pm i p_y$ chiral $p$-wave state previously proposed for SRO would classify as such triplet.

These two types of SCOP, spin-singlet and spin-triplet, are often mentioned separately in the literature because most superconducting models are single-orbital and/or spin-diagonal.
In SRO however, there have been increasing discussions about $^-O$ states ~\cite{kaba_group-theoretical_2019,romer_knight_2019, ramires_superconducting_2019,roising_superconducting_2019, suh_stabilizing_2020, lindquist_distinct_2020, chen_interorbital_2020, zhang_possible_2021}.
Additionally, a few papers have studied the frequency dependence of the gap function and found purely $^-T$ states when neglecting SOC~\cite{gingras_superconducting_2019, kaser_inter-orbital_2021}.
Considering SOC, the coexistence between $^+T$ and $^-T$ solutions should be ubiquitous~\cite{black-schaffer_odd-frequency_2013,triola_role_2019}.
One work actually linked the ubiquitous presence of $^-T$ correlations in multi-orbital SRO with its observed finite Kerr effect~\cite{komendova_odd-frequency_2017}.

\subsection{\label{sec:supercond_group}Group theory.}

In the Ginzburg-Landau paradigm, ordered states are characterized by an order parameter, which breaks at least one symmetry of the normal state.
The SCOP breaks $U(1)$ gauge symmetry, associated with breaking conservation of the total number of particles.
In BCS superconductors~\cite{bardeen_theory_1957,tinkham_introduction_1996}, it is the only broken symmetry.
In unconventional superconductors, the complex mechanisms mediating pairing usually lead to additional symmetry breaking~\cite{scalapino_common_2012}.
The different symmetries that are broken in an ordered phase have a specific label attached, referred to as an irrep.
In Appendix~\ref{sec:group}, we detail the symmetries and irreps of the $D_{4h}$ space group.

As the irrep characterizing a SCOP involves all quantum numbers at the same time, it is instructive to look at how it decomposes in each of them separately.
The total irrep is obtained from the direct product of the irreps of each independent space; spin, orbital, wave-vector.
In \sref{sec:supercond_group_spin-orb}, we discuss the spin-orbital components themselves, without including $\textbf{k}$-dependence.
First in the case without SOC, we classify the basis functions in spin and orbital spaces separately.
Then, we introduce SOC, which entangles them.
Certain products of irreps become reducible while maintaining pseudospin symmetry.
In \sref{sec:supercond_group_time-rev}, we generalize the classification of Ref.~\citenum{geilhufe_symmetry_2018}, but for systems with multiple $SPOT$ contributions.
We explain why time-reversal operation $\hat{\mathcal{T}}$ would be preferable to $\hat{T}^*$ in systems with multiple $SPOT$ contributions.
However, $\hat{\mathcal{T}}$ has a phase ambiguity that prevents actually using it.
As an in-between, we use the frequency dependence of the intra-orbital components of the SCOP in our classification.
Even-frequency (odd-frequency) intra-orbital components are denoted by a $+$ ($-$) superscript in the irrep label.

\subsubsection{\label{sec:supercond_group_spin-orb}Spin-orbital basis.}
\paragraph*{Without SOC.}
The electrons in SRO transform like irreps of the double group $\tilde{D}_{4h}$~\cite{altmann_point-group_1994}.
An irrep of this group $\text{D}_{l\sigma} \equiv \text{D}_l \otimes \text{E}_{1/2,g}$ can decomposed into the orbital part $\text{D}_l$ and the spin part $\text{E}_{1/2,g}$.
The gap function describing Cooper pairs depends on two such electrons, thus it transforms like $\text{D}_{l_1\sigma_1} \otimes \text{D}_{l_2\sigma_2}$.
Without SOC, the normal state Hamiltonian is diagonal in the spin-basis and can be mapped to a doubly degenerate spin-diagonal Hamiltonian.
In that case, the spin part of the gap function is separable into $\text{E}_{1/2,g} \otimes \text{E}_{1/2,g} = \text{A}_{1g} \oplus \text{A}_{2g} \oplus \text{E}_g$ where $\text{A}_{1g}$ corresponds to the spin singlet $\Delta^{\uparrow\downarrow} - \Delta^{\downarrow\uparrow}$ with total spin $0$ while the three others are spin-triplet with total  spin $1$.
The $\text{A}_{2g}$ one is given by $\Delta^{\uparrow\downarrow} + \Delta^{\downarrow\uparrow}$ with spin projection $0$ and the $\text{E}_g$ ones are $\Delta^{\uparrow\uparrow}$ and $\Delta^{\downarrow\downarrow}$ with spin projection $\pm 1$.
These four components are represented by Pauli matrices shown in Table~\ref{tab:basis_gap_functions_spin}.

\begin{table}[ht!]
    \centering
    \caption{Basis functions of the SCOP in spin space expressed as irreps. $S$ is the parity under spin exchange.}
    \begin{tabular}{c|c|c}
        \label{tab:basis_gap_functions_spin}
        Irrep & Spin basis function & $S$ \\
        \hline \hline
        A$_{1g}$    &   $\sigma_2$    &   - \\
        A$_{2g}$    &   $\sigma_1$    &   + \\
        E$_g$       &   $\sigma_0$, $\sigma_3$    &   +
    \end{tabular}
\end{table}

%\paragraph*{\label{sec:group_orb}Orbital basis.}
On the other hand, the orbital part depends in this case on the $t_{2g}$  basis.
Those transform intrinsically like two independent subsets: the $xy$ orbital transforms like the one-dimensional $\text{D}_{xy} = \text{B}_{2g}$ irrep while the $yz$ and $zx$ orbitals transform like the two-dimensional $\text{D}_{yz/zx} = \text{E}_g$ irrep.
In other words, the electrons states forming the Cooper pairs transform non-trivially depending on the orbital that hosts them.
The orbital part of the SCOP transforms like $\text{D}_{l_1} \otimes \text{D}_{l_2}$ with three distinct possibilities:
\begin{itemize}
    \item The $l_1 = l_2 = xy$ sector that involves a single component transforming like $\text{B}_{2g} \otimes \text{B}_{2g} = \text{A}_{1g}$.
    \item The $l_1, l_2 \in \{yz, zx\}$ sector that involves four components transforming like $\text{E}_g \otimes \text{E}_g = \text{A}_{1g} \oplus \text{A}_{2g} \oplus \text{B}_{1g} \oplus \text{B}_{2g}$.
    \item The $l_1 = xy$ and $l_2 \in \{yz, zx\}$ sectors (and vice-versa), involving four components transforming like $\text{B}_{2g} \otimes \text{E}_g = \text{E}_g \otimes \text{B}_{2g} = \text{E}_g$.
\end{itemize}
The orbital basis functions written in terms of the irreps are shown in Table~\ref{tab:basis_gap_functions_orb}.
See how $\text{E}_g \otimes \text{E}_g$ is reducible.
Moreover, these basis functions can be either even $^+O$ or odd $^-O$ under exchange of the two orbitals forming the pair.

Neglecting SOC makes $\bm \Delta$ diagonal in spins, implying it is an eigenvector of $\hat{S}$.
It is also block diagonal in the orbital basis with the three independent sectors given above.
Note again that we are neglecting the wave-vector dependence of the gap functions.
If we were considering a non-trivial irrep in wave-vector space, it would have to be multiplied by the other irreps.
For example, a spin-singlet component for $xy;xy$ orbitals transforms like $\text{A}_{1g}$ in spin-orbit basis.
If, in $\textbf{k}$-space, it transforms like the $d_{x^2-y^2}$ function that is $\text{B}_{1g}$, this component globally transforms like $\text{A}_{1g} \otimes \text{B}_{1g} = \text{B}_{1g}$.

\begin{table}[ht!]
    \centering
    \caption{Basis functions of the SCOP in orbital space expressed as irreps. $O$ is the parity under orbital exchange. $\ket{1\pm}$ and $\ket{2\pm}$ are notations that are helpful later.}  
\begin{tabular}{c|c|c}
    \label{tab:basis_gap_functions_orb}
    \centering
    Irrep  & Orbital basis function      & $O$ \\
    \hline
    \hline
    \ A$_{1g}$ \   & $\ket{xy;xy}$     & \ + \ \\
    \hline
    A$_{1g}$  & $\ket{yz;yz} + \ket{zx;zx}$   & + \\
    B$_{1g}$  & $\ket{yz;yz} - \ket{zx;zx}$   & + \\
    A$_{2g}$  & $\ket{yz;zx} - \ket{zx;yz}$   & - \\
    B$_{2g}$  & $\ket{yz;zx} + \ket{zx;yz}$   & + \\
    \hline
    E$_{g}$   & \ $\ket{1+} \equiv \ket{xy;yz} + \ket{yz;xy}$ \  & + \\
              & $\ket{2+} \equiv \ket{xy;zx} + \ket{zx;xy}$   & + \\
    E$_{g}$  & $\ket{1-} \equiv \ket{xy;yz} - \ket{yz;xy}$   & - \\
              & $\ket{2-} \equiv \ket{xy;zx} - \ket{zx;xy}$   & -
\end{tabular}
\end{table}

\paragraph*{\label{sec:group_spinorb}With SOC.}
Introducing SOC in the Hamiltonian of the non-interacting system has the effect that spin and orbital are no longer good quantum numbers at the one-particle level.
As a result, an electron is in a superposition of different orbitals and spins, with coefficients depending on its wave-vector.
In other words, the Hamiltonian is no longer block diagonal in those spaces, although it stays diagonal in pseudospin space.
Consequently, the three orbital sectors are coupled together and the gap functions involve all of the orbitals through combinations of $\hat{S}\hat{P}^*\hat{O}\hat{T}^*=-1$ contributions.
All these contributions however have to globally transform like a single irrep. 

In spin-orbital space, the whole space of $t_{2g}$ orbitals transforms like $\text{D}_{t_{2g}\sigma} = (\text{B}_{2g}\oplus \text{E}_g)\otimes \text{E}_{1/2,g}$.
The product ${\text{D}_{t_{2g}\sigma}} \otimes \text{D}_{t_{2g}}$ involved in the SCOP gives new basis functions that entangle spins and orbitals.
Most of these are trivially obtained, like a singlet state in the $xy;xy$ orbitals transforms like $\ket{xy;xy} \otimes \sigma_2$, that is A$_{1g} \otimes \text{A}_{1g} = \text{A}_{1g}$.
The only non-trivial basis functions are those involving the 2D irreps in both orbital and spin spaces, that is the $\{yz, zx\}$ orbitals with $\{\sigma_0, \sigma_3\}$ spins.
This $\text{E}_g\otimes \text{E}_g$ product is reduced by forming linear combinations that transform like the $\text{A}_{1g}$, $\text{A}_{2g}$, $\text{B}_{1g}$ and $\text{B}_{2g}$ irreps.
These combinations are given in Table~\ref{tab:basis_gap_function_soc}.

\begin{table}[ht!]
    \centering
    \caption{Non-trivial basis functions of the SCOP in spin-orbital space expressed as irreps. The orbital and spin parts are respectively defined in Tables~\ref{tab:basis_gap_functions_spin} and \ref{tab:basis_gap_functions_orb}.
    $O$ ($S$) is the parity under orbital (spin) exchange.}
\begin{tabular}{c|c|c|c}
    \label{tab:basis_gap_function_soc}
    \centering
    \text{Irrep}    & Basis function    & $S$ & $O$  \\
    \hline
    \hline
%    $\Delta_z \otimes \sigma_2$ &   $B_{1g}$ \\
%    $\Delta_z \otimes \sigma_1$ &   $B_{2g}$ \\
%    $\Delta_z \otimes (\sigma_{0} \pm \sigma_{3})$  &   $E_g$ \\
%    \hline
    \ A$_{1g}$ \    &   \ $\ket{1+} \otimes \sigma_0 + i \ket{2+} \otimes \sigma_3$ \   & \ + \ & \ + \ \\
    B$_{1g}$    &   $\ket{1+} \otimes \sigma_0 - i \ket{2+} \otimes \sigma_3$   & + &  + \\
    A$_{2g}$    &   $\ket{1+} \otimes \sigma_3 + i \ket{2+} \otimes \sigma_0$   & + &  + \\
    B$_{2g}$    &   $\ket{1+} \otimes \sigma_3 - i \ket{2+} \otimes \sigma_0$   & + &  + \\
    \hline
    A$_{1g}$    &   $\ket{1-} \otimes \sigma_0 + i \ket{2-} \otimes \sigma_3$   & + &  - \\
    B$_{1g}$    &   $\ket{1-} \otimes \sigma_0 - i \ket{2-} \otimes \sigma_3$   & + &   -\\
    A$_{2g}$    &   $\ket{1-} \otimes \sigma_3 + i \ket{2-} \otimes \sigma_0$   & + &   - \\
    B$_{2g}$    &   $\ket{1-} \otimes \sigma_3 - i \ket{2-} \otimes \sigma_0$   & + &   - 
\end{tabular}
\end{table}

Although SOC couples orbitals of different sectors, the effective reduction in  pseudospin space given in \eref{eq:effective_eliashberg} implies the gap functions are purely intra- or inter-pseudospin.
In this work, we only study the inter-pseudospin solutions as explained in \sref{sec:spin-charge_pp}.

\subsubsection{\label{sec:supercond_group_time-rev}Time-reversal symmetry.}
In Ref.~\citenum{geilhufe_symmetry_2018}, the group theory classification for the $D_{4h}$ space group was extended to the Shubnikov group of the second kind to include odd-frequency superconductivity.
They do so by including $\hat{T}^*$, the exchange of relative-time, to the symmetry operations.
However, $\hat{T}^*$ is not necessarily a symmetry of the normal state, since the actual symmetry of the normal state is TR $\hat{\mathcal{T}}$.
It involves all electronic degrees of freedom while $\hat{T}^*$ only affects $i\omega_n$. 
%The SCOP is time-reversal invariant when we have $\hat{\mathcal{T}} \bm \Delta = e^{i\delta} \bm \Delta$, where $\delta$ is an arbitrary real number. .

As shown with two proofs in Appendix~\ref{sec:one-p-G_prop}, under $\hat{\mathcal{T}}$, the Gorkov function $\bm F$ transforms like 
\begin{align}
    \label{eq:gorkov-time-rev_main}
    \left[ \hat{\mathcal{T}} \bm F \right]^{\sigma_1\sigma_2}_{\textbf{k}l_1l_2}(i\omega_m) = \epsilon_{\sigma_1\sigma_2} \bm F^{-{\sigma}_1-{\sigma}_2}_{-\textbf{k}l_1l_2}(-i\omega_m)^*
\end{align}
where $\epsilon = \delta_{\sigma_1\sigma_2} - \delta_{-\sigma_1\sigma_2}$.
In the ordered phase, TR symmetry is either conserved or broken by the Gorkov function, with $\hat{\mathcal{T}}\bm F = \pm \bm F$.
On the other hand, $\bm F$ is not necessarily an eigenstate of $\hat{T}^*$ and a multi-orbital state that respects TR symmetry can still exhibit the coexistence of even- and odd-frequency correlations in the presence of SOC.
%However, even is not invariant under a global phase shift $\bm F \rightarrow e^{i\phi} \bm F$ and leads to $\hat{\mathcal{T}}\bm F \rightarrow e^{-2i\phi} \hat{\mathcal{T}} \bm F$. 
%Unfortunately, taking $\phi \rightarrow \phi + \frac{\pi}{2}$ leads to $\hat{\mathcal{T}}\bm F = - \bm F$, which prohibits the actual calculation of the sign associated to time-reversal symmetry.
%A consequence of this global phase ambiguity implies 
Our solutions to the Eliashberg equation all satisfy \eref{eq:gorkov-time-rev_main}, up to a global phase discussed in Appendix~\ref{sec:one-p-G_prop}.
Such a global phase has no consequence of observables.
% Because of this phase, we do not know whether the eigenvalue of $\bm F$ to $\hat{\mathcal{T}}$ is $+$ or $-1$.
%While we verify that for each solution of \eref{eq:reduced_eliashberg} we can find a global phase shift $\phi$ for which $\hat{\mathcal{T}}\bm F = \bm F$, we cannot use this relation to verify whether we have even-time-reversal or odd-time-reversal states.

Since SCOPs in multi-orbital systems with SOC are not purely even or odd under the $\hat{T}^*$ operation, we use an in-between criterion to label states.
The intra-orbital components of the SCOPs are always pure in Matsubara frequencies.
Thus, in the spirit of Ref.~\citenum{geilhufe_symmetry_2018}, we use their $SPOT$ character to label irreps with the superscript $+$ ($-$), given they have a $^+T$ ($^-T$) character.

\begin{figure}[b]
    \centering
    \includegraphics[width=\linewidth]{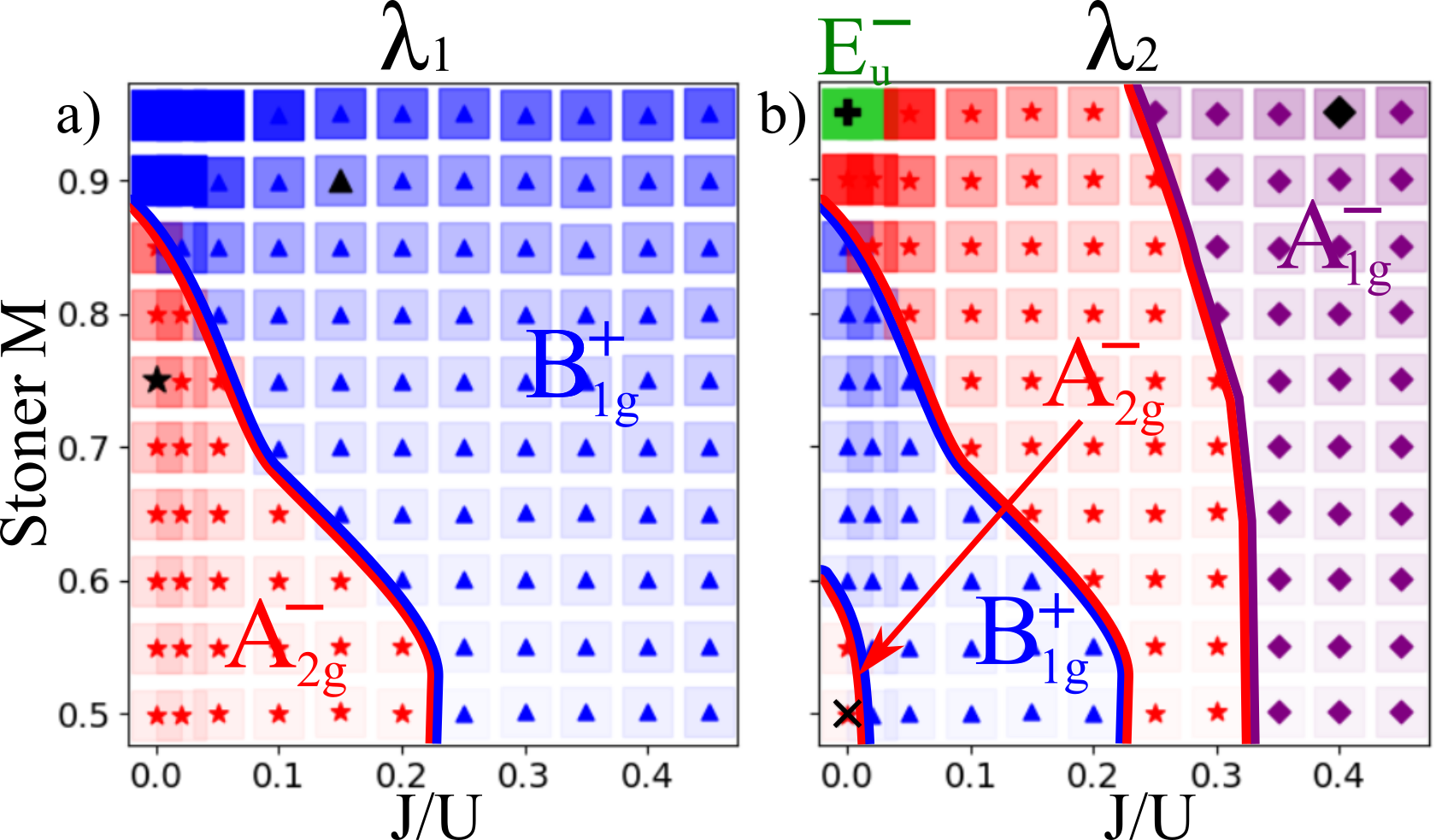}
    \caption{(Color online) Global irrep of the a) Leading and b) first subleading eigenvectors at $T=250$~K and for a $24\times24\times2$ $\textbf{q}$-grid.
    The parameter space is constructed as a function of the dressed parameters $J/U$ and $S^m$ given in \fref{fig:stoners}~c).
    The transparency of the background square represents the size of the eigenvalue between zero and one. Notice the $\blacktriangle$, $\bigstar$, \ding{53}, $\blacklozenge$ and \ding{58} symbols associated to eigenvectors with global irreps B$_{1g}^+$, two different A$_{2g}^-$, A$_{1g}^-$ and E$_u^-$ respectively. These specific solutions will be discussed further in the text.}
    \label{fig:new_phase_diag}
\end{figure}

\begin{figure*}
    \centering
    \includegraphics[width=\linewidth]{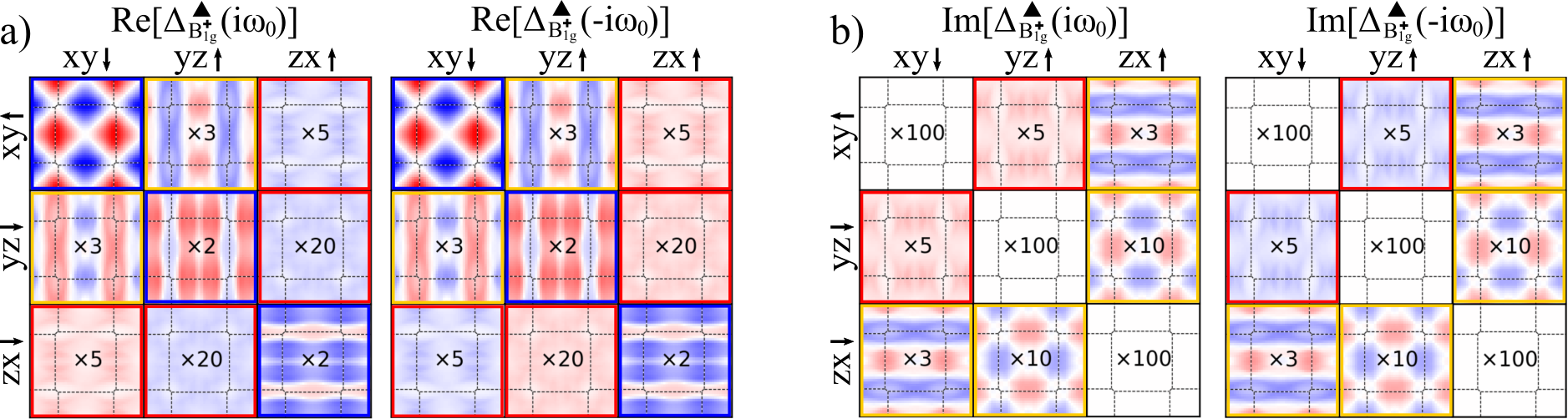}
    \caption{a) Real and b) imaginary parts at the first positive $i\omega_0$ and negative $-i\omega_0$ Matsubara frequencies of the leading B$_{1g}^+$ gap function denoted  $\blacktriangle$ in \fref{fig:new_phase_diag}. Each entry of a $3\times 3$ matrix is the momentum distribution in the first Brillouin zone for k$_z=0$ and k$_z=2\pi/c$ of a component of the inter-pseudospin gap function. Colors go from $-1$ (blue) to $1$ (red) and components were rescaled with a coefficient printed at their $\Gamma$ point. The colors around the squares show the $SPOT$ character of each component: blue, red and orange correspond to $^-S^+P^+O^+T$, $^+S^+P^+O^-T$ and $^+S^+P^-O^+T$ respectively, as in Table~\ref{tab:spot_contributions_b1g_a2g}.}
    \label{fig:gaps_b1g} 
\end{figure*}

\section{\label{sec:supercond_lead}Leading and subleading eigenvectors}
We now present solutions to the effective Eliashberg equation \eref{eq:effective_eliashberg} in the inter-pseudospin channel.
The \textit{p-h} channel is dressed using the RPA vertex, as explained in \sref{sec:spin-charge_ph}, at $T=250$~K, on a $24\times24\times2$ $\textbf{q}$-grid and with parameters shown in \fref{fig:stoners}~c).
The resulting leading eigenvector symmetries in terms of global irreps are shown in \fref{fig:new_phase_diag}~a) while in b) we show the first subleading symmetries.
They are labelled using irreps defined in \sref{sec:supercond_group} and the transparency of each square is proportional to the size of the eigenvalue, between zero and one.
We find various possibilities, but only two leading instabilities.
For each different irrep, we selected a point in parameter space where we will present actual gap functions in the following discussion.
In particular, we identify the $SPOT$ decomposition the following way: from the solution of \eref{eq:effective_eliashberg} in a given parity channel, we can express it in pseudospin-orbital space using \eref{eq:gap_function_parity}.
We then rotate it to the spin-orbital space and apply the $\hat{S}$, $\hat{P}^*$, $\hat{O}$ and $\hat{T}^*$ operators.
We find that many different $SPOT$ contributions are associated to different real and imaginary parts of the spin-orbital components, although they all transform like the same unique irrep.
The interference between different $SPOT$ contributions lead to complex numbers with a non-trivial distribution of angles in the complex plane.

This analysis is performed for the B$_{1g}^+$ state in \sref{sec:lead_b1g} and for the A$_{2g}^-$ in \sref{sec:lead_a2g}.
The phase diagram naturally suggests the possibility of an accidental degeneracy between these two, which could explain the experimental discrepancies.
We then discuss the temperature dependence of the solutions in \sref{sec:lead_temp}, along with subleading candidates and reasons why we do not find some of the other SCOPs proposed for SRO in \sref{sec:lead_sub}.

\begin{table}[b]
    \centering
    \caption{$SPOT$ decompositions of the gap functions denoted by $\blacktriangle$ and $\bigstar$ in \fref{fig:new_phase_diag}. Here, $\mathcal{P}^{SPOT}\bm \Delta$ is the ratio of the absolute value of the projected gap function $\bm \Delta$ for a specific $SPOT$ on the total one. Each $SPOT$ is specified by a color that is used in Figs~\ref{fig:gaps_b1g} to \ref{fig:a2g_phase}.}
\begin{tabular}{c|c|c|c|c|c}
    \label{tab:spot_contributions_b1g_a2g}
    \centering
    \ $S$ \ & \ $P$ \ & \ $O$ \ & \ $T$ \ & $\mathcal{P}^{SPOT} \bm \Delta^{\blacktriangle}({\text{B}_{1g}^+})$ & $\mathcal{P}^{SPOT} \bm \Delta^{\bigstar}({\text{A}_{2g}^-})$  \\
    \hline
    \hline
    \rowcolor{blue!30} - & + & + & + & 95\% & $<$1\% \\
    \rowcolor{red!30} + & + & + & - & 10\% & 81\% \\
    \rowcolor{orange!30} + & + & - & + & 29\% & 58\% \\
    \rowcolor{Green!30} - & + & - & - &     0\% & $<$1\%
\end{tabular}
\end{table}

\subsubsection{\label{sec:lead_b1g}The B$_{1g}^+$ irrep.}
Denoted by a $\blacktriangle$ in \fref{fig:new_phase_diag}, the $J/U = 0.15$ and $S^m = 0.9$ parameter set has a B$_{1g}^+$ leading irrep.
This symmetry is a prime candidate that could form an accidental degeneracy in SRO.
It is usually discussed as having a purely $^-S^+P^+O^+T$ character, neglecting the entanglement of $SPOT$ characters.
However, SOC leads to entanglement of the electronic quantum numbers and the $SPOT$ decomposition has $^-S^+P^+O^+T$, $^+S^+P^-O^+T$ and $^+S^+P^+O^-T$ characters, given in Table~\ref{tab:spot_contributions_b1g_a2g}.
Those can be understood by studying the real and imaginary parts of the spin-orbital components of $\bm \Delta^{\blacktriangle}({\text{B}^+_{1g}})$, shown in \fref{fig:gaps_b1g}~a) for the first positive and negative Matsubara frequencies.
Globally, this state transforms like a single global irrep with ubiquitous even- and odd-frequency correlations, as expected in such a multi-orbital system~\cite{black-schaffer_odd-frequency_2013}.
Although arguably responsible for the finite polar Kerr effect~\cite{komendova_odd-frequency_2017}, these odd-frequency correlations are insufficient to explain the two-component signatures in SRO, which motivates revisiting the polar Kerr experiment under uniaxial pressure. 

The intra-orbital parts have $^-S^+P^+O^+T$ character.
We make them real by selecting the appropriate global phase and the gap function transforms like $\hat{\mathcal{T}} \bm \Delta = \bm \Delta$ under TR symmetry.
These components are the largest and because they are spin-singlet $^-S$, they are characterized by the $\ket{l;l} \otimes \sigma_2$ basis functions defined in \sref{sec:supercond_group}.
While the $xy;xy$ component transforms like A$_{1g}$ in spin-orbital space, it transforms like a B$_{1g}$ $d_{x^2-y^2}$ function in $\textbf{k}$-space, making it globally B$_{1g}$.
The $yz;yz$ and $zx;zx$ components transform into each other and their phase difference implies that they transform like the B$_{1g}$ orbital basis function in \tref{tab:basis_gap_functions_orb}.
In $\textbf{k}$-space, they transform like A$_{1g}$ making them globally B$_{1g}$.
Because of the $^+T$ character of these intra-orbital components, this gap function is labelled B$_{1g}^+$.

These intra-orbital components involve opposite spins.
Because of SOC, the orbital sectors are coupled by spin-flip processes as follows: one of the paired $xy$ electrons of a $xy;xy$ pair can propagate to the $\{xz,yz\}$ orbitals by flipping a spin (yet preserving pseudospin).
The resulting inter-orbital pair involving one $xy$ electrons with one $yz$ or $zx$ must now have equal spins, {\it i.e.} form a triplet pair with $^+S$.
Having a fixed $^+P$, there are two possible $SPOT$ characters that can entangle: $^+S^+P^-O^+T$ and $^+S^+P^+O^-T$.
Inspecting \fref{fig:gaps_b1g}, we see that the real parts of $xy;yz$ and $yz;xy$ and imaginary parts of $xy;zx$ and $zx;xy$ have $^+S^+P^-O^+T$ character and can be expressed using the B$_{1g}$ $\ket{1-} \otimes \sigma_0 - i\ket{2-} \otimes \sigma_3$ basis function.
On the other hand, the imaginary parts of $xy;yz$ and $yz;xy$ and real parts of $xy;zx$ and $zx;xy$ have $^+S^+P^+O^-T$ character and can be expressed using the B$_{1g}$ $\ket{1+} \otimes \sigma_0 - i\ket{2+} \otimes \sigma_3$ basis function.
These $xy;\{yz, zx\}$ inter-orbital components are smaller yet similar in magnitude to the $xy;xy$ one and also transform like B$_{1g}^+$.

Moreover, the other $xy$ electron of the initial pair can also flip its spin and propagate to $\{yz, zx\}$ orbitals.
All orbital sectors are connected and the intra-orbital $\bm \Delta_{yz,yz}^{-\sigma,\sigma}$ and $\bm \Delta_{zx, zx}^{-\sigma,\sigma}$ are the second largest components.
They globally transform like the $\bm \Delta_{xy, xy}^{\sigma, -\sigma}$ component as B$_{1g}^+$ with $^-S^+P^+O^+T$ character.
The inter-orbital components $yz;zx$ and $zx;yz$ are much smaller than the others.
Their real (imaginary) parts have $^+S^+P^+O^-T$ ($^+S^+P^-O^+T$) character expressed using $\left[ \ket{yz;zx} + \ket{zx;yz}\right] \otimes \sigma_1$ ($\left[ \ket{yz;zx} - \ket{zx;yz}\right] \otimes \sigma_1$).
The irreps associated to those components are B$_{2g}$ (A$_{2g}$) in orbital space, A$_{2g}$ (A$_{2g}$) in spin space and A$_{1g}$ (B$_{1g}$) in $\textbf{k}$-space, thus globally transforming like B$_{1g}$.

\begin{figure}[t]
    \centering
    \includegraphics[width=.6\linewidth]{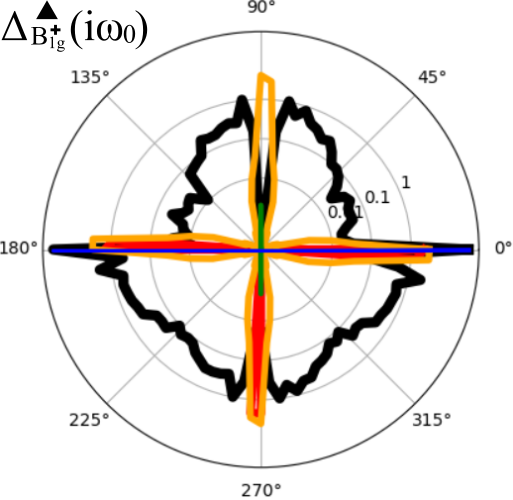}
    \caption{(Color online) Weighted distributions of the phases of the complex numbers entering the $\blacktriangle$ gap function at $i\omega_0$, defined in Eq.~(\ref{eq:phase_distribution}). The black line is associated to the total gap. The other colors are the distributions of each $SPOT$ contribution of Table~\ref{tab:spot_contributions_b1g_a2g}, with corresponding colors.}
    \label{fig:b1g_phase}
\end{figure}

Although there are multiple $SPOT$ contributions to the state, it still satisfies the TR condition \eref{eq:gorkov-time-rev_main}.
However, the transformation properties of individual components involve complex numbers, which means that the gap function represented by the $3\times 3$ matrices cannot simply be written with real numbers. 
To illustrate this fact, we show in \fref{fig:b1g_phase} a polar plot of the weighted distribution of phases in the complex plane of the $\blacktriangle$ B$_{1g}^+$ gap function \fref{fig:gaps_b1g} at $i\omega_0$, on a semi-log scale.
It is defined by 
\begin{equation}
    \label{eq:phase_distribution}
    \mathcal{D}(\phi) = \sum_{\textbf{k} \mu_1\mu_2}| \bm \Delta^{\mu_1\mu_2}_{K}| \ \text{with} \ \phi \ \text{in} \ \bm \Delta_K^{\mu_1\mu_2} = |\bm \Delta_K^{\mu_1\mu_2}| e^{i\phi}.
\end{equation}
The black line corresponds to the distribution of phases when we include all the matrix elements entering the gap functions. The colored distributions are associated to a single $SPOT$ characters, with colors associated to characters as in Table~\ref{tab:spot_contributions_b1g_a2g} and the contours of \fref{fig:gaps_b1g}.
For a given $SPOT$ contribution, the distribution has mostly $\phi \in \{0, \frac{\pi}{2}, \pi, \frac{3\pi}{2} \}$ angles, in other words a given SPOT contribution can be represented by purely real or imaginary numbers.
The purely imaginary ones are imposed by the $C_4$ symmetry operation detailed in Appendix~\ref{sec:group}, acting on the spin-space of inter-orbital components.
Including all $SPOT$ contributions creates the wild phase distribution seen in black in \fref{fig:b1g_phase}.
This kind of distribution emerging from the interference of different $SPOT$ contributions is unavoidable in multi-orbital systems with SOC.
It is not clear how these phases affect physical observables.

\begin{figure*}
    \centering
    \includegraphics[width=\linewidth]{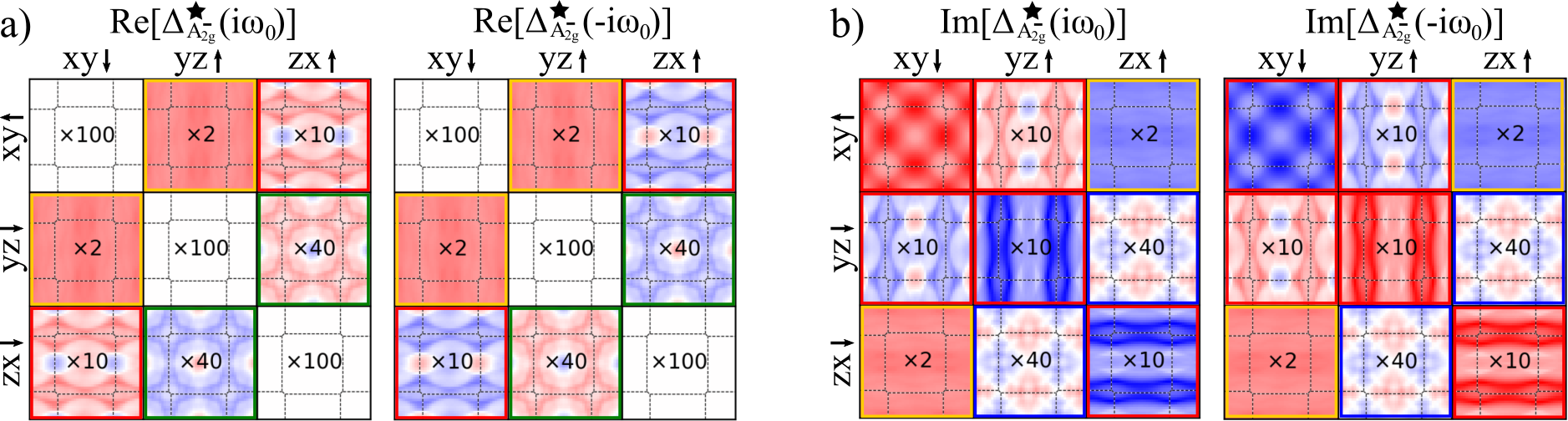}
    \caption{a) Real and b) imaginary parts at the first positive and negative Matsubara frequencies of the leading A$_{2g}^-$ gap function denoted  $\bigstar$ in \fref{fig:new_phase_diag}. See \fref{fig:gaps_b1g} for more details of what is shown. The additional $^-S^+P^-O^-T$ component is highlighted in green.}
    \label{fig:gaps_a2g}
\end{figure*}

\subsubsection{\label{sec:lead_a2g}The A$_{2g}^-$ irrep.}
The other leading irrep in \fref{fig:new_phase_diag} is the A$_{2g}^-$ gap function. 
We study the one found at $J/U=0$ and $S^m=0.75$ and denoted by the star ($\bigstar$) symbol, where its real and imaginary parts are shown in \fref{fig:gaps_a2g} for the first positive and negative Matsubara frequencies.
It has $^+S^+P^+O^-T$ intra-orbital components, which explains the minus superscript in our notation.
This state is a generalization of regular odd-frequency to a multi-orbital system where SOC leads to multiple coexisting $SPOT$ contributions.
In this case, the dominant characters are $^+S^+P^-O^+T$ and $^+S^+P^+O^-T$ in the proportions given in Table~\ref{tab:spot_contributions_b1g_a2g}, with negligible contributions from $^-S^+P^-O^-T$ and $^-S^+P^+O^+T$ characters.

The dominant components are intra-orbital and transform like $^+S^+P^+O^-T$.
If we make them real by choosing the appropriate global phase, we find that the whole gap transforms like $\hat{\mathcal{T}} \bm \Delta = \bm \Delta$ under TR symmetry.
If instead we apply an additional $\pi/2$ phase shift, these intra-orbital components are purely imaginary and the whole gap now transforms like $\hat{\mathcal{T}} \bm \Delta = - \bm \Delta$.
This global phase invariance is detailed in \sref{sec:supercond_group_time-rev}.

The intra-orbital $xy;xy$ component is the largest and is characterized by the $\ket{xy;xy} \otimes \sigma_1$ basis function defined in \sref{sec:supercond_group_spin-orb}, which transforms like A$_{1g}$ in orbital space and is a spin zero A$_{2g}$ triplet in spin space.
In $\textbf{k}$-space, it is almost uniform, transforming like A$_{1g}$ so that overall this component transforms like A$_{2g}$.
The other intra-orbital components are much smaller and have the same phase.
They transform like $\left[\ket{yz;yz} + \ket{zx;zx} \right] \otimes \sigma_1$ in spin-orbital space and they are almost uniform in $\textbf{k}$-space, thus they globally transform like A$_{2g}$ as well. 

Now, because the intra-orbital components are $^+S$ for opposite spins $\sigma_1 = - \sigma_2$, a single spin-flip induced by SOC generates inter-orbital components that are necessarily $^+S$ as well.
This explains why the dominant $SPOT$ contributions are both triplet.
Those are of the same order of magnitude as the dominant $xy;xy$ one.
The real part of $xy;yz$ and imaginary part of $xy;zx$ can be expressed as the $\ket{1+} \otimes \sigma_3 + i \ket{2+}\otimes \sigma_0$ basis function with $^+S^+P^-O^+T$ character, while the imaginary part of $xy;yz$ and real part of $xy;zx$ is instead of the form $\ket{1-} \otimes \sigma_3 + i \ket{2-} \otimes \sigma_0$ with $^+S^+P^+O^-T$ character.
Both of these transform like A$_{2g}$.
Looking at the small $yz;zx$ components, their real (imaginary) parts have $^-S^+P^-O^-T$ ($^-S^+P^+O^+P$) characters.
All components consistently transform like A$_{2g}$ globally.

\begin{figure}[b]
    \centering
    \includegraphics[width=.6\linewidth]{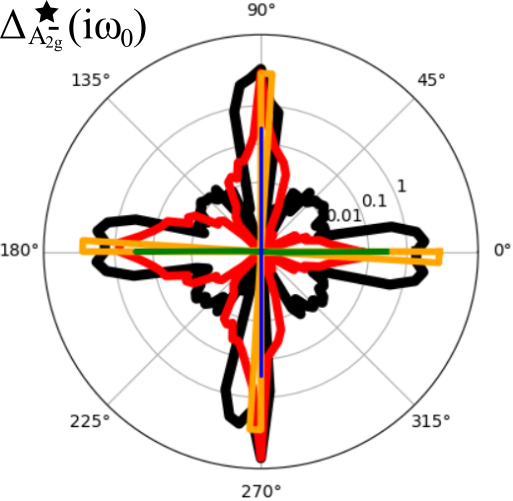}
    \caption{(Color online) Weighted distributions of the phases of the complex numbers entering the $\bigstar$ gap function at $i\omega_0$. The black line is associated to the total gap. The other colors are the distributions of each $SPOT$ contribution of Table~\ref{tab:spot_contributions_b1g_a2g}, with corresponding colors.}
    \label{fig:a2g_phase}
\end{figure}

Again, we can look at the weighted distribution Eq.~(\ref{eq:phase_distribution}) of phases in the complex plane shown in \fref{fig:a2g_phase}, for all components of the gap function in black and for the individual $SPOT$ contributions are shown in \fref{fig:gaps_a2g} with colors as in Table~\ref{tab:spot_contributions_b1g_a2g}.
In this case, most individual $SPOT$ contributions have only complex phases as $\phi \in \{0, \frac{\pi}{2}, \pi, \frac{3\pi}{2}\}$, but the $^+S^+P^+O^-T$ contribution has additional interference coming from the inter-orbital components.
We cannot explain why that is.
Overall, all $SPOT$ contributions again interfere, leading to a wild distribution of complex number phases for which it is not clear whether a signature can be observed in experiments, motivating further investigations.
% When isolating a $SPOT$ contribution, it then has a distribution that has mostly $\phi \in \{0, \frac{\pi}{2}, \pi, \frac{3\pi}{2} \}$ angles, that is purely real or imaginary phases.
% The purely imaginary ones are imposed by the $C_4$ symmetry operation acting on spin-space of inter-orbital components.
% Combining different $SPOT$ contributions generates interference which creates a wild phase distribution.
% Because it is caused by SOC, this distribution is intrinsic and will always be observed.
% It is not clear if those phase differences lead to time-reversal symmetry breaking signatures.

Now, because this state has spin-triplet intra-orbital components in the inter-pseudospin channel, it is expected to have nearly degenerate intra-pseudospin solutions with a similar structure.
This is what unconverged calculations confirm.
Because the A$_{2g}$ character of the different components comes from the spin degrees of freedom, the intra-pseudospin solutions instead have E$_{g}$ spin irreps and globally transform like two degenerate E$_g^-$ states.

The interest in this A$_{2g}^-$ SCOP is two-fold.
First, it transforms like A$_{2g}$, consistent with ultrasound experiments.
Second, intra-orbital odd-frequency SCOP are gapless at the Fermi level~\cite{balatsky_new_1992}.
This property implies that their effect on the Fermi surface of SRO could be negligible enough that it has a negligible signature in specific heat.
A similar behavior is expected for the even-frequency components since, being inter-orbital, they open gaps where the bands with the same orbital characters cross, which is mostly away from the Fermi level for different characters.
Consequently, an accidental degeneracy between B$_{1g}^+$ and A$_{2g}^-$ in the system at ambient pressure could appear as having two components and breaking TR symmetry.
Applying uniaxial strain, one can imagine that the B$_{1g}^+$ acquires a larger critical temperature while the A$_{2g}^-$ remains almost constant.
As a result, one would observe a first transition when entering the B$_{1g}^+$ state and a second one when entering the B$_{1g}^+\pm e^{i\phi} \text{A}_{2g}^-$, which could suddenly break TR symmetry, with a negligible signature in specific heat.
Such a scenario should be further investigated.

% {\color{Green}
% Another possibility is the accidental degeneracy of two one-dimensional irreps in the unstrain case. Such a coincidence should be easily perturbed, yet isotropic strain was unable to produce the same result as uniaxial strain~\cite{grinenko_unsplit_2021}. Nevertheless, this is not a definitive evidence as isotropic strain do not affect much the electronic structure of SRO and could preserve the accidental degeneracy within error bars. Thus, many such proposals were made. The most discussed is the $d+ig$ solution~\cite{kivelson_proposal_2020, yuan_strain-induced_2021, clepkens_higher_2021}
% but also proposals for $d+s'$~\cite{romer_knight_2019, romer_theory_2020, romer_fluctuation-driven_2020}. Even \textit{e-p} + \textit{o-p}~\cite{scaffidi_degeneracy_2020}. Non-magnetic impurity should lift degeneracy~\cite{zinkl_impurity-induced_2021}.

\begin{figure}[b]
    \centering
    \includegraphics[width=\linewidth]{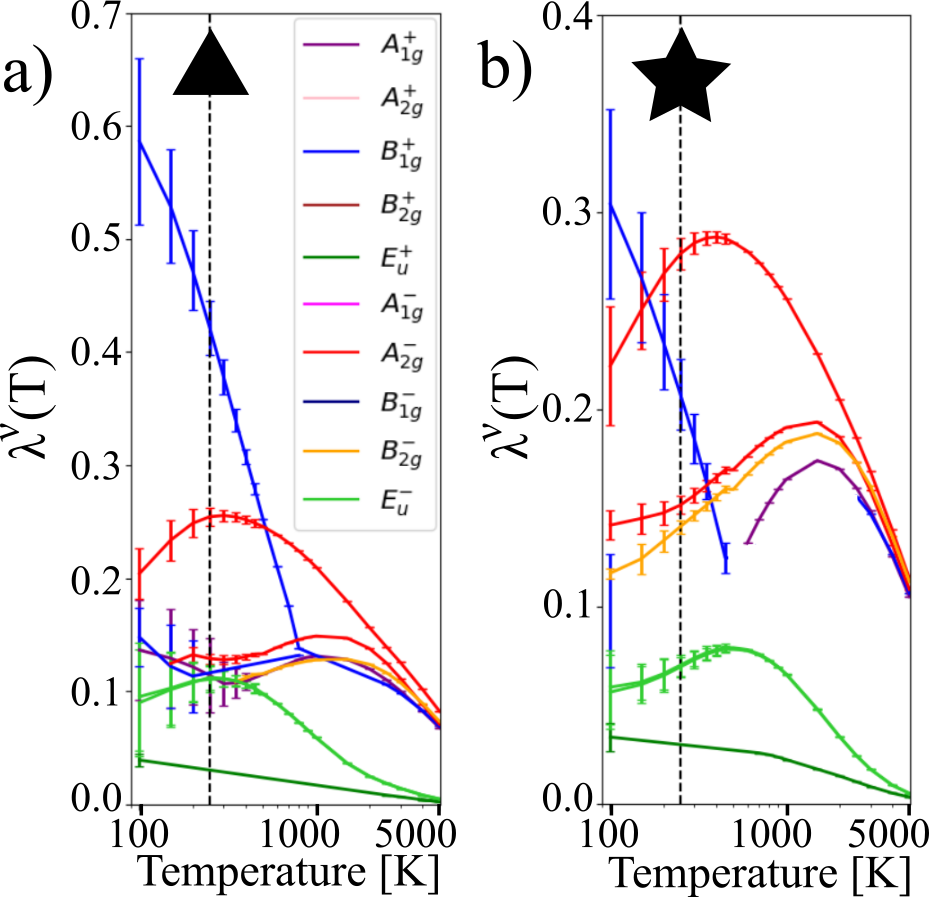}
    \caption{(Color online) Eeigenvalues' temperature dependence at the a) $\blacktriangle$ and b) $\bigstar$ points of \fref{fig:new_phase_diag}. Error bars explained in Appendix~\ref{sec:eig_conv}.}
    \label{fig:b1g_a2g_temp_dep}
\end{figure}

\subsubsection{\label{sec:lead_temp}Temperature dependence.}
While the eigenvectors of \fref{fig:new_phase_diag} have largest eigenvalues, these are lower than one, meaning that the system remains in the normal state.
%This is because we are looking at instabilities from the normal state.
The assumption is that these eigenvalues are enhanced by lowering temperature and at some point become unity, signaling a phase transition.
The temperature dependence of the eigenvalue is very important as a subleading eigenvalue might rise faster than the leading one.
We studied this possibility for the $\blacktriangle$ and $\bigstar$ points of the phase diagram, which have leading eigenvectors $\bm \Delta^{\blacktriangle}({\text{B}^+_{1g}})$ and $\bm \Delta^{\bigstar}({\text{A}^-_{2g}})$.
The results are shown in \fref{fig:b1g_a2g_temp_dep}~a) and b), respectively.
The eigenvalues are very challenging to calculate at lower temperatures where the convergence in $\textbf{k}$-points and fermionic frequencies becomes increasingly difficult.
The way the error bars are obtained is explained in Appendix~\ref{sec:eig_conv}.

\begin{figure*}
    \centering
    \includegraphics[width=.88\linewidth]{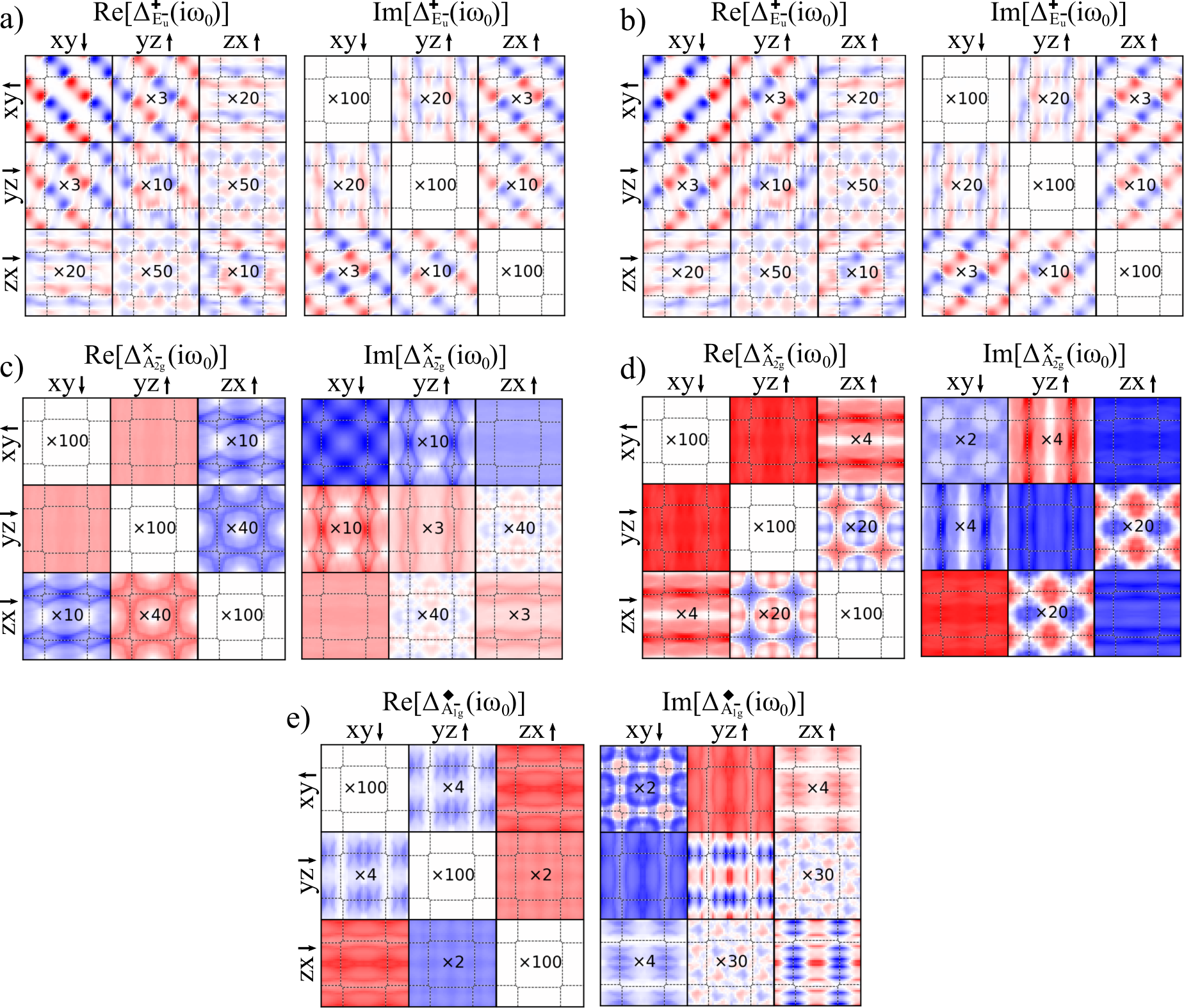}
    \caption{(Color online) Other gap functions at the points denoted by the \ding{58}, \ding{53} and $\blacklozenge$ symbols on \fref{fig:new_phase_diag}. In a) and b), eigenvectors with second and third largest eigenvalues at the \ding{58} point. They are degenerate and correspond to the two-dimensional E$_u^-$ irrep. In c) and d), first and second eigenvectors at the \ding{53} point. They both transform like different A$_{2g}^-$ irreps. In e), second largest eigenvector at the $\blacklozenge$ point, which have A$_{1g}^-$ symmetry. For more details about what is shown, see \fref{fig:gaps_b1g}.}
    \label{fig:other_gaps}
\end{figure*}

We observe that the eigenvalues of states with odd-frequency intra-orbital components, denoted by the minus superscript in our notation, are not monotonous in temperature.
In other words, they have a maximal value at a finite temperature.
This re-entrance property has been predicted in odd-frequency superconductors and could lead to a metal-superconductor-metal transition induced by temperature~\cite{abrahams_properties_1995, kusunose_possible_2011}.
In our case, the maxima appear at high temperatures, with eigenvalues far from unity.
Because this has not been studied a lot, it is unclear whether this behavior can be avoided with better choices of interactions and normal states, leading to enhanced maximum eigenvalues at lower temperatures.
Otherwise, these states might never trigger a phase transition.
However, the B$_{1g}^+$ state behaves normally and rises as temperature is lowered, guaranteeing its importance as a potential state to form an accidental degeneracy.

\subsubsection{\label{sec:lead_sub}Subleading gap functions.}
As the temperature dependence suggests, it is not impossible that subleading gap functions become leading when lowering temperature further.
We briefly present a few of the other symmetries present in the phase diagram, denoted by the \ding{58}, \ding{53} and $\blacklozenge$ symbols on  \fref{fig:new_phase_diag}~b) and shown in \fref{fig:other_gaps}.

At the \ding{58} symbol, there are both large amounts of spin and charge fluctuations, with both Stoner factors around 95\%.
In this parameter regime, the second and third largest eigenvalues are degenerate and transform like the E$_u^-$ irrep.
The largest component in pseudospin-orbital basis is the intra-orbital $xy;xy$, classified as a $^-S^-P^+O^-T$.
The second largest are the inter-orbital $xy;yz/zx$ ones.
These gap functions are shown in \fref{fig:other_gaps}~a) and b).
One can see that they are complementary, exhibiting the fact that they represent the same two-dimensional irrep.
The symmetry protected degeneracy between these states would naturally be broken under uniaxial conditions. However, they would have either zero or two temperature transitions under uniaxial strain.

At the \ding{53} symbol, both the spin and charge fluctuations are relatively low around 50\%.
In this region of parameter space, both the first and second eigenvectors transform like A$_{2g}^-$, a one-dimensional irrep.
They are shown in \fref{fig:other_gaps}~c) and d).
These two gaps should be linearly independent.
The first one has similar relative amplitudes and momentum structure as the $\Delta^{\bigstar}({\text{A}_{2g}^-})$ gap function.
This is expected as they are part of the same region in the phase diagram \fref{fig:new_phase_diag}.
The second gap is similar in structure, but the magnitude between different components is different, and most strickingly, the phase between the intra-orbital $xy;xy$ component and the $yz;yz$ and $zx;zx$ ones is opposite, which makes it orthogonal to the A$_{2g}^-$ state.

At the $\blacklozenge$ symbol, spin fluctuations are very large with a magnetic Stoner factor of 95\%, while charge fluctuations are unenhanced by correlations.
At this point in parameter space, the second eigenvector transforms like the A$_{1g}^-$ irrep, shown in \fref{fig:other_gaps}~e).
Its intra-orbital components have a lot of gradients and sharp values due to extremely sharp peaks in the dressed \textit{p-h} susceptibility.
Equivalently important components are the uniform inter-orbital $xy;yz$ and $xy;zx$ ones.

We notice that all these second leading eigenvectors have odd-frequency intra-orbital components.
We do not find the even-frequency solutions that could've been expected for SRO.
First, the $d_{zx}\pm id_{yz}$ state that transforms like the E$_g$ irrep and could originate from $\textbf{k}$-SOC~\cite{suh_stabilizing_2020, clepkens_shadowed_2021} is vanishing at k$_z=0$ and k$_z=\frac{2\pi}{c}$.
To study this state, we need to include more k$_z$ dispersion and cannot use the pseudospin reduction of the Eliashberg equation \eref{eq:reduced_eliashberg}.
However, we do not expect this state to be dominant, since the $\textbf{k}$-SOC obtained from DFT is negligibly small and SRO is well known to be quasi-2D~\cite{hussey_normal-state_1998}.
Second, the odd-frequency inter-orbital E$_g$ found when computing the dressed \textit{p-h} susceptibility using DMFT~\cite{kaser_inter-orbital_2021} was obtained when neglecting SOC.
Restoring it, the gap function would need to involve all orbital sectors, in which case intra-orbital E$_g$ would necessarily involve a strong $k_z$ dispersion like the $d_{zx} \pm i d_{yz}$ state or higher order harmonics, which is improbable.
It might however be compatible with the E$_g^-$ intra-pseudospin state analogous to $\bm \Delta^\bigstar$.
Third, although the extended $s$-wave proposed in some works~\cite{raghu_theory_2013, romer_knight_2019, romer_fluctuation-driven_2020} was found when neglecting SOC~\cite{gingras_superconducting_2019}, it appears only in the $yz/zx$ orbital sector because it is originating from fluctuations caused by the quasi-one-dimensional nesting vector.
Again, considering strong SOC implies that all sectors are active, but the A$_{1g}$ irrep is incompatible with the $xy$ orbital sector due to the repulsive nature of the singlet Coulomb interaction.
Fourth, various works have proposed the $g_{xy(x^2-y^2)}$ A$_{2g}$ state~\cite{kivelson_proposal_2020, yuan_strain-induced_2021, clepkens_higher_2021, Wagner_GL_2021}.
As a higher angular momentum version of $d_{x^2-y^2}$, this state might involve too many gradients in $\textbf{k}$-space, thus reducing its eigenvalue.

From our results, it seems that the only state with even-frequency intra-orbital components that has a large enough eigenvalue to be considered a leading eigenvector for SRO is the B$_{1g}^+$ state.
Our calculations suggest that the other dominant states have odd-frequency intra-orbital components, making this kind of state an important case that should be further studied.

\section{\label{sec:concl}Conclusion}
Unconventional superconductivity in strongly correlated systems is a phenomenon involving many degrees of freedom.
Consequently, the space of possible order parameters is broad and identifying the right symmetry is a challenging process.
In this paper, we present a general formalism to study frequency-dependent correlation-enhanced pairing, starting from realistic electronic structures that characterize the normal states of multi-orbital systems with SOC.
We apply it to the archetypal unconventional superconductor SRO for which the normal state is extremely well understood, while the superconducting one continues to evade clear theoretical propositions with seemingly contradicting experimental evidences.

We start by projecting DFT wave functions on a localized basis set of $4d$-$t_{2g}$ orbitals of the Ru atom. 
We discuss how SOC couples orbitals with different spins, yet preserves a pseudospin symmetry for intra-layer wave-vectors.
%Strong local correlations are incorporated through \textit{p-h} vertex corrections and the Bethe-Salpeter equation.
We use the RPA approximation to obtain dressed \textit{p-h} susceptibilities.
While SOC entangles the spin and charge channels, the magnetic and density Stoner factors remain relevant indicators of the amount of spin and charge fluctuations.

In \textit{p-h} fluctuation mediated pairing, correlation-enhanced pairing is studied by inspecting \textit{p-p} vertex corrections through the Eliashberg equation.
It is extremely difficult to solve when considering all momenta, orbitals, spins and frequencies, the relevant quantum numbers.
Using pseudospin symmetry and Pauli principle, we map the inter-pseudospin channel to an equation that does not involve spins, thus greatly reduces the numerical costs of the Eliashberg equation.

Because of SOC, we find complex and rich SCOPs that involve all quantum numbers.
The ubiquitous coexistence between even and odd contributions in the exchange of spins, relative momenta, orbitals and relative frequencies invites us to define $SPOT$ decomposition to characterize SCOPs.
Moreover, we show under which irreps of the $D_{4h}$ space group each quantum number transforms.
 We use the pure $SPOT$ character of the intra-orbital components to extend the space of irreps to include states with odd-frequency intra-orbital components.

In SRO, we find two different leading symmetries to the Eliashberg equation: a B$_{1g}^+$ and an A$_{2g}^-$ that have even- and odd-frequency intra-orbital components, respectively.
We discuss how the leading spin-orbital components naturally lead to coexistence of $SPOT$ contributions due to SOC.
We study the temperature dependence of these leading states and find that gap functions with odd-frequency intra-orbital components have a maximum eigenvalue at a finite temperature.
We also observe that SOC imposes all orbital sectors to be active, which greatly reduces possible SCOPs.
The only state with even-frequency intra-orbital components is the B$_{1g}^+$ state, while all other dominant states have odd-frequency intra-orbital components.
We contend that this generalized picture with extended possible SCOP symmetries is an important step in understanding completely the phenomenon of unconventional superconductivity and that it has the potential to explain the mysterious experimental signatures of SRO.

\begin{acknowledgments}
    We are grateful for discussions with Michel Ferrero, Sékou-Oumar Kaba and David Sénéchal.
    This work has been supported by the Fonds de Recherche du Qu\'ebec—Nature et Technologie (FRQNT), the Hydro-Qu\'ebec fellowship, and the Université de Montréal (OG), the Research Chair in the Theory of Quantum Materials, the Canada First Research Excellence Fund, the Natural Sciences and Engineering Research Council of Canada (NSERC) under Grants No. RGPIN-2014-04584, No. RGPIN-2019-05312 (AMST), and No. RGPIN-2016-06666 (MC).
    Simulations were performed on computers provided by the Canada Foundation for Innovation, the Minist\`ere de l'\'Education, du Loisir et du Sport (MELS) (Qu\'ebec), Calcul Qu\'ebec, and Compute Canada. 
    The Flatiron Institute is a division of the Simons Foundation.
    The authors are members of the Regroupement qu\'ebécois sur les matériaux de pointe (RQMP).
\end{acknowledgments}

\appendix

\begin{widetext}
\section{\label{sec:free-energy}Derivation from free-energy.}

Consider a system ruled by the grand canonical Hamiltonian $\hat{K} = \hat{H} - \mu \hat{N}$. Each electron is characterized by a set of quantum numbers $\bm 1 \equiv (\textbf{k}_1, \sigma_1, \tau_1, l_1)$ with respective labels for its momentum, spin, imaginary time and orbital degrees of freedom. The more usual basis is in position space instead of momentum, but for our purpose, momentum is more convenient. Creation and annihilation operators for such a particle are written $\psi^{\dag}(\bm 1)$ and $\psi(\bm 1)$.  In systems with inversion $\hat{P}^*$ and time-reversal $\hat{\mathcal{T}}$ symmetries, superconducting pairs form between degenerate states related by $\hat{\mathcal{T}}$. Acted on $\psi/\psi^{\dag}$, this operation flips momentum and spin~\cite{linder_odd-frequency_2019}. 
% In this general case, we have
% \begin{equation}
%     \label{eq:time-rev_psi}
%     \hat{\mathcal{T}} \psi^{\dag}(\textbf{k}_1, \sigma_1, l_1, \tau_1) = i\sigma^y_{\sigma_1\bar{\sigma}_2}\psi^{\dag}(-\textbf{k}_1, \bar{\sigma}_2, l_1, \tau_1)
%     , \quad 
%     \hat{\mathcal{T}} \psi(\bm 1) = - \left[\hat{\mathcal{T}} \psi^{\dag}(\bm 1) \right]^{\dag}
    % \hat{\mathcal{T}} \left\{ 
    %     \begin{array}{c}
    %         \psi(\textbf{k}_1, \uparrow, \tau_1, l_1) \\
    %         \psi(\textbf{k}_1, \downarrow, \tau_1, l_1)
    %     \end{array} \right.
    % =
    % \left\{
    %     \begin{array}{c}
    %         \ \ \ \psi(-\textbf{k}_1, \downarrow, \tau_1, l_1) \\
    %         -\psi(-\textbf{k}_1, \uparrow, \tau_1, l_1)
    %     \end{array} \right.
% \end{equation}
% where the bar over $\sigma_2$ implies that this quantum number is summed over and $\sigma^y$ is the second Pauli matrix in spin space.
% ({\color{red}Is time right?})
% In spin-space, this 
We will consider general spin states, thus we employ a time and multi-orbital generalization of the Barlian-Werthammer particle-hole spinor~\cite{balian_superconductivity_1963, coleman_introduction_2015}, leading to
\begin{equation}
    \label{eq:free-en_ph-basis}
    \bm \Psi(\bm 1) = \left(
        \begin{array}{c}
            \psi(\bm 1) \\
            \psi^{\dag}(\bm 1^*)
        \end{array} \right)
    \ , \quad \quad
    \bm \Psi^{\dag}(\bm 1) = \Big( \psi^{\dag}(\bm 1) \ \ \psi(\bm 1^*) \Big)
    % \quad \text{with} \
    % \psi_{\textbf{k}}(\bm 1) = \left(
    %     \begin{array}{c}
    %         \psi_{\textbf{k}\uparrow}(\bm 1) \\
    %         \psi_{\textbf{k}\downarrow}(\bm 1)
    %     \end{array} \right)
\end{equation}
where $\bm 1^* \equiv (-\textbf{k}_1, -\sigma_1, l_1, \tau_1)$ and in this two-dimensional particle-hole (\textit{p-h}) representation, we label the first and second components with indices as $\bm \Psi_{1}(\bm 1) \equiv \psi(\bm 1)$ and $\bm \Psi_{2}(\bm 1) \equiv \psi^{\dag}(\bm 1^*)$.
% With equation \eref{eq:time-rev_psi}, we can write
% \begin{equation}
%     \bm \Psi_{\alpha}(\bm 1) = \rho^x_{\alpha\bar{\beta}}  \bm \Psi^{\dag}_{\bar{\beta}}(\bm 1^*)
%     , \quad
%     \bm \Psi^{\dag}_{\alpha}(\bm 1) =  \rho^x_{\alpha\bar{\beta}} \bm \Psi_{\bar{\beta}}(\bm 1^*)
% \end{equation}
% where again the bar over $\beta$ implies a summation in particle-hole space with $\rho^x$ the first Pauli matrix.
% Because of these connections, the components actually have the following anticommutation relations:
% \begin{equation}
%     \label{eq:anticomm_nambu_psi}
%     \left\{ \bm \Psi_{\alpha}(\bm 1), \bm \Psi^{\dag}_{\beta}(\bm 2) \right\} = \delta_{\alpha\beta}\delta(\bm 1; \bm 2)
%     \ , \quad
%     \left\{ \bm \Psi_{\alpha}(\bm 1), \bm \Psi_{\beta}(\bm 2) \right\} = \rho^x_{\alpha\beta}  \delta(\bm 1, \bm 2^*).
% \end{equation}
In order to keep canonical anticommutation relations, we restrict ourselves to $k_x>0$ so that
\begin{equation}
    \label{eq:anticomm_nambu_psi}
    \left\{ \bm \Psi_{\alpha}(\bm 1), \bm \Psi^{\dag}_{\beta}(\bm 2) \right\} = \delta_{\alpha\beta}\delta(\bm 1; \bm 2)
    \ , \quad
    \left\{ \bm \Psi_{\alpha}(\bm 1), \bm \Psi_{\beta}(\bm 2) \right\} = 0.
\end{equation}
We now consider source terms $ \bm  \phi(\bm 1; \bm 2)$ that couple as follows: $\bm \Psi^{\dag}_{\alpha}(\bm 1) \bm  \phi_{\alpha\beta}(\bm 1; \bm 2) \bm \Psi_{\beta}(\bm 2)$.
In this \textit{p-h} basis, $\phi_{11}$ and $\phi_{22}$ conserve the number of particles while $\phi_{12}$ and $\phi_{21}$ do not. 
The partition function characterizing this system including source terms is
\begin{equation}
    \label{eq:free-en_partition}
    Z[ \bm \phi] 
    = \text{Tr} \left[ T_{\tau} S[ \bm  \phi] \right] 
    \quad \text{with} \quad S[ \bm  \phi] \equiv 
    e^{-\beta \hat{K}} e^{-\bm \Psi_{\bar{\alpha}}^{\dag}(\bar{\bm 1})  \bm  \phi_{\bar{\alpha}\bar{\beta}} (\bar{\bm 1}; \bar{\bm 2}) \bm \Psi_{\bar{\beta}}(\bar{\bm 2})}
\end{equation}
with $T_\tau$ the time-ordering operator and where the bar over the quantum numbers implies that the continuous variables are integrated and the discrete ones are summed over. Similarly, the bar over the \textit{p-h} indices implies summation. Above, this integration is performed in the argument of the exponential. Even if we sum over only half of the Brillouin zone, the contributions from $\bm 1$ and $\bm 1^*$ compensate. 
Moreover, each field is independent. In other words,
\begin{equation}
    \frac{\delta  \bm  \phi_{\alpha\beta}(\bm 1; \bm 2)}{\delta  \bm  \phi_{\gamma\delta}(\bm 3; \bm 4)} = \delta(\bm 1; \bm 3) \delta(\bm 2; \bm 4)\delta_{\alpha\gamma}\delta_{\beta\delta}.
\end{equation}

The Helmholtz free-energy is $\mathcal{F}[ \bm  \phi] = - \frac{1}{\beta} \ln Z[ \bm  \phi]$ and the generalized Nambu Green function in the presence of the source fields is defined as
\begin{equation}
    \mathcal{\bm G}(\bm 1; \bm 2;  \bm  \phi) = - \frac{1}{Z[ \bm  \phi]} \text{Tr} \left[ T_{\tau} S[ \bm  \phi] \bm \Psi(\bm 1) \bm \Psi^{\dag}(\bm 2) \right].
\end{equation}
Each component of the Green function is related to the free-energy by
\begin{align}
    \mathcal{\bm G}_{\alpha\beta}(\bm 1; \bm 2;  \bm  \phi)
     = \beta \frac{\delta \mathcal{F}[ \bm  \phi]}{\delta  \bm  \phi_{\beta\alpha}(\bm 2; \bm 1)}
    = - \frac{\delta \ln Z[ \bm  \phi]}{\delta  \bm  \phi_{\beta\alpha}(\bm 2; \bm 1)}
    % = - \frac{1}{\beta Z} \frac{\delta Z[ \bm  \phi]}{\delta  \bm  \phi_{\beta\alpha}(2, 1)}    = & - \delta_{\alpha\beta} \delta(\bm 1, \bm 2)
     = \frac{1}{Z[ \bm  \phi]} \text{Tr} \Bigg[ T_{\tau} S[ \bm  \phi]
        \bm \Psi_{\bar{\alpha}}^{\dag}(\bar{\bm 1}) \frac{\delta  \bm  \phi_{\bar{\alpha}\bar{\beta}}(\bar{\bm 1}; \bar{\bm 2})}{\delta  \bm  \phi_{\beta\alpha}(\bm 2; \bm 1)} \bm \Psi_{\bar{\beta}}(\bar{\bm 2}) \Bigg].
\end{align}
Without source terms, we recover the system's normal and anomalous Green functions
\begin{align}
    \mathcal{\bm G}(\bm 1; \bm 2;  \bm  \phi)\Big\rvert_{ \bm  \phi=0} = - \langle T_{\tau} \bm \Psi(\bm 1) \bm \Psi^{\dag}(\bm 2) \rangle 
    % {\color{red} = \left(
    %     \begin{array}{cc}
    %         - \langle T_{\tau} \psi(\bm 1) \psi^{\dag}(\bm 2) \rangle   &
    %         - \langle T_{\tau} \psi(\bm 1) \psi(\bm 2) \rangle \\
    %         - \langle T_{\tau} \psi^{\dag}(\bm 1) \psi ^{\dag}(\bm 2) \rangle &
    %         - \langle T_{\tau} \psi^{\dag} (\bm 1) \psi(\bm 2) \rangle
    %     \end{array}
    % \right)
    % = \left(
    %     \begin{array}{cc}
    %         \bm G(\bm 1, \bm 2)   &
    %         \bm F(\bm 1, \bm 2) \\
    %         \bm F^{\dag}(\bm 1, \bm 2) &
    %         - \bm G(\bm 2, \bm 1)
    %     \end{array}
    % \right) }
    = \left(
        \begin{array}{cc}
            - \langle T_{\tau} \psi(\bm 1) \psi^{\dag}(\bm 2) \rangle
            &
            - \langle T_{\tau} \psi(\bm 1) \psi(\bm 2^*) \rangle
            \\
            - \langle T_{\tau} \psi^{\dag}(\bm 1^*) \psi^{\dag}(\bm 2) \rangle
            &
            - \langle T_{\tau} \psi^{\dag}(\bm 1^*) \psi(\bm 2^*) \rangle
        \end{array}
    \right)
    \equiv \left(
        \begin{array}{cc}
            \bm G(\bm 1; \bm 2)   &
            - \bm F(\bm 1; \bm 2) \\
            - \bar{\bm F}(\bm 1; \bm 2) &
            \bar{\bm G}(\bm 1; \bm 2)
        \end{array}
    \right)
\end{align}
where we define the particle-particle and hole-hole Gorkov functions $\bm F$ and $\bar{\bm F}$, which vanish in the normal state, along with the particle and hole propagators $\bm G$ and $\bar{\bm G}~$\cite{kaser_inter-orbital_2021}. These two are related by
\begin{equation}
    \label{eq:hole_propagator}
    \bar{\bm G}(\bm 1; \bm 2) = -\bm G(\bm 2^*, \bm 1^*).
\end{equation}

The equations of motion are given by $\mathcal{\bm G}^{-1}_{\alpha\beta}( \bm  \phi) = {\mathcal{\bm G}^0}^{-1}_{\alpha\beta} - \bm  \phi_{\alpha\beta} - \bm \Sigma_{\alpha\beta}(\phi)$ with $\mathcal{\bm G}_{\alpha\bar{\gamma}} \mathcal{\bm G}^{-1}_{\bar{\gamma}\beta} = \delta_{\alpha\beta}$ since
\begin{equation}
    \left[ \underbrace{\left( -\frac{\partial}{\partial\tau_1} - \bm H_0 - \mu\right) \delta_{\alpha\bar{\beta}}\delta(\bm 1, \bar{\bm 2})}_{[\mathcal{\bm G}_0^{-1}]_{\alpha\bar{\beta}}(\bm 1, \bar{\bm 2})} - \bm \Sigma_{\alpha \bar{\beta}}(\bm 1; \bar{\bm 2};  \bm  \phi) - \bm  \phi_{\alpha\bar{\beta}}(\bm 1; \bar{\bm 2}) \right] \mathcal{\bm G}_{\bar{\beta}\beta}(\bar{\bm 2}; \bm 2;  \bm  \phi) 
    =
    \delta_{\alpha\beta}\delta(\bm 1; \bm 2)
\end{equation}
% where the self-energy was defined by ({\color{red} How to generalize Eq. 36.12/36.13? Define $\bm V$ and $^+$.})
% \begin{equation}
%     \bm \Sigma_{\alpha\bar{\beta}}(\bm 1; \bar{\bm 2};  \bm  \phi)\mathcal{\bm G}_{\bar{\beta}\gamma}(\bar{\bm 2}; \bm 3;  \bm  \phi)
%     \equiv
%     {\color{red} - \bm V(\bm 1 - \bar{\bm 2})\Big\langle T_{\tau} S\left[\phi\right] \bm \Psi^{\dag}(\bar{\bm 2^+}) \bm \Psi(\bar{\bm 2}) \bm \Psi(\bm 1) \bm \Psi^{\dag}(\bm 3) \Big\rangle }.
% \end{equation}
where $\bm H_0$ are matrix elements of the Hamiltonian's non-interacting part and $\bm \Sigma$ is the self-energy, which accounts for the electronic correlations at the one-particle level.

The particle-hole and particle-particle susceptibilities are obtained from
\begin{equation}
    \bm \chi_{ph}(\bm 1; \bm 2; \bm 3; \bm 4)
    \equiv - {\beta}
    \frac{\delta^2 \mathcal{F}[ \bm  \phi]}
    {\delta  \bm  \phi_{11}(\bm 2; \bm 1) \delta  \bm  \phi_{11}(\bm 3; \bm 4)}
    \Big\lvert_{ \bm  \phi = 0}
    % = \frac{1}{\beta^2} \text{Tr} \left[ \frac{\partial \bm G_{12}}{\partial \mathcal{U}_{43}} \right]
    , \quad 
    \bm \chi_{pp}(\bm 1; \bm 2; \bm 3; \bm 4)
    \equiv - \frac{\beta}{2}
    \frac{\delta^2 \mathcal{F}[ \bm  \phi]}
    {\delta  \bm  \phi_{21}(\bm 2; \bm 1) \delta  \bm  \phi_{12}(\bm 3; \bm 4)}
    \Big\lvert_{ \bm  \phi = 0}
    % \equiv \frac{1}{\beta^2}
    % \text{Tr} \left[ \frac{\partial \bm F_{12}}{\partial \mathcal{V}_{34}} \right].
\end{equation}
where the factor $\frac{1}{2}$ in the \textit{p-p} channel is necessary to avoid double counting due to the indiscernibility of electrons.
These susceptibilities can be expressed as four-point correlation functions. In the normal state, we find, leaving the time-ordering operator implicit,
\begin{align}
    \bm \chi_{ph}(\bm 1; \bm 2; \bm 3; \bm 4)
    % \frac{1}{\beta} \frac{\delta^2 \mathcal{F}[ \bm  \phi]}{\delta  \bm  \phi_{11}(\bm 2; \bm 1) \delta  \bm  \phi_{11}(\bm 3; \bm 4)} 
    % {\color{blue}
    % = \frac{1}{\beta} \frac{\delta \mathcal{\bm G}_{11}(\bm 1; \bm 2;  \bm  \phi)}{\delta  \bm  \phi_{11}(\bm 3; \bm 4)} 
    % = \frac{1}{\beta}\frac{\delta}{\delta  \bm  \phi_{11}(\bm 3; \bm 4)}\frac{1}{Z[ \bm  \phi]} \text{Tr} \left[ T_{\tau} S[ \bm  \phi]
    %     \psi^{\dag}(\bm 2) \psi(\bm 1) \right] }
    % \\
    % {\color{blue}= -\frac{1}{\beta Z^2[ \bm  \phi]} 
    % \frac{\delta Z[ \bm  \phi]}{\delta  \bm  \phi_{11}(\bm 3; \bm 4)} 
    % \text{Tr} \left[ 
    %     T_{\tau} S[ \bm  \phi] \psi^{\dag}(\bm 2) \psi(\bm 1) 
    % \right] - 
    % \frac{1}{Z} 
    % \text{Tr} \left[ T_{\tau}
    %     S[ \bm  \phi]
    %     \Psi_{\bar{\alpha}}^{\dag}(\bar{\bm 1}) \frac{ \delta   \bm  \phi_{\bar{\alpha}\bar{\beta}}(\bar{\bm 1}; \bar{\bm 2})}{\delta  \bm  \phi_{11}(\bm 3; \bm 4)} \Psi_{\bar{\beta}}(\bar{\bm 2}) \psi^{\dag}(\bm 2) \psi(\bm 1) 
    % \right]}
    % \\
    % {\color{blue}\text{Blue is superfluous?} \quad } 
    & =
    \langle T_{\tau} \psi^{\dag}(\bm 3) \psi(\bm 4) \psi^{\dag}(\bm 2) \psi(\bm 1)  \rangle 
    - \langle T_{\tau} \psi^{\dag}(\bm 3)\psi(\bm 4) \rangle \langle T_{\tau} \psi^{\dag}(\bm 2) \psi(\bm 1) \rangle 
    , \\
    \bm \chi_{pp}(\bm 1; \bm 2; \bm 3; \bm 4) 
    % \frac{1}{\beta} \frac{\delta^2 \mathcal{F}[ \bm  \phi]}{\delta  \bm  \phi_{12}(\bm 2; \bm 1) \delta  \bm  \phi_{12}(\bm 3; \bm 4)} 
    & =
    \frac{1}{2} \langle T_{\tau} \psi^{\dag}(\bm 3) \psi^{\dag}(\bm 4^*) \psi(\bm 2^*) \psi(\bm 1)  \rangle.
\end{align}

From $\frac{\delta \mathcal{G}}{\delta \phi} = - \mathcal{G} \frac{\delta \mathcal{G}^{-1}}{\delta \mathcal{G}} \mathcal{G} = 0$ and the equations of motion,
\begin{align}
    \frac{\delta \mathcal{\bm G}_{11}(\bm 1; \bm 2;  \bm  \phi)}{\delta  \bm  \phi_{11}(\bm 3; \bm 4)} 
    % & = - \bm G_{\alpha\bar{\alpha}}(1, \bar{1})[ \bm  \phi] \frac{\delta \bm G^{-1}_{\bar{\alpha}\bar{\beta}}(\bar{1}, \bar{2})}{\delta  \bm  \phi_{11}(3, 4)} \bm G_{\bar{\beta}\beta}(\bar{2}, 2) 
    & = - \mathcal{\bm G}_{1\bar{\alpha}}(\bm 1; \bar{\bm 1};  \bm  \phi)
    \overbrace{
        \Bigg[
        % \underbrace{
            -
            \frac{\delta  \bm  \phi_{\bar{\alpha}\bar{\beta}}(\bar{\bm 1}; \bar{\bm 2})}{\delta  \bm  \phi_{11}(\bm 3; \bm 4)}
        % }_{
        %     \delta_{\bar{\alpha}1}\delta_{\bar{\beta}1}\delta(\bar{\bm 1}, \bm 3)\delta(\bar{\bm 2}, \bm 4)
        % } 
        - \underbrace{
            \frac{\delta \bm \Sigma_{\bar{\alpha}\bar{\beta}}(\bar{\bm 1}; \bar{\bm 2};  \bm  \phi)}
            {\delta \mathcal{\bm G}_{\bar{\gamma}\bar{\delta}}(\bar{\bm 3}; \bar{\bm 4};  \bm  \phi)}
        }_{ 
            \bm \Gamma^{irr}_{\bar{\alpha}\bar{\beta}\bar{\gamma}\bar{\delta}}(\bar{\bm 1}; \bar{\bm 2}; \bar{\bm 3}; \bar{\bm 4};  \bm  \phi)
        }
        \frac{\delta \mathcal{\bm G}_{\bar{\gamma}\bar{\delta}}(\bar{\bm 3}; \bar{\bm 4};  \bm  \phi)}{\delta  \bm  \phi_{11}(\bm 3; \bm 4)}
        \Bigg]
        }^{
            \frac{\delta \mathcal{\bm G}^{-1}_{\bar{\alpha}\bar{\beta}}(\bar{\bm 1}; \bar{\bm 2};  \bm  \phi)}{\delta  \bm  \phi_{11}(\bm 3; \bm 4)}
        }
    \mathcal{\bm G}_{\bar{\beta}1}(\bar{\bm 2}; \bm 2;  \bm  \phi)  \\
    & = \bm G(\bm 1; \bm 3;  \bm  \phi) \bm G(\bm 4; \bm 2;  \bm  \phi) 
        + \mathcal{\bm G}_{1\bar{\alpha}}(\bm 1; \bar{\bm 1};  \bm  \phi)
        \mathcal{\bm G}_{\bar{\beta}1}(\bar{\bm 2}; \bm 2;  \bm  \phi)
        \bm \Gamma^{irr}_{\bar{\alpha}\bar{\beta}\bar{\gamma}\bar{\delta}}(\bar{\bm 1}; \bar{\bm 2}; \bar{\bm 3}; \bar{\bm 4};  \bm  \phi) \frac{\delta \mathcal{\bm G}_{\bar{\gamma}\bar{\delta}}(\bar{\bm 3}; \bar{\bm 4};  \bm  \phi)}{\delta  \bm  \phi_{11}(\bm 3; \bm 4)}. \label{eq:dGdp}
\end{align}
In this expression, we have defined the irreducible vertex $\bm \Gamma^{irr}$. Since we are looking at instabilities from the normal state, taking the $ \bm  \phi \rightarrow 0$ limit means the off-diagonal components of the Nambu Green function vanish. Thus the following equalities and the ones with $\phi_{12}\rightarrow\phi_{21}$ are satisfied:
\begin{align}
    % \frac{\delta \bm G_{11}(\bm 1, \bm 2;  \bm  \phi)}{\delta  \bm  \phi_{12}(\bm 3, \bm 4)} \Big\lvert_{ \bm  \phi=0}
    % = \bm G_{11}(\bm 1, \bar{\bm 1})
    %     \bm G_{11}(\bar{\bm 2}, \bm 2)
    %     \bm \Gamma^{11\bar{\gamma}\bar{\delta}}(\bar{\bm 1}, \bar{\bm 2}, \bar{\bm 3}, \bar{\bm 4}) \frac{\delta \bm G_{\bar{\gamma}\bar{\delta}}(\bar{\bm 3}, \bar{\bm 4};  \bm  \phi)}{\delta  \bm  \phi_{12}(\bm 3, \bm 4)} \Big\lvert_{ \bm  \phi=0}.
    \frac{\delta \mathcal{\bm G}_{\alpha\alpha}(\bm 1; \bm 2;  \bm  \phi)}{\delta  \bm  \phi_{12}(\bm 3, \bm 4)} \Big\lvert_{ \bm  \phi=0}
    = \frac{\delta \mathcal{\bm G}_{12}(\bm 1, \bm 2;  \bm  \phi)}{\delta  \bm  \phi_{\beta\beta}(\bm 3, \bm 4)} \Big\lvert_{ \bm  \phi=0} 
    = 0.
    % = \bm G_{11}(\bm 1, \bar{\bm 1})
    %     \bm G_{11}(\bar{\bm 2}, \bm 2)
    %     \bm \Gamma^{11\bar{\gamma}\bar{\delta}}(\bar{\bm 1}, \bar{\bm 2}, \bar{\bm 3}, \bar{\bm 4}) \frac{\delta \bm G_{\bar{\gamma}\bar{\delta}}(\bar{\bm 3}, \bar{\bm 4};  \bm  \phi)}{\delta  \bm  \phi_{12}(\bm 3, \bm 4)} \Big\lvert_{ \bm  \phi=0}.
\end{align}
By conservation of momentum and of particle number when $\phi_{12}=\phi_{21}=0$, we have $\bm \Gamma^{irr}_{11\gamma\delta} = \delta_{\gamma\delta}\delta_{\gamma 1}\bm \Gamma^{irr}_{1111}$. 
Defining the bare \textit{p-h} susceptibility $\bm \chi^0_{ph}(\bm 1; \bm 2; \bm 3; \bm 4)$ as $- \bm G(\bm 1; \bm 3) \bm G(\bm 4; \bm 2)$, \eref{eq:dGdp} becomes
\begin{align}
    - \frac{\delta \mathcal{\bm G}_{11}(\bm 1; \bm 2;  \bm  \phi)}{\delta  \bm  \phi_{11}(\bm 3; \bm 4)} \Big\lvert_{ \bm  \phi=0}
     = \bm \chi^0_{ph}(\bm 1; \bm 2; \bm 3; \bm 4)
    - \bm \chi^0_{ph}(\bm 1; \bm 2; \bar{\bm 1}; \bar{\bm 2})
         \bm \Gamma_{ph}(\bar{\bm 1}; \bar{\bm 2}; \bar{\bm 3}; \bar{\bm 4})
    \frac{\delta \mathcal{\bm G}_{11}(\bar{\bm 3}; \bar{\bm 4};  \bm  \phi)}{\delta  \bm  \phi_{11}(\bm 3; \bm 4)}\Big\lvert_{ \bm  \phi=0}
\end{align}
\begin{align}
    \text{where} \ \bm \Gamma_{ph}(\bar{\bm 1}; \bar{\bm 2}; \bar{\bm 3}; \bar{\bm 4}) \equiv \bm \Gamma^{irr}_{1111}(\bar{\bm 1}; \bar{\bm 2}; \bar{\bm 3}; \bar{\bm 4}).
    % - \bm \Gamma^{irr}_{1122}(\bar{\bm 1}; \bar{\bm 2}; \bar{\bm 4^*}; \bar{\bm 3^*}) 
    % = 2 \bm \Gamma^{irr}_{1111}(\bar{\bm 1}; \bar{\bm 2}; \bar{\bm 3}; \bar{\bm 4}).
\end{align}
The above can also be written in the form
\begin{equation}
    \bm \chi_{ph}(\bm 1; \bm 2; \bm 3; \bm 4) = \bm \chi^0_{ph}(\bm 1; \bm 2; \bm 3; \bm 4) + \bm \chi^0_{ph}(\bm 1; \bm 2; \bar{\bm 1}; \bar{\bm 2}) \bm \Gamma_{ph}(\bar{\bm 1}; \bar{\bm 2}; \bar{\bm 3}; \bar{\bm 4}) \bm \chi_{ph}(\bar{\bm 3}; \bar{\bm 4}; \bm 3; \bm 4).
\end{equation}

Now looking at functional derivatives of off-diagonal components, we have
\begin{align}
    \frac{\delta \mathcal{\bm G}_{12}(\bm 1; \bm 2;  \bm  \phi)}{\delta  \bm  \phi_{12}(\bm 3; \bm 4)}
     = \mathcal{\bm G}_{11}(\bm 1; \bm 3; \phi) \mathcal{\bm G}_{22}(\bm 4; \bm 2; \phi)
     +  \mathcal{\bm G}_{11}(\bm 1; \bar{\bm 1}; \phi)
        \mathcal{\bm G}_{22}(\bar{\bm 2}; \bm 2; \phi)
        \bm \Gamma^{irr}_{12\bar{\gamma}\bar{\delta}}(\bar{\bm 1}; \bar{\bm 2}; \bar{\bm 3}; \bar{\bm 4}; \phi) \frac{\delta \mathcal{\bm G}_{\bar{\gamma}\bar{\delta}}(\bar{\bm 3}; \bar{\bm 4};  \bm  \phi)}{\delta  \bm  \phi_{12}(\bm 3, \bm 4)}.
\end{align}
Momentum and particle number conservation in the vertex when $\phi_{12}=\phi_{21}=0$ imposes $(\bar{\gamma}, \bar{\delta}) = (1, 2)$.
Replacing the hole propagator for a particle one using \eref{eq:hole_propagator} and defining $\bm \chi^0_{pp}(\bm 1; \bm 2; \bm 3; \bm 4)$ as $\frac{1}{2}\bm G(\bm 1; \bm 3) \bm G(\bm 2^*; \bm 4^*)$, we use the antisymmetry of the \textit{p-p} susceptibility under the exchange of both particles to write
\begin{equation}
    \bm \chi_{pp}(\bm 1; \bm 2; \bm 3; \bm 4) 
    = \bm \chi^0_{pp}(\bm 1; \bm 2; \bm 3; \bm 4) 
    + \bm \chi^0_{pp}(\bm 1; \bm 2; \bar{\bm 1}; \bar{\bm 2}) 
    % \underbrace{\left[\bm \Gamma^{irr}_{1212}(\bar{\bm 1}; \bar{\bm 2}; \bar{\bm 3}; \bar{\bm 4}) - \bm \Gamma^{irr}_{1221}(\bar{\bm 1}; \bar{\bm 2}; \bar{\bm 4^*}; \bar{\bm 3^*}) \right]
    % }_{\bm \Gamma_{pp}(\bar{\bm 1}; \bar{\bm 2}; \bar{\bm 3}; \bar{\bm 4}) = 2 \bm \Gamma^{irr}_{1212}(\bar{\bm 1}; \bar{\bm 2}; \bar{\bm 3}; \bar{\bm 4})}
    \underbrace{\left[\bm \Gamma^{irr}_{1212}(\bar{\bm 1}; \bar{\bm 2}; \bar{\bm 3}; \bar{\bm 4}) - \bm \Gamma^{irr}_{1212}(\bar{\bm 1}; \bar{\bm 2}; \bar{\bm 4}; \bar{\bm 3})
    \right]}_{\bm \Gamma_{pp}(\bar{\bm 1}; \bar{\bm 2}; \bar{\bm 3}; \bar{\bm 4})}
    \bm \chi_{pp}(\bar{\bm 3}; \bar{\bm 4}; \bm 3; \bm 4)
\end{equation}
and $\bm \Gamma_{pp}$ is defined such that it is antisymmetrized and to avoid double counting.

For normal state Green function that are diagonal in momentum space, we have $\bm G(\bm 1; \bm 2) = \bm G(\bm 1; \bm 2)\delta_{\textbf{k}_1 , \textbf{k}_2}$. It also satisfies time-translation symmetry. Matsubara transforming it yields
\begin{equation}
    \bm G_{K}^{\mu_1\mu_2} \equiv \int_0^{\beta} e^{i\omega_m\tau} \bm G^{\mu_1\mu_2}(\textbf{k}_1,\tau; \textbf{k}_1,0) \ d\tau 
\end{equation}
where $ \tau = \tau_1-\tau_2,$ while $K \equiv (\textbf{k}=\textbf{k}_1, i\omega_m)$ is the momentum-energy quadrivector and we use $\mu_i \equiv (\sigma_i, l_i)$ for compactness. We write the bare susceptibilities accordingly as
\begin{align}
    \left[ \bm \chi^0_{ph}(Q) \right]^{\mu_1\mu_2\mu_3\mu_4}_{KK'} & = \int_0^{\beta} d\tau \ e^{(i\omega_m + i\nu_n)\tau} \int_0^{\beta} d\tau' \ e^{i\omega_m\tau'} \bm \chi^0_{ph}(\bm 1; \bm 2; \bm 3; \bm 4) = - \frac{1}{\beta} \bm G^{\mu_1\mu_3}_{K+Q} \bm G^{\mu_4\mu_2}_{K} \delta_{KK'},
    \\
    \left[ \bm \chi^0_{pp}(Q) \right]^{\mu_1\mu_2^*\mu_3\mu_4^*}_{KK'} & = \int_0^{\beta} d\tau \ e^{(i\omega_m + i\nu_n)\tau} \int_0^{\beta} d\tau' \ e^{-i\omega_m\tau'} \bm \chi^0_{pp}(\bm 1; \bm 2; \bm 3; \bm 4) = \frac{1}{2\beta} \bm G^{\mu_1\mu_3}_{K+Q} \bm G^{\mu_2^*\mu_4^*}_{-K} \delta_{KK'}
\end{align}
where we used $\tau = \tau_1-\tau_3$ and $\tau' = \tau_4-\tau_2$ and $\mu^* \equiv (-\sigma, l)$. The spin $\sigma$ in the latter label needs to be replaced by a pseudospin index $\rho$ in the presence of spin-orbit coupling. The $1/\beta$ factor is associated to the Kronecker delta in Matsubara frequencies. 
The bare susceptibilities characterize two non-interacting particles. In the \textit{p-h} channel, a particle is created at $\bm 1$ ($\bm 4$) and removed at $\bm 3$ ($\bm 2$). It has a frequency $i\omega_m + i\nu_n$ ($i\omega_m$) and a momentum $\textbf{k} + \textbf{q}$ ($\textbf{k}$). The vertex encodes an interaction between these particles. Since they conserve energy-momentum, we take $K_1 = K+Q$, $K_2 = K$, $K_3 = K'+Q$ and $K_4 = K'$. In the \textit{p-p} channel, $K'$ and $K$ are changed for $-K'$ and $-K$. A number of additional relations between the vertices of both channels are derived in Appendix~\ref{sec:spin-charge_vertices}. The susceptibilities read
\begin{align}
    \left[ \bm \chi_{ph}(Q) \right]^{\mu_1\mu_2\mu_3\mu_4}_{KK'} =
     \left[ \bm \chi^0_{ph}(Q)
     \right]^{\mu_1\mu_2\mu_3\mu_4}_{KK'}
     +
     \sum_{K''K'''} \sum_{\mu_5...\mu_8} 
     \left[ \bm \chi^0_{ph}(Q)
     \right]^{\mu_1\mu_2\mu_5\mu_6}_{KK''}
     \left[ \bm \Gamma_{ph} (Q) \right]^{\mu_5\mu_6\mu_7\mu_8}_{K''K'''}
     \left[ \bm \chi_{ph}(Q)
     \right]^{\mu_7\mu_8\mu_3\mu_4}_{K'''K'}
     \\
    \left[ \bm \chi_{pp}(Q) \right]^{\mu_1\mu_2\mu_3\mu_4}_{KK'} =
     \left[ \bm \chi^0_{pp}(Q)
     \right]^{\mu_1\mu_2\mu_3\mu_4}_{KK'}
     +
     \sum_{K''K'''} \sum_{\mu_5...\mu_8} 
     \left[ \bm \chi^0_{pp}(Q)
     \right]^{\mu_1\mu_2\mu_5\mu_6}_{KK''}
     \left[ \bm \Gamma_{pp} (Q) \right]^{\mu_5\mu_6\mu_7\mu_8}_{K''K'''}
     \left[ \bm \chi_{pp}(Q)
     \right]^{\mu_7\mu_8\mu_3\mu_4}_{K'''K'}.
\end{align}

\section{\label{sec:spin-charge_vertices}Two-particle vertices.}

The completely reducible particle-hole and particle-particle vertices are given by~\cite{bickers_self-consistent_2004}
\begin{align}
	\bm \Gamma = \frac{1}{2} \hspace{-.2cm} \sum_{\tiny \begin{array}{c}KK'Q \\ \mu_1 ... \mu_4 \end{array}}  \hspace{-.2cm} \left[ \bm \Gamma(Q) \right]^{\mu_1\mu_2\mu_3\mu_4}_{K;K'} \hat{c}^{\dag,\mu_1}_{K+Q}\hat{c}^{\mu_2}_{K} \hat{c}^{\dag,\mu_4}_{K'} \hat{c}^{\mu_3}_{K'+Q}
% \end{align}
\quad \text{and} \quad
% \begin{align}
	\bm \Gamma_P = \frac{1}{2} \hspace{-.2cm} \sum_{\tiny \begin{array}{c}KK'Q \\ \mu_1 ... \mu_4 \end{array}}  \hspace{-.2cm} \left[ \bm \Gamma_P(Q) \right]^{\mu_1\mu_2\mu_3\mu_4}_{K;K'} \hat{c}^{\dag,\mu_1}_{K+Q} \hat{c}^{\dag,\mu_2}_{-K} \hat{c}^{\mu_4}_{-K'} \hat{c}^{\mu_3}_{K'+Q}
\end{align}
where $\mu\equiv(\sigma,l)$ with as usual spin replaced by pseudospin in the presence of SOC. Using the anticommutation relations of the $c$-operators, one can show that inside a time-ordered product it implies the crossing relations
\begin{align}
    \label{eq:crossing_rels}
	\left[ \bm \Gamma(Q) \right]^{\mu_1\mu_2;\mu_3\mu_4}_{K;K'} 
	= - \left[ \bm \Gamma(K'-K) \right]^{\mu_4\mu_2;\mu_3\mu_1}_{K;K+Q}
	= \left[ \bm \Gamma(-Q) \right]^{\mu_4\mu_3;\mu_2\mu_1}_{K'+Q;K+Q} 
	= -\left[ \bm \Gamma_P(K'+K+Q) \right]^{\mu_1\mu_4; \mu_3\mu_2}_{-K';-K}.
\end{align}

The usual perturbative approach is to define irreducible vertices $\bm \Gamma^{ph}, \bar{\bm\Gamma}^{ph}, \bm \Gamma^{pp}$, which are irreducible in either the horizontal channel ($\bm \Gamma^{ph}$ and $\bm \Gamma^{pp}$) or the vertical channel ($\bar{\bm \Gamma}^{ph}$) and to construct the full vertices using the Bethe-Salpeter equations
\begin{align}
	\left[ \bm \Gamma_P(Q) \right]^{\mu_1\mu_2;\mu_3\mu_4}_{K;K'} = \left[ \bm \Gamma_{pp}(Q) \right]^{\mu_1\mu_2; \mu_3\mu_4}_{K;K'}
	+ \left[ \bm \Gamma_P(Q) \bm \chi^{0}_{pp}(Q) \bm \Gamma_{pp}(Q) \right]^{\mu_1\mu_2;\mu_3\mu_4}_{K;K'} \\
	\left[ \bm \Gamma(Q) \right]^{\mu_1\mu_2;\mu_3\mu_4}_{K;K'} = \left[ \bm \Gamma_{ph}(Q) \right]^{\mu_1\mu_2;\mu_3\mu_4}_{K;K'}
	+ \left[ \bm \Gamma(Q) \bm \chi^0_{ph}(Q) \bm \Gamma_{ph}(Q) \right]^{\mu_1\mu_2;\mu_3\mu_4}_{K;K'}.
\end{align}

Using the crossing relations Eq.~\ref{eq:crossing_rels}, one finds the Parquet equations, which can be reduced to
\begin{align}
    \label{eq:ph_vertex}
	\left[ \bm \Gamma_{ph}(Q) \right]^{\mu_1\mu_2;\mu_3\mu_4}_{K;K'} 
	& = \left[ \bm \Lambda_{ph}(Q) \right]^{\mu_1\mu_2;\mu_3\mu_4}_{K;K'} - \left[ \bm \Phi(K'-K) \right]^{\mu_4\mu_2;\mu_3\mu_1}_{K;K+Q} + \left[ \bm \Psi(K'+K+Q) \right]^{\mu_4\mu_1;\mu_3\mu_2}_{-K-Q;-K} \\
    \label{eq:pp_vertex}
	\left[\bm \Gamma_{pp}(Q) \right]^{\mu_1\mu_2;\mu_3\mu_4}_{K;K'}
	& = - \left[ \bm \Lambda_{ph}(K'+K+Q) \right]^{\mu_1\mu_4;\mu_3\mu_2}_{-K';-K} + \left[ \bm \Phi(K'-K) \right]^{\mu_2\mu_4;\mu_3\mu_1}_{-K';K+Q} - \left[ \bm \Phi(K'+K+Q)\right]^{\mu_1\mu_4;\mu_3\mu_2}_{-K';-K}
\end{align}

% \begin{equation}
%     \left[ \bm \Gamma^0_{pp} \right]^{\sigma_1\sigma_2;\sigma_3\sigma_4}_{l_1l_2;l_3l_4} = \left[ \bm \Gamma^0_{ph} \right]^{\sigma_1\sigma_4;\sigma_3\sigma_2}_{l_1l_4;l_3l_2} 
%     \ ; \quad
%     \bm \Gamma^{0,s/t}_{pp} = \left[\bm \Gamma^0_{pp} \right]^{\uparrow\downarrow;\uparrow\downarrow} \mp [\bm \Gamma^0_{pp}]^{\uparrow\downarrow; \downarrow\uparrow}
% \end{equation}

where $\bm \Lambda_{ph}$ is the vertex which is irreducible in all channels given by \eref{eq:kanamori_vertex} and the vertex corrections are characterized by the \textit{p-h} and \textit{p-p} ladder functions $\bm \Phi$ and $\bm \Psi$, respectively given by
\begin{align}
    \label{eq:ladder_ph}
	\left[ \bm \Phi(Q) \right]^{\mu_1\mu_2;\mu_3\mu_4}_{K;K'} = \left[ \bm \Gamma_{ph} \bm \chi_{ph} \bm \Gamma_{ph}(Q) \right]^{\mu_1\mu_2;\mu_3\mu_4}_{K;K'}
	\quad \text{and} \quad
	\left[ \bm \Psi(Q) \right]^{\mu_1\mu_2;\mu_3\mu_4}_{K;K'} = \left[ \bm \Gamma_{pp} \bm \chi_{pp} \bm \Gamma_{pp}(Q) \right]^{\mu_1\mu_2;\mu_3\mu_4}_{K;K'}.
\end{align}
In the approximation that we use, the fully reducible vertices appearing in the latter equations are replaced by, respectively, the particle-hole irreducible vertex and the particle-particle irreducible vertex.

\section{Kanamori vertex function.}
\label{sec:kanamori}
In this work, we assume that local interactions between electrons mediate unconventional superconductivity. These local correlations on individual sites labelled $i$ are modeled by the Kanamori-Slater Hamiltonian~\cite{georges_strong_2013}, that is
\begin{equation}
    \label{eq:kanamori-ham}
        \bm \hat{H}_{\text{int}} = 
        \underbrace{\sum_{il} U \hat{n}_{il}^{\uparrow} \hat{n}_{il}^{\downarrow}}_{\hat{H}_U} +
        \underbrace{\sum_{il_1\neq l_2} U' \hat{n}_{il_1}^{\uparrow} \hat{n}_{il_2}^{\downarrow}}_{\hat{H}_{U'}} + 
        \underbrace{\sum_{i\sigma l_1 \neq l_2} U'' \hat{n}_{il_1}^{\sigma} \hat{n}_{il_2}^{\sigma}}_{\hat{H}_{U''}} \underbrace{-\sum_{il_1 \neq l_2} J \hat{c}^{\uparrow, \dag}_{il_1} \hat{c}_{il_1 }^{\downarrow} \hat{c}^{\downarrow,\dag}_{il_2}\hat{c}_{il_2}^{\uparrow}}_{\hat{H}_{\text{sf}}}
        +
        \underbrace{\sum_{il_1 \neq l_2} J' \hat{c}^{\uparrow,\dag}_{il_1} \hat{c}^{\downarrow,\dag}_{il_1} \hat{c}_{il_2}^{\downarrow} \hat{c}_{il_2}^{\uparrow}}_{\hat{H}_{\text{ph}}}.
\end{equation}
In this expression, $\sigma$ and $l$ are electronic spin and orbital respectively and $U$, $U'$ and $U''$ are the intra-orbital, opposite spin inter-orbital and same spin inter-orbital Coulomb repulsion, respectively.
$J$ is the spin-flip (sf) term and $J'$ the pair-hopping (ph) term.
In the vertex formulation, this interacting Hamiltonian can be recasted as
\begin{equation}
    \bm \hat{H}_{\text{int}} = \frac{1}{2}\sum_{i}\sum_{l_1...l_4}\sum_{\sigma_1...\sigma_4} \bm I ^{\sigma_1\sigma_2;\sigma_3\sigma_4}_{l_1l_2;l_3l_4} \hat{c}^{\sigma_1,\dag}_{il_1} \hat{c}^{\sigma_2,\dag}_{il_2} \hat{c}_{il_4}^{\sigma_4} \hat{c}_{il_3}^{\sigma_3}.
\end{equation}

In this formalism, $\bm I$ is the antisymmetrized Coulomb interaction and here it is given in the particle-particle channel. We take it as the bare particle-particle vertex $\bm \Lambda_{pp} = \bm I$. To rotate it to the particle-hole channel, we use the second crossing relation $\left[\bm \Lambda_{ph} \right]^{\sigma_1\sigma_2\sigma_3\sigma_4}_{l_1l_2l_3l_4} = - \left[ \bm \Lambda_{pp}\right]^{\sigma_1\sigma_4\sigma_3\sigma_2}_{l_1l_4l_3l_2}$. Then, the bare particle-hole vertex is given by
\begin{align}
    \label{eq:kanamori_vertex}
    \left[\bm \Lambda_{ph} \right]^{\sigma_1\sigma_2\sigma_3\sigma_4}_{l_1l_2l_3l_4} = \left\{
        \begin{array}{ccc}
            U       & l_1 = l_2 = l_3 = l_4     & \multirow{4}{*}{$\Bigg]$ \footnotesize $\sigma_1 = \sigma_2 = -{\sigma}_3 = -{\sigma}_4$} \\
            U'      & l_1 = l_2 \neq l_3 = l_4  & \\
            J       & l_1 = l_3 \neq l_2 = l_4  & \\
            J'      & l_1 = l_4 \neq l_2 = l_3  & \\
            U''     & l_1 = l_2 \neq l_3 = l_4  & - \  \  \ \multirow{1}{*}{\footnotesize $\sigma_1 = \sigma_2 = \sigma_3 = \sigma_4$} \quad \quad
        \end{array}
    \right.
    + \left\{
        \begin{array}{ccc}
            -U      & l_1 = l_2 = l_3 = l_4     & \multirow{4}{*}{$\Bigg]$ \footnotesize $\sigma_1 = \sigma_3 = -{\sigma}_2 = -{\sigma}_4$} \\
            -U'     & l_1 = l_3 \neq l_2 = l_4  & \\
            -J      & l_1 = l_2 \neq l_3 = l_4  & \\
            -J'     & l_1 = l_4 \neq l_2 = l_3  & \\
            -U''    & l_1 = l_3 \neq l_2 = l_4  & - \ \ \ \multirow{1}{*}{\footnotesize $\sigma_1 = \sigma_2 = \sigma_3 = \sigma_4$} \quad \quad
        \end{array}
    \right.
\end{align}
\end{widetext}

Half of its elements have a minus sign difference with the other half as a consequence of the first crossing relation
\begin{equation}
    \left[ \bm \Lambda_{ph} \right]
        ^{\sigma_1\sigma_2\sigma_3\sigma_4}_{l_1l_2l_3l_4} 
    = - \left[ \bm \Lambda_{ph} \right]
        ^{\sigma_1\sigma_3\sigma_2\sigma_4}_{l_1l_3l_2l_4}.
\end{equation}

Assuming the rotationally invariant formulation of this interaction, we have $U' = U - 2J$, $U''=U-3J$ and $J'=J$. Thus, the vertex $\bm \Lambda_{ph}\left[U, J\right]$ depends only on two parameters, the on-site Coulomb repulsion $U$ and the Hund's coupling $J$ and one finds the relation
\begin{equation}
    \label{eq:scaling_kanamori}
    \bm \Lambda_{ph} \left[U, J\right] = U \cdot \bm \Lambda_{ph} \left[1, J/U\right].
\end{equation}

\begin{figure}[h]
    \centering
    \includegraphics[width=1\linewidth]{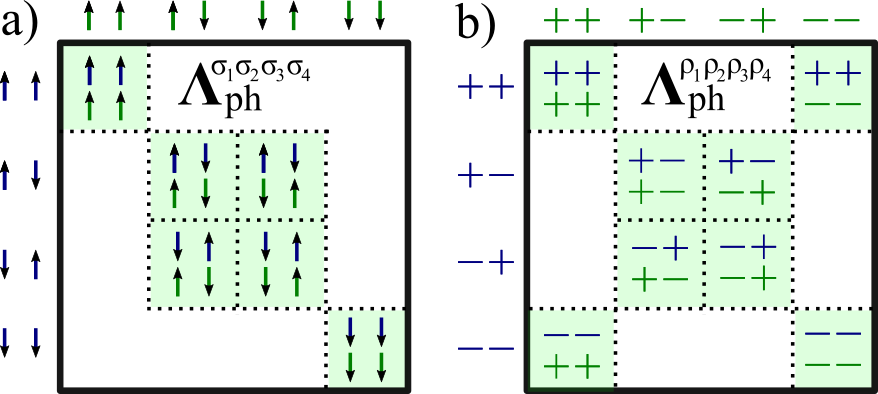}
    \caption{Structure of the \textit{p-h} vertex in a) spin and b) pseudospin basis. In each small square of this representation, the upper (lower) labels are associated to spins/pseudospins labelled 1 and 2 (3 and 4).}
    \label{fig:ph_vertex_spin}
\end{figure}

Moreover, because it preserves the spin projection, the vertex can be interpreted as an exchange of either spin 0 or 1. It is spin-diagonalized~\cite{esirgen_fluctuation_1998, nourafkan_nodal_2016} from the $\frac{1}{2} \otimes \frac{1}{2}$ basis represented in \fref{fig:ph_vertex_spin}~a) into the $0 \oplus 1 \sim \text{A}_{1g} \oplus \text{A}_{2g} \oplus \text{E}_g$ one in spin-space. The spin 0 (A$_{1g}$) bare vertex $\bm \Lambda_d$ is the density channel while the spin 1 (A$_{2g}\oplus \text{E}_g$) bare vertices $\bm \Lambda_{m,-1}$, $\bm \Lambda_{m,0} \equiv \bm \Lambda_m$ and $\bm \Lambda_{m,+1}$ are form the magnetic channels. Because of rotational invariance, the choice of axis for the projection is arbitrary and the magnetic channels are degenerate. One can check that $\bm \Lambda_{m,+1} \equiv \bm \Lambda^{\uparrow\downarrow\uparrow\downarrow} = \bm \Lambda_{m,-1} \equiv \bm \Lambda^{\downarrow\uparrow\downarrow\uparrow} = \bm \Lambda_{m}$ and that
\begin{align}
    \label{eq:kanamori_spindiag}
    \left[\bm \Lambda_{d/m} \right]_{l_1l_2l_3l_4} 
        \equiv \bm \Lambda^{\uparrow\uparrow\uparrow\uparrow}_{l_1l_2l_3l_4} 
            +/- \bm \Lambda^{\uparrow\uparrow\downarrow\downarrow}_{l_1l_2l_3l_4} \quad \quad \quad \quad \\
        = \left\{
            \begin{array}{cc}
                2U'-J   \ / \   -J      & l_1 = l_2 \neq l_3 = l_4 \\
                -U'+ 2J \ / \   -U'     & l_1 = l_3 \neq l_2 = l_4 \\
                J'      \ / \   -J'     & l_1 = l_4 \neq l_2 = l_3 \\
                U       \ / \   -U      & l_1 = l_2 = l_3 = l_4 \\
                0                       & \text{otherwise.}
            \end{array}
        \right.
\end{align}

Writing the \textit{p-h} vertex in the pseudospin basis introduces off-block diagonal elements, for example the diagram of \fref{fig:spin_flip} in the main text.
Its structure in this basis is represented in \fref{fig:ph_vertex_spin}~b).

% \section{Green and Gorkov functions and time-reversal}
\section{\label{sec:one-p-G_conn}Connection beween the general Green functions and the Green functions in model space.}
The general Matsubara Green function of a system in thermodynamic equilibrium is given by
\begin{equation}
    \bm G^{\sigma_1\sigma_2}(\textbf{r}_1, \textbf{r}_2, \tau) \equiv - \langle T_{\tau} \hat{\bm \Psi}^{\sigma_1}(\textbf{r}_1, \tau) \hat{\bm \Psi}^{\sigma_2,\dag}(\textbf{r}_2) \rangle
\end{equation}
where $\hat{\bm \Psi}$ is a field operator and $\hat{A}(\tau) \equiv e^{H\tau}\hat{A}e^{-H\tau}$. Note that the hat distinguishes this operator from the Nambu spinor of \sref{sec:free-energy}.
We follow the derivations of Ref.~\citenum{nourafkan_effect_2017}.
Using the relation $\bm G(\textbf{r}_1, \textbf{r}_2) = \frac{1}{N} \sum_{\textbf{R}_0} \bm G(\textbf{r}_1 + \textbf{R}_0, \textbf{r}_2 + \textbf{R}_0)$ for $\textbf{R}_0$ a translational vector of the unit cell, the Fourier transform leads to
\begin{align}
    \bm G^{\sigma_1\sigma_2}_{\textbf{Q}\textbf{Q}'}(\textbf{q}, \tau) = \frac{1}{V} & \int d\textbf{r}_1 d\textbf{r}_2  \times \\
    & e^{-i(\textbf{q} + \textbf{Q}) \cdot \textbf{r}_1} \bm G^{\sigma_1\sigma_2}(\textbf{r}_1, \textbf{r}_2, \tau) e^{i(\textbf{q} + \textbf{Q}')\cdot \textbf{r}_2} \nonumber
\end{align}
where $\textbf{q}$ is taken within the first Brillouin zone and $\textbf{Q}, \textbf{Q}'$ are reciprocal lattice vectors.
 
 For an interaction well-defined in an atomiclike orbital basis set $\{\phi_{\textbf{R}_il}^{\sigma}(\textbf{r})\}$ where $l$ and $\sigma$ are orbital and spin indices respectively and $\textbf{R}_i$ is the position of the $i^{\text{th}}$ atom, it is suitable to expand the field operator in this basis with destruction operators $\psi$ as
\begin{equation}
    \hat{\bm \Psi}^{\sigma}(\textbf{r}) = \sum_{\textbf{R}_il} \phi^{\sigma}_{\textbf{R}_il}(\textbf{r}) \psi^{\sigma}_{\textbf{R}_il}.
\end{equation}

Defining one-body oscillator matrix elements
\begin{equation}
    \bm O^{\sigma_1}_{\textbf{R}_1l_1}(\textbf{q}) = \frac{1}{\sqrt{V}} \int d\textbf{r}_1 \ e^{-i \textbf{q} \cdot \textbf{r}_1} \phi^{\sigma_1}_{\textbf{R}_1l_1}(\textbf{r}_1),
\end{equation}
we can write
\begin{align}
    \bm G^{\sigma_1\sigma_2}_{\textbf{Q}\textbf{Q}'}(\textbf{q}, \tau) = -& \hspace{-.2cm} \sum_{l_1l_2 \textbf{R}_1 \textbf{R}_2} \hspace{-.2cm} \bm O^{\sigma_1}_{\textbf{R}_1l_1}(\textbf{q} + \textbf{Q}) \ \times \\
    & \langle T_{\tau} \psi^{\sigma_1}_{\textbf{R}_1l_1}(\tau) \psi^{\sigma_2, \dag}_{\textbf{R}_2l_2} \rangle \bm O^{\sigma_2, *}_{\textbf{R}_2 l_2}(\textbf{q} + \textbf{Q}'). \nonumber
\end{align}

Introducing Fourier tranforms
\begin{align}
    \psi^{\sigma}_{\textbf{R}l} & = \frac{1}{\sqrt{N}} \sum_{\textbf{k}} e^{i\textbf{k}\cdot \textbf{R}} \psi^{\sigma}_{\textbf{k}l} \quad \text{and} \\
    \bm O^{\sigma}_{\textbf{k}l}(\textbf{q}) & = \frac{1}{\sqrt{N}} \sum_{\textbf{R}} e^{i\textbf{k} \cdot \textbf{R}} \bm O^{\sigma}_{\textbf{R}l}(\textbf{q}),
\end{align}
one has
% \begin{equation}
%     \label{eq:wavefunction}
%     \hat{\bm \Psi}^{\sigma}(\textbf{r}) = \sum_{\textbf{k}l} \phi^{\sigma}_{\textbf{k}l}(\textbf{r}) \psi^{\sigma}_{\textbf{k}l}.
% \end{equation}
\begin{align}
    \bm G^{\sigma_1\sigma_2}_{\textbf{Q} \textbf{Q}'}(\textbf{q}) = \hspace{-.2cm} \sum_{l_1l_2\textbf{k}\textbf{k}'} \hspace{-.2cm} \bm O^{\sigma_1}_{\textbf{k}l_1}(\textbf{q} + \textbf{Q}) \bm G^{\sigma_1\sigma_2}_{\textbf{k} \textbf{k}' l_1l_2} \bm O^{\sigma_2, *}_{\textbf{k}'l_2}(\textbf{q} + \textbf{Q}')
\end{align}
% The Green function becomes
% \begin{equation}
%     \bm G^{\sigma_1\sigma_2}(\textbf{r}_1, \textbf{r}_2, \tau) = \sum_{\textbf{k}\textbf{k}'l_1l_2} \phi^{\sigma_1}_{\textbf{k}l_1}(\textbf{r}_1) {\phi^{\sigma_2}_{\textbf{k}'l_2}(\textbf{r}_2)}^* \bm G^{\sigma_1\sigma_2}_{\textbf{k}\textbf{k}'l_1l_2}(\tau)
% \end{equation}
where there $\tau$ dependence was omitted and
\begin{equation}
    \label{eq:green_orb}
    \bm G^{\sigma_1\sigma_2}_{\textbf{k}\textbf{k}'l_1l_2}(\tau) \equiv - \langle T_{\tau} \psi^{\sigma_1}_{\textbf{k}l_1}(\tau) \psi^{\sigma_2, \dag}_{\textbf{k}'l_2} \rangle.
\end{equation}
is the Green function in the model space.
The oscillator matrix elements are needed to make contact with observables such as the local density, but for calculations of the Green function in the model space with the model Hamiltonian, they are not necessary.
For detailed arguments on the relation between observable susceptibilities and susceptibilities calculated in the model space, see Ref.~\citenum{nourafkan_effect_2017}.
This is particularly important for nonsymmorphic groups. 

Similarly, the Gorkov (or anomalous Green) functions of a stable system are defined as
\begin{align}
    \bm F^{\sigma_1\sigma_2}(\textbf{r}_1, \textbf{r}_2, \tau) & \equiv \langle T_{\tau} \hat{\bm \Psi}^{\sigma_1}(\textbf{r}_1, \tau) \hat{\bm \Psi}^{\sigma_2}(\textbf{r}_2) \rangle \quad \text{and} \\
    \bar{\bm F}^{\sigma_1\sigma_2}(\textbf{r}_1, \textbf{r}_2, \tau) & \equiv \langle T_{\tau} \hat{\bm \Psi}^{\sigma_2, \dag}(\textbf{r}_2, \tau) \hat{\bm \Psi}^{\sigma_1, \dag}(\textbf{r}_1) \rangle.
\end{align}
% and using \ref{eq:wavefunction}, we obtain
% \begin{equation}
%     \bm F^{\sigma_1\sigma_2}(\textbf{r}_1, \textbf{r}_2, \tau) = \sum_{\textbf{k}\textbf{k}'} \sum_{l_1l_2} \phi^{\sigma_1}_{\textbf{k}l_1}(\textbf{r}_1) \phi^{\sigma_2}_{\textbf{k}'l_2}(\textbf{r}_2) \bm F^{\sigma_1\sigma_2}_{\textbf{k}\textbf{k}'l_1l_2}(\tau)
% \end{equation}
% and
% \begin{equation}
%     \bm F^{\sigma_1\sigma_2}(\textbf{r}_1, \textbf{r}_2, \tau)^* = \sum_{\textbf{k}\textbf{k}'}\sum_{l_1l_2} \phi^{\sigma_1}_{\textbf{k}l_1}(\textbf{r}_1)^* \phi^{\sigma_2}_{\textbf{k}'l_2}(\textbf{r}_2)^* \bar{\bm F}^{\sigma_1\sigma_2}_{\textbf{k}\textbf{k}'l_1l_2}(\tau)
% \end{equation}
Following the same procedure as for the normal Green function, we find
\begin{align}
    \bm F^{\sigma_1\sigma_2}_{\textbf{Q} \textbf{Q}'}(\textbf{q}) & = \hspace{-.2cm} \sum_{l_1l_2\textbf{k}\textbf{k}'} \hspace{-.2cm} \bm O^{\sigma_1}_{\textbf{k}l_1}(\textbf{q} + \textbf{Q}) \bm F^{\sigma_1\sigma_2}_{\textbf{k} \textbf{k}' l_1l_2} \bm O^{\sigma_2}_{\textbf{k}'l_2}(-\textbf{q} - \textbf{Q}') \quad \text{and} \\
    \bar{\bm F}^{\sigma_1\sigma_2}_{\textbf{Q} \textbf{Q}'}(\textbf{q}) & = \hspace{-.2cm} \sum_{l_1l_2\textbf{k}\textbf{k}'} \hspace{-.2cm} \bm O^{\sigma_2,*}_{\textbf{k}'l_2}(-\textbf{q} - \textbf{Q}) \bm F^{\sigma_1\sigma_2}_{\textbf{k} \textbf{k}' l_1l_2} \bm O^{\sigma_1, *}_{\textbf{k}l_1}(\textbf{q} + \textbf{Q}')
\end{align}
where
\begin{align}
    \bm F^{\sigma_1\sigma_2}_{\textbf{k}\textbf{k}'l_1l_2}(\tau) \equiv
    \bm F^{\mu_1\mu_2}_{\textbf{k}\textbf{k}'}(\tau) & 
        \equiv \langle T_{\tau} \psi^{\mu_1}_{\textbf{k}}(\tau) \psi^{\mu_{2}}_{\textbf{k}'} \rangle, \\
    \bar{\bm F}^{\sigma_1\sigma_2}_{\textbf{k}\textbf{k}'l_1l_2}(\tau) \equiv
    \bar{\bm F}^{\mu_1\mu_2}_{\textbf{k}\textbf{k}'}(\tau) &
        \equiv \langle T_{\tau} \psi^{\mu_2, \dag}_{\textbf{k}'}(\tau) \psi^{\mu_{1}, \dag}_{\textbf{k}} \rangle.
\end{align}

\section{\label{sec:one-p-G_prop}Properties of model Green functions.}
In this work, we consider a local model Hamiltonian that is invariant under the crystal translation symmetry, which allows to simply take $\textbf{k} = \textbf{k}'$.

In the presence of spin-orbit coupling, it is useful to employ spinors $ \Psi_{\textbf{k}l}^{\dag}(\tau) \equiv \left( {\psi^{\uparrow,\dag}_{\textbf{k}l}}(\tau) \quad {\psi^{\downarrow,\dag}_{\textbf{k}l}}(\tau) \right)$ and electrons can propagate flipping spins. Again, this spinor is different from the Nambu spinor defined in \sref{sec:free-energy}. The probability amplitude is related to the Green function propagator
\begin{equation}
    \bm G^{\sigma_1\sigma_2}_{\textbf{k}l_1l_2}(\tau) = -\langle T_{\tau} \left[ \Psi_{\textbf{k}l_1}(\tau) \Psi_{\textbf{k}l_2}^{\dag} \right]^{\sigma_1\sigma_2} \rangle.
    \label{eq:general_green}
\end{equation}
Applying complex conjugation on each side, we find the relation
\begin{equation}
    \label{eq:green_conj_t}
    \bm G^{\mu_1\mu_2}_{\textbf{k}}(\tau)^* = \bm G^{\mu_2\mu_1}_{\textbf{k}}(\tau)
\end{equation}
where $\mu_i \equiv (\sigma_i, l_i)$.

For a symmetry of the system characterized by the operator $\hat{S}$, the matrix element of \eref{eq:general_green} is equal to
\begin{equation}
    \left[\hat{S} \bm G\right]_{\textbf{k}l_1l_2}(\tau) = -\langle T_{\tau} S^{-1} \Psi_{\textbf{k}l_1}(\tau) \Psi_{\textbf{k}l_2}^{\dag} S \rangle.
\end{equation}

For time-reversal symmetry, $S = \mathcal{T} = iK \sigma_y $ with $\sigma_y$ the second Pauli matrix in spin-space and $K$ is the right-side conjugation operator which acts as $\langle K^{-1}\mathcal{O}K \rangle = \langle {\mathcal{O}^{\dag}}^*\rangle$. Thus
\begin{equation}
    \left[\hat{\mathcal{T}} \bm G\right]_{\textbf{k}l_1l_2}(\tau) = -\langle T_{\tau} \sigma_y [\Psi_{\textbf{k}l_2}^*(\tau) \Psi_{\textbf{k}l_1}^{\dag *}] \sigma_y \rangle.
\end{equation}
Using $\left[ \sigma_y A \sigma_y \right]^{\sigma_1\sigma_2} = \epsilon_{\sigma_1\sigma_2} A^{-{\sigma}_1-{\sigma}_2}$ with $\epsilon_{\sigma_1\sigma_2} \equiv \delta_{\sigma_1\sigma_2} - \delta_{\sigma_1-{\sigma}_2}$ and ${\psi_{\textbf{k}l}^{\sigma*}} = \psi_{-\textbf{k}l}^{\sigma}$ where orbitals are taken to be time-reversal invariant~\cite{nourafkan_correlation-enhanced_2016}, we find
\begin{equation}
    \label{eq:identity_green_t1}
    \left[\hat{\mathcal{T}} \bm G\right]^{\mu_1\mu_2}_{\textbf{k}}(\tau) = \epsilon_{\sigma_1\sigma_2} \bm G^{\mu_2^*\mu_1^*}_{-\textbf{k}}(\tau),
\end{equation}
where $\mu^* \equiv (-\sigma, l)$.

Given that spins are invariant under inversion symmetry $\hat{S}=\hat{I}$, systems with $\hat{I}$ and orbitals that are even under $\hat{I}$ satisfy
\begin{equation}
    \label{eq:identity_green_t2}
    \bm G^{\mu_1\mu_2}_{\textbf{k}}(\tau) = \bm G^{\mu_1\mu_2}_{-\textbf{k}}(\tau).
\end{equation}

Transforming from imaginary time $\tau$ to Matsubara frequencies $\omega_m$, complex conjugation  \eref{eq:green_conj_t}, time-reversal \eref{eq:identity_green_t1} and inversion symmetry \eref{eq:identity_green_t2} become respectively
\begin{align}
    \bm G^{\mu_1\mu_2}_{\textbf{k}}(i\omega_m) 
        & = \bm G^{\mu_2\mu_1}_{\textbf{k}}(-i\omega_m)^* \label{eq:green_conj_w}\\
        & = \epsilon_{\sigma_1\sigma_2} \bm G^{\mu_2^*\mu_1^*}_{-\textbf{k}}(i\omega_m) \label{eq:identity_green_w1} \\
        & = \bm G^{\mu_1\mu_2}_{-\textbf{k}}(i\omega_m) \label{eq:identity_green_w2}.
\end{align}
We write the Green function as $\bm G^{\mu_1\mu_2}_{Kl_1l_2} \equiv \bm G^{\mu_1\mu_2}_{\textbf{k}}(i\omega_m)$ with the fermionic four-momentum $K\equiv (\textbf{k}, i\omega_m)$.

Without magnetic field or applied current, the composite bosons described by the Gorkov function usually favours a vanishing center of mass momentum which leads to $\textbf{k}' = - \textbf{k}$. We define
\begin{align}
    \label{eq:gorkov_def}
    \bm F^{\mu_1\mu_2}_{\textbf{k}}(\tau)  
        & \equiv \langle T_{\tau} \Psi_{\textbf{k}}^{\mu_1}(\tau) \Psi_{-\textbf{k}}^{\mu_2} \rangle \quad \text{and} \\
    \bar{\bm F}^{\mu_1\mu_2}_{\textbf{k}}(\tau)  
        & \equiv \langle T_{\tau} \Psi^{\mu_2, \dag}_{-\textbf{k}}(\tau) \Psi^{\mu_1, \dag}_{\textbf{k}} \rangle.
\end{align}
We find relations analogous to those found above for the normal Green function for Pauli, complex conjugation and time-reversal operation, respectively
% \begin{align}
%     \bm F^{\sigma_1\sigma_2}_{\textbf{k}l_1l_2}(\tau)^* 
%         = \bm F^{\dag,\sigma_1\sigma_2}_{\textbf{k}l_1l_2}(\tau) \quad \text{and} \\
%     \left[\hat{\mathcal{T}} \bm F\right]^{\sigma_1\sigma_2}_{\textbf{k}l_1l_2}(\tau) 
%         = \epsilon_{\sigma_1\sigma_2} \bm F^{\dag,\bar{\sigma}_2\bar{\sigma}_1}_{-\textbf{k}l_1l_2}(\tau).
% \end{align}
% Pauli
% \begin{equation}
%     \bm F^{\sigma_1\sigma_2}_{\textbf{k}l_1l_2}(\tau) = - \bm F^{\sigma_2\sigma_1}_{-\textbf{k}l_2l_1}(-\tau)
% \end{equation}
%
% In frequencies:
\begin{align}
    \bm F^{\mu_1\mu_2}_{\textbf{k}}(i\omega_m) = - \bm F^{\mu_2\mu_1}_{-\textbf{k}}(-i\omega_m), \quad \quad \\
    \bm F^{\mu_1\mu_2}_{\textbf{k}}(i\omega_m)^{*} 
         = \bar{\bm F}^{\mu_1\mu_2}_{\textbf{k}}(-i\omega_m) \quad \text{and} \\
    \left[ \hat{\mathcal{T}} \bm F \right]^{\mu_1\mu_2}_{\textbf{k}}(i\omega_m) = \epsilon_{\sigma_1\sigma_2} \bar{\bm F}^{\mu_1^*\mu_2^*}_{-\textbf{k}}(i\omega_m).
\end{align}

Combining these last two gives
\begin{align}
    \label{eq:gorkov-time-rev}
    \left[ \hat{\mathcal{T}} \bm F \right]^{\mu_1\mu_2}_{\textbf{k}}(i\omega_m) = \epsilon_{\sigma_1\sigma_2} \bm F^{\mu_1^*\mu_2^*}_{-\textbf{k}}(-i\omega_m)^*.
\end{align}

Using the fact that field operators $\psi$ can be multiplied by a global phase $e^{-i\frac{\delta}{2}}$, we can also work with $\tilde{\bm F} = e^{-i\delta}\bm F$ so that in this gauge, time-reversal symmetry becomes
\begin{equation}
    \label{eq:gorkov-time-rev-phase}
    \left[ \hat{\mathcal{T}} \tilde{\bm F} \right]^{\mu_1\mu_2}_{\textbf{k}}(i\omega_m) = e^{2i\delta} \epsilon_{\sigma_1\sigma_2} \tilde{\bm F}^{\mu_1^*\mu_2^*}_{-\textbf{k}}(-i\omega_m)^*.
\end{equation}
Our solutions to the linearized Eliashberg equation satisfy the above equation with $\delta=0$.  The eigenvectors  have components that pick up phases that may differ depending on quantum numbers (components of the eigenvectors) so that there is no way to choose a global phase that would set all elements of the eigenvectors to have the same phase. %That makes the sign of the $\hat{T}$ operation ill-defined, as discussed in the main text.

\subsection{Other proof for time-reversal.}

Here is another proof of \eref{eq:gorkov-time-rev}. We start from the spectral weight in the Nambu representation, given by
\begin{equation}
    \bm A(\textbf{k}, t) = \left[
        \begin{array}{cc}
            \langle \{ \bm \Psi_1(\textbf{k}, t), \bm \Psi_1^{\dag}(\textbf{k}) \} \rangle & \langle \{ \bm \Psi_1(\textbf{k}, t), \bm \Psi_2^{\dag}(\textbf{k}) \} \rangle \\
            \langle \{ \bm \Psi_2(\textbf{k}, t), \bm \Psi_1^{\dag}(\textbf{k}) \} \rangle & \langle \{ \bm \Psi_2(\textbf{k}, t), \bm \Psi_2^{\dag}(\textbf{k})\} \rangle
        \end{array}
    \right].
\end{equation}

We look at the $12$ component and write spins and orbitals indices:
\begin{equation}
    \bm A_{12}^{\mu_1 \mu_2}(\textbf{k}, t) = \langle \{ \psi_{\textbf{k}}^{\mu_1}(t), \psi_{-\textbf{k}}^{\mu_2^*} \} \rangle.
\end{equation}

Upon complex conjugation it becomes,
\begin{equation}
    \bm A_{12}^{\mu_1 \mu_2}(\textbf{k}, t)^* = \langle \{ {\psi^{\mu_2^*}_{-\textbf{k}}}^{\dag}, {\psi^{\mu_1}_{\textbf{k}}}^{\dag}(t) \} \rangle.
\end{equation}

Applying time-reversal,
\begin{align}
    \hat{\mathcal{T}}\left[ \bm A_{12}^{\mu_1 \mu_2}(\textbf{k}, t) \right] & = \langle \{ \Theta {\psi^{\mu_2^*,\dag}_{-\textbf{k}}}^{} \Theta^{-1}, \Theta {\psi_{\textbf{k}}^{\mu_1,\dag}}^{}(t) \Theta^{-1} \} \rangle \nonumber \\
    & = \epsilon_{\sigma_1\sigma_2} \langle \{ {\psi_{\textbf{k}}^{\mu_2,\dag}}, {\psi_{-\textbf{k}}^{\mu_1^*,\dag}}(-t) \} \rangle.
\end{align}

Comparing the last two equations, we have
\begin{equation}
    \hat{\mathcal{T}} \left[ \bm A_{12}^{\mu_1\mu_2}(\textbf{k}, t) \right] = \epsilon_{\sigma_1\sigma_2} \bm A_{12}^{\mu_1^*\mu_2^*}(- \textbf{k}, -t)^*.
\end{equation}

From this point of view, time-reversal symmetry takes a simple form, namely time-reversal symmetry is satisfied if
\begin{equation}
      \bm A_{12}^{\mu_1\mu_2}(\textbf{k}, t) = \epsilon_{\sigma_1\sigma_2} \bm A_{12}^{\mu_1^*\mu_2^*}(- \textbf{k}, -t)^*.
\end{equation}
which is similar to what one would require from wave functions. The above does not require that $\bm A_{12}^{\mu_1\mu_2}(\textbf{k}, t)$ be real. Fourier transforming back to space however, if time-reversal is satisfied $\bm A_{12}^{\mu_1\mu_2}(\textbf{r}, 0)$ is real when the spins are identical, or pure imaginary if the spins are anti-parallel (assuming throughout that the orbitals are invariant under time-reversal). In other words, for a given spin configuration, there is a global phase that can make $\bm A_{12}^{\mu_1\mu_2}(\textbf{r}, 0)$ real. However, seen as a matrix in the spin-indices, there is no way that matrix can be made purely real with a global phase.

How this translates in frequency is less familiar, but is not difficult to find. Using
\begin{equation}
    \int dt \ e^{i\omega t} f^*(-t) = f^*(\omega),
\end{equation}
we have
\begin{equation}
    \hat{\mathcal{T}} \left[ \bm A_{12}^{\mu_1\mu_2} (\textbf{k}, \omega) \right] = \epsilon_{\sigma_1\sigma_2} \bm A_{12}^{\mu_1^*\mu_2^*} (-\textbf{k}, \omega)^*.
\end{equation}

We can also obtain the Gorkov function in Matsubara frequency as follows
\begin{equation}
    \mathcal{G}_{12}(\textbf{k}, i\omega_n) \equiv \bm F_{\textbf{k}}^{\mu_1 \mu_2} (i\omega_n) = \int \frac{d\omega}{2\pi} \frac{\bm A_{12}^{\mu_1 \mu_2} (\textbf{k}, \omega)}{i\omega_n - \omega}.
\end{equation}

We thus recover our previous results \eref{eq:gorkov-time-rev} or \eref{eq:gorkov-time-rev_main} since
\begin{align}
    \hat{\mathcal{T}}\bm F^{\mu_1\mu_2}_{\textbf{k}} (i\omega_n) 
    & = \epsilon_{\sigma_1\sigma_2} \int \frac{ d\omega}{2\pi} \frac{\bm A^{\mu_1^*\mu_2^*}_{12}(-\textbf{k}, \omega)^*}{i\omega_n-\omega} \\
    % & = \epsilon_{\sigma_1\sigma_2} \left[\int \frac{d\omega}{\pi} \frac{A^{\mu_1^*\mu_2^*}_{12}(-\textbf{k}, \omega)}{-i\omega_n-\omega}\right]^* \\
    & = \epsilon_{\sigma_1\sigma_2} \bm F^{\mu_1^*\mu_2^*}_{-\textbf{k}}(-i\omega_n)^*.
\end{align}
Note that $\bm F^{\mu_1\mu_2}_{\textbf{k}} (i\omega_n)$ does not need to be real, nor to be even or odd in frequency to satisfy this equation. 

However, in the presence of time-reversal symmetry, the usual singlet combination $\bm F^{\sigma,-\sigma}_{\textbf{k}} (i\omega_n)-\bm F^{-\sigma,\sigma}_{\textbf{k}} (i\omega_n) $ for a single real orbital is even in Matsubara frequency when there is also inversion symmetry since
\begin{align}
    \hat{\mathcal{T}}\bm F^{\sigma,-\sigma}_{\textbf{k}} (i\omega_n)-\hat{\mathcal{T}}\bm F^{-\sigma,\sigma}_{\textbf{k}} (i\omega_n) \\
    =\bm F^{\sigma,-\sigma}_{-\textbf{k}}(-i\omega_n)-\bm F^{-\sigma,\sigma}_{-\textbf{k}}(-i\omega_n).
\end{align}
This is the familiar BCS result.

\section{\label{sec:susc_props}Properties of susceptibilities.}

Within the RPA approximation, an element of the bare susceptibility in the particle-hole channel is given by

\begin{equation}
    [\tilde{\bm \chi}^0_{ph}(Q)]^{\mu_1\mu_2\mu_3\mu_4} \equiv - \left( \frac{k_BT}{N} \right) \sum_K \bm G^{\mu_1\mu_3}_{K+Q} \bm G^{\mu_4\mu_2}_{K}
\end{equation}
where $Q\equiv(\textbf{q}, i\nu_n)$ is a bosonic four-momentum. Because of the sum over fermionic four-momentum, one can show it equals
\begin{equation}
    [\tilde{\bm \chi}^0_{ph}(-Q)]^{\mu_4\mu_3\mu_2\mu_1}.
\end{equation}
Using the property of Green functions under complex conjugation \eref{eq:green_conj_w}, we have
\begin{equation}
    [\tilde{\bm \chi}^0_{ph}(\textbf{q}, i\nu_n)]^{\mu_1\mu_2\mu_3\mu_4} =
        {[\tilde{\bm \chi}^0_{ph}(\textbf{q},-i\nu_n)]^{\mu_3\mu_4\mu_1\mu_2}}^*
\end{equation}
and with time-reversal and inversion \eref{eq:identity_green_w2}, one can show that
\begin{equation}
    [\tilde{\bm \chi}^0_{ph}(Q)]^{\mu_1\mu_2\mu_3\mu_4} = [\tilde{\bm \chi}^0_{ph}(-Q)]^{\mu_1\mu_2\mu_3\mu_4}.
\end{equation}

Using time reversal symmetry \eref{eq:identity_green_w1} leads to the following relations:
\begin{align}
    [\tilde{\bm \chi}^0_{ph}]^{\mu_1\mu_2\mu_3\mu_4}
         \nonumber & = \epsilon_{\sigma_1\sigma_3} [\tilde{\bm \chi}^0_{ph}]^{\mu_3^*\mu_2\mu_1^*\mu_4} \\
        & = \epsilon_{\sigma_2\sigma_4} [\tilde{\bm \chi}^0_{ph}]^{\mu_1\mu_4^*\mu_3\mu_2^*} \\
        \nonumber & = \epsilon_{\sigma_1\sigma_3} \epsilon_{\sigma_2\sigma_4} [\tilde{\bm \chi}^0_{ph}]^{\mu_3^*\mu_4^*\mu_1^*\mu_2^*},
\end{align}
where the $Q$-dependence is implicit.

\begin{figure}[b]
    \centering
    \includegraphics[width=\linewidth]{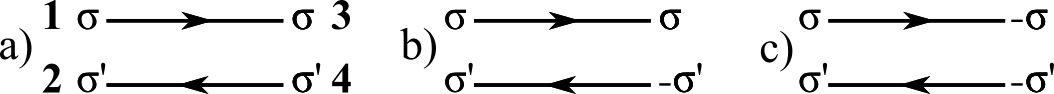}
    \caption[Three categories of diagrams representing bare particle-hole susceptibilities.]{Three categories of diagrams representing bare particle-hole susceptibilities. They are characterized by having a) zero, b) one and c) two spin-flips where the spin $\bar{\sigma} = - \sigma$.}
    \label{fig:bare_chi_ph}
\end{figure}

Time-reversal symmetry then connects different spin-sectors into three categories, diagrammatically represented on \fref{fig:bare_chi_ph}. The first one has no spin-flips (a) and satisfies
\begin{align}
    & [\tilde{\bm \chi}^0_{ph}]^{\sigma\sigma\sigma\sigma}_{l_1l_2l_3l_4} = \\
    & \quad \quad
    [\tilde{\bm \chi}^0_{ph}]^{\bar{\sigma}\sigma\bar{\sigma}\sigma}_{l_3l_2l_1l_4} =
    [\tilde{\bm \chi}^0_{ph}]^{\sigma\bar{\sigma}\sigma\bar{\sigma}}_{l_1l_4l_3l_2} = 
    [\tilde{\bm \chi}^0_{ph}]^{\bar{\sigma}\bar{\sigma}\bar{\sigma}\bar{\sigma}}_{l_3l_4l_1l_2},
    \nonumber
\end{align}
the second has one spin-flip (b) with
\begin{align}
   & [\tilde{\bm \chi}^0_{ph}]^{\sigma\sigma\sigma\bar{\sigma}}_{l_1l_2l_3l_4} = \\
   & \quad \quad \quad [\tilde{\bm \chi}^0_{ph}]^{\bar{\sigma}\sigma\bar{\sigma}\bar{\sigma}}_{l_3l_2l_1l_4} =
    - [\tilde{\bm \chi}^0_{ph}]^{\sigma\sigma\sigma\bar{\sigma}}_{l_1l_4l_3l_2} = 
    - [\tilde{\bm \chi}^0_{ph}]^{\bar{\sigma}\sigma\bar{\sigma}\bar{\sigma}}_{l_3l_4l_1l_2} 
    \nonumber
\end{align}
and the last category has two spin-flips (c) with
\begin{align}
    & [\tilde{\bm \chi}^0_{ph}]^{\sigma\sigma\bar{\sigma}\bar{\sigma}}_{l_1l_2l_3l_4} = \\
    & \quad \quad \quad
    - [\tilde{\bm \chi}^0_{ph}]^{\sigma\sigma\bar{\sigma}\bar{\sigma}}_{l_3l_2l_1l_4} =
    - [\tilde{\bm \chi}^0_{ph}]^{\sigma\sigma\bar{\sigma}\bar{\sigma}}_{l_1l_4l_3l_2} = 
    [\tilde{\bm \chi}^0_{ph}]^{\sigma\sigma\bar{\sigma}\bar{\sigma}}_{l_3l_4l_1l_2}.
    \nonumber
\end{align}

The particle-particle bare susceptibility is given by
\begin{equation}
    \left[ \bm \chi^0_{pp}(Q) \right]_{KK'}^{\mu_1\mu_2\mu_3\mu_4} = \frac{1}{2\beta} \bm G_{K+Q}^{\mu_1\mu_3} \bm G_{-K}^{\mu_2\mu_4} \delta_{KK'}.
\end{equation}
In spin and charge fluctuation-mediated superconductivity, it only involves $Q=0$ in \eref{eq:eliashberg} and we change notation for $\bm\chi_{pp}^0(K)$. The following relations hold from the definition and from the property of Green functions under complex conjugation \eref{eq:green_conj_w}:
\begin{align}
    \label{eq:chipp0_1221}
    \left[ \bm \chi^0_{pp}(K) \right]^{\mu_1\mu_2\mu_3\mu_4} 
    = \left[ \bm \chi^0_{pp}(-K) \right]^{\mu_2\mu_1\mu_4\mu_3} \\
    =\left[{ \left[ \bm \chi^0_{pp}(K^*) \right]^{\mu_3\mu_4\mu_1\mu_2}}\right]^*,
\end{align}
where $K^* \equiv (\textbf{k}, -i\omega_n)$.
Again, using time reversal \eref{eq:identity_green_w1} and inversion \eref{eq:identity_green_w2} leads to
\begin{align}
    [\bm \chi^0_{pp}(K)]^{\mu_1\mu_2\mu_3\mu_4}
         \nonumber & = \epsilon_{\sigma_1\sigma_3} [\bm \chi^0_{pp}(K)]^{\mu_3^*\mu_2\mu_1^*\mu_4} \\
        & = \epsilon_{\sigma_2\sigma_4} [\bm \chi^0_{pp}(K)]^{\mu_1\mu_4^*\mu_3\mu_2^*} \\
        \nonumber & = \epsilon_{\sigma_1\sigma_3} \epsilon_{\sigma_2\sigma_4} [\bm \chi^0_{pp}(K)]^{\mu_3^*\mu_4^*\mu_1^*\mu_2^*}.
\end{align}

\section{Group theory and $D_{4h}$ space group.}
\label{sec:group}

\begin{table}[b]
    \caption[Space group symmetry generators $g$ and corresponding transformation matrices in momentum $T_{\textbf{k}}$, orbital $T_l$ ($\{xy, yz, zx\}$) and spin $T_{\sigma}$ basis.]{Space group symmetry generators $g$ and corresponding transformation matrices in momentum $T_{\textbf{k}}$, orbital $T_l$ ($\{xy, yz, zx\}$) and spin $T_{\sigma}$ basis. $C_4$ is a rotation by $\pi/2$ around the z-axis, while $\sigma_x$, $\sigma_y$ and $\sigma_z$ are mirrors with respect to the $yz$, $zx$ and $xy$ planes.}
\begin{tabular}{cccc}
    \label{tab:d4h_generators}
    g       & $T_{\textbf{k}}$ & $T_l$     & $T_{\sigma}$ \\
    \hline
    $C_4$
        & $\left( \begin{array}{ccc} 
            0 & -1 & 0 \\ 1 & 0 & 0 \\ 0 & 0 & 1 \end{array}
        \right)$
        & $\left( \begin{array}{ccc} 
            -1 & 0 & 0 \\ 0 & 0 & 1 \\ 0 & -1 & 0 \end{array}
        \right)$
        & $\left( \begin{array}{cc}
            \frac{1+i}{\sqrt{2}} & 0 \\ 0 & \frac{1-i}{\sqrt{2}} \end{array} \right)$ \\
    $\sigma_x$
        & $\left( \begin{array}{ccc} 
            -1 & 0 & 0 \\ 0 & 1 & 0 \\ 0 & 0 & 1 \end{array}
        \right)$
        & $\left( \begin{array}{ccc} 
            -1 & 0 & 0 \\ 0 & 1 & 0 \\ 0 & 0 & -1 \end{array}
        \right)$
        & $\left( \begin{array}{cc}
            0 & i \\ i & 0 \end{array} \right)$ \\
    $\sigma_y$
        & $\left( \begin{array}{ccc} 
            1 & 0 & 0 \\ 0 & -1 & 0 \\ 0 & 0 & 1 \end{array}
        \right)$
        & $\left( \begin{array}{ccc} 
            -1 & 0 & 0 \\ 0 & -1 & 0 \\ 0 & 0 & 1 \end{array}
        \right)$
        & $\left( \begin{array}{cc}
            0 & 1 \\ -1 & 0 \end{array} \right)$ \\
    $\sigma_z$
        & $\left( \begin{array}{ccc} 
            1 & 0 & 0 \\ 0 & 1 & 0 \\ 0 & 0 & -1 \end{array}
        \right)$
        & $\left( \begin{array}{ccc} 
            1 & 0 & 0 \\ 0 & -1 & 0 \\ 0 & 0 & -1 \end{array}
        \right)$
        & $\left( \begin{array}{cc}
            i & 0 \\ 0 & -i \end{array} \right)$ \\
    \hline
\end{tabular}
\end{table}

The normal state of the system is invariant under a set $G$ of  symmetry operations $g$. They define the space group of the system and can all be constructed from a set of generators. In the case of SRO and as discussed in previous works~\cite{geilhufe_symmetry_2018, kaba_group-theoretical_2019, ramires_superconducting_2019, suh_stabilizing_2020}, the generators are written in Table~\ref{tab:d4h_generators}, along with their representations in momentum, orbital and spin basis.

Thus, the annihilation operators transform like
\begin{equation}
    \left[\hat{g} \psi_{\textbf{k}}\right]^{\mu_1} 
        = \left[ T(g)\right]^{\mu_1\mu_2} \left[\psi_{T_{\textbf{k}}(g^{-1})\textbf{k}}\right]^{\mu_2}
\end{equation}
where $T(g) = T_l(g) \otimes T_{\sigma}(g)$. It follows that gap functions transform like
\begin{equation}
    \left[\hat{g}\bm \Delta \right]_{\textbf{k}}^{\mu_1\mu_2}
        = \left[ T(g) \right]^{\mu_1 \mu_3} \left[ T(g) \right]^{\mu_2 \mu_4} \bm \Delta_{T_{\textbf{k}}(g^{-1}){\textbf{k}}}^{\mu_3\mu_4}.
\end{equation}

\begin{figure*}
    \centering
    \includegraphics[width=\linewidth]{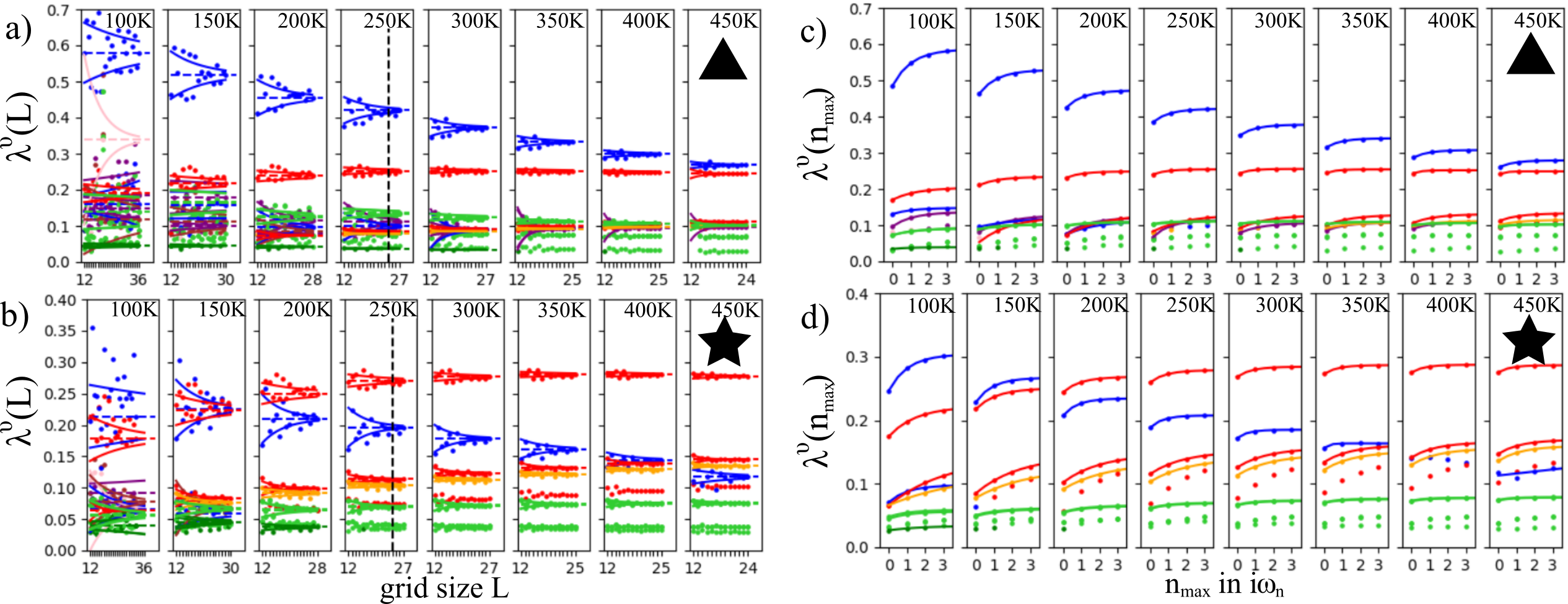}
    \caption{(Color online) Grid size dependence of the eigenvalues  for various temperatures a) for the triangle point in parameter space where the dominant eigenvalue is B$_{1g}^+$ and b) for the star point where the dominant eigenvalue is A$_{2g}^-$. Convergence of the eigenvalues in the number of fermionic frequencies $n_{\text{max}}$ for the $15\times15\times2$ $\textbf{q}$-grid for c) the triangle point in parameter space and d) for the star point.}
    \label{fig:k_w_conv}
\end{figure*}

\begin{table}[t]
    \caption[Character table for the $D_{4h}$ space group.]{Character table for the $D_{4h}$ space group. Each row is assigned to an irrep. Each column is a class of symmetry operations with dimension as the prefactor. Each operations can be expressed in terms of the generators of Table~\ref{tab:d4h_generators}. $E$ is identity, $C_4$ is a $\pi/2$ rotation around the z-axis, $C_2$, $C_2'$ and $C_2''$ are $\pi$ rotations around the $z$-axis, the $x$- or $y$-axis and the ($x+y$)- or ($x-y$)-axis, $i$ is inversion and $S_4$, $\sigma_h$, $\sigma_v$, $\sigma_d$ are $C_4$, $C_2$, $C_2'$, $C_2''$ times $i$, respectively.}
\begin{tabular}{c| cccccccccc}
    \label{tab:character_table}
    $D_{4h}$    & $E$ & $2C_4$    & $C_2$ & $2C^{'}_2$   & $2C^{''}_2$  & $i$ & $2S_4$    & $\sigma_h$  & 2$\sigma_v$   & $2\sigma_d$  \\
    \hline
    A$_{1g}$ & 1 & 1 & 1 & 1 & 1 & 1 & 1 & 1 & 1 & 1 \\
    A$_{2g}$ & 1 & 1 & 1 & -1 & -1 & 1 & 1 & 1 & -1 & -1 \\
    B$_{1g}$ & 1 & -1 & 1 & 1 & -1 & 1 & -1 & 1 & 1 & -1 \\
    B$_{2g}$ & 1 & -1 & 1 & -1 & 1 & 1 & -1 & 1 & -1 & 1 \\
    E$_g$ & 2 & 0 & -2 & 0 & 0 & 2 & 0 & -2 & 0 & 0 \\
    A$_{1u}$ & 1 & 1 & 1 & 1 & 1 & -1 & -1 & -1 & -1 & -1 \\
    A$_{2u}$ & 1 & 1 & 1 & -1 & -1 & -1 & -1 & -1 & 1 & 1 \\
    B$_{1u}$ & 1 & -1 & 1 & 1 & -1 & -1 & 1 & -1 & -1 & 1 \\
    B$_{2u}$ & 1 & -1 & 1 & -1 & 1 & -1 & 1 & -1 & 1 & -1 \\
    E$_u$ & 2 & 0 & -2 & 0 & 0 & -2 & 0 & 2 & 0 & 0 
\end{tabular}
\end{table}

Any space group can be decomposed into a set of irreducible representations (irreps). They characterize the fundamental ways objects transform under the operations of the group. For an object $\phi^p$ which transforms like the one-dimensional irrep $p$, we have $\hat{g}\phi^p = \chi^p(g)\phi^p$ where $\chi^p(g)$ is the character of the operation $g$ for the $p$ irrep. 
The $n$-dimensional irreps are characterized by $n$ independent objects $\{\phi^p_i\}_{i\in \mathbb{N}}$, which transforms like
\begin{equation}
    \hat{g} \phi^p_i = \sum_j \chi^p_{ij}\phi^p_j.
\end{equation}
In these cases, the character of the operation $g$ for the $p$ irrep is given as $\chi^p(g) = \text{Tr}\left[\chi^p_{ij} \right]$. For the $D_{4h}$ space group, the character table is printed in Table~\ref{tab:character_table}. 

The projector operator $\hat{\mathcal{P}}^p$ selects only the contribution associated with the $p$ irrep. It is defined and acts as
\begin{equation}
    \hat{\mathcal{P}}^p = \frac{1}{N_G} \sum_{g \in G} \left[ \chi^p(g)\right]^* \hat{g}, \quad
    \hat{\mathcal{P}}^p \bm \phi^{q} = \delta_{pq} \bm \phi^{p}
\end{equation}
where $N_G$ is the number of symmetry operations of the group. To characterize the symmetry of an order parameter, we find the irrep that represents its transformation properties under the operations of the group.
There is a set of basis functions for each of the four quantum numbers characterizing a gap function. These basis functions can be classified  with the irrep that represents how they transform.

\section{\label{sec:eig_conv}Convergence of the gap functions.}
While the gap functions considered have orbital and spin basis that have a fixed number of states, the relative momentum and frequency are discretized so their convergence needs to be studied.
In momentum space, temperature introduces a broadening in the Green functions, thus high temperatures do not necessitate large numbers of $\textbf{k}$-points.
Lowering temperatures however, the Green functions become increasingly sharp and better resolutions are required.
In \fref{fig:k_w_conv}~a) and b), we study the $\blacktriangle$ (B$_{1g}^+$ leading) and $\bigstar$  (A$_{2g}^-$ leading) points of \fref{fig:new_phase_diag}, respectively.
For temperatures going from 100 to 450~K, we present the eigenvalues of the five leading eigenvectors as a function of the grid size $L$ characterizing $L \times L \times 2$ $\textbf{q}$-grids for the susceptibilities.

These $\textbf{q}$-grid convergence calculations are all performed with two fermionic frequencies.
The eigenvalues are classified by their global irrep, with the same colors as in \fref{fig:b1g_a2g_temp_dep}.
For each irreps $\nu$ and each temperature $T$, the eigenvalue at the largest $L$ is presumed converged and we fit an exponential on the other values to define the error $\Delta \lambda_{\textbf{q}}^{\nu}(T)$ associated to $\textbf{q}$-point convergence.
At high temperatures than $450$~K, we simply take the eigenvalues for $L=24$ as they are well converged.
The phase diagram of \fref{fig:new_phase_diag} is using irreps at 250~K for $L=24$, indicated by the dash lines.

Since the pairing interactions are dynamical and delayed in time, the fermionic frequencies should also be converged.
However, solving the Eliashberg equation for a large number of fermionic frequencies in a multi-orbital and strongly $\bm q$-dependent system like SRO is very challenging.
Because Matsubara frequencies are spaced proportionnally to temperature, they are very spreaded at large temperature and the first fermionic Matsubara frequencies $\pm i\omega_0$ are sufficient to capture all the dynamics of the gap functions.
At lower temperature, again, the proximity of the frequencies to the origin implies that a lot of frequencies are required to have accurate eigenvalues.
In \fref{fig:k_w_conv}~c) and d), we present frequency convergences of the superconducting eigenvalues with respect to temperatures ranging from 100 to 450~K.
Each eigenvector is again assigned a global irrep with color defined in \fref{fig:b1g_a2g_temp_dep}.
For each temperature $T$ and each irrep $\nu$, we fit an exponential function and look at the value at infinity to define the error $\Delta \lambda_{i\omega_n}^{\nu}(T)$ associated to the frequency convergence.
The error bars in \fref{fig:b1g_a2g_temp_dep} are given as the sum of the errors $\Delta\lambda_{\textbf{q}}^{\nu}(T) + \Delta\lambda_{i\omega_n}^{\nu}(T)$.

\end{document}